\documentclass[fleqn,usenatbib]{mnras}

\usepackage{newtxtext,newtxmath}

\usepackage[T1]{fontenc}
\usepackage{ae,aecompl}

\usepackage{graphicx}	
\usepackage{amsmath}	
\usepackage{amssymb}	
\usepackage{multicol}        
\usepackage{bm}		
\usepackage{pdflscape}	
\usepackage{soul}
\usepackage{enumitem}
\usepackage{pdflscape}	
\usepackage[dvipsnames]{xcolor}
\usepackage[normalem]{ulem}

\newcommand{\Mpc}{\mathrm{Mpc}}
\newcommand{\Myr}{\mathrm{Myr}}

\newcommand{\kpc}{\mathrm{kpc}}
\newcommand{\pc}{\mathrm{pc}}
\newcommand{\kmpers}{\mathrm{km} \, \mathrm{s}^{-1}}
\newcommand{\cmcube}{\mathrm{cm}^{-3}}

\newcommand{\protonmass}{m_\mathrm{p}}
\newcommand{\mppercmcube}{\protonmass \, \cmcube}

\newcommand{\Msol}{\textup{M}_\mathrm{\sun}}
\newcommand{\Zsol}{\textup{Z}_\mathrm{\sun}}

\newcommand{\K}{\mathrm{K}}
\newcommand{\xMsol}[2]{\ensuremath{{#1}\times 10^{#2} \,\Msol}}
\newcommand{\xScientific}[2]{\ensuremath{{#1} \times 10^{#2}}}
\newcommand{\xScientificErrorBar}[4]{\ensuremath{{#1}^{+#2}_{-#3} \times 10^{#4}}}

\newcommand{\mdm}{m_{\mathrm{DM}}}
\newcommand{\softening}{\Delta x}
\newcommand{\epsilonff}{\epsilon_{\text{ff}}}
\newcommand{\tff}{t_{\text{ff}}}
\newcommand{\SFR}{\mathrm{SFR}}

\newcommand{\Mvir}{M_{200}}

\newcommand{\Mstar}{M_{\star}}
\newcommand{\rvir}{r_{200}}
\newcommand{\magv}{\mathcal{M}_V}
\newcommand{\rhalflight}{r_{1/2, \text{V}}}
\newcommand{\vcirc}{v_{\text{circ}}}

\newcommand{\feh}{[\rm Fe / H]}
\newcommand{\averagefeh}{\langle [\rm Fe / H] \rangle}

\newcommand{\averageOmetallicity}{\langle 12+\log({\rm O/H}) \rangle}
\newcommand{\massloading}{\eta_{M}}

\newcommand{\nh}{n_{\mathrm{H}}}

\newcommand{\Mhi}{M_{\mathrm{\hi}}}

\newcommand{\hi}{\text{H}\,\textsc{i}}
\newcommand{\hmol}{\mathrm{H}_\mathrm{2}}
\newcommand{\hii}{\mathrm{H}\,{\textsc{ii}}}
\newcommand{\hei}{\mathrm{He}\,{\textsc{i}}}
\newcommand{\heii}{\mathrm{He}\,{\textsc{ii}}}
\newcommand{\heiii}{\mathrm{He}\,{\textsc{iii}}}
\newcommand{\MstarMhi}{\Mstar - \Mhi}
\newcommand{\MstarMhalo}{\Mstar - \Mvir}

\graphicspath{{./}{figs/}}

\title[EDGE2]
{\textsc{edge}: the emergence of dwarf galaxy scaling relations from cosmological radiation-hydrodynamics simulations}

\author[M. P. Rey et al.]
{Martin P. Rey,$^{1, 2}$\thanks{E-mail: \href{mpr47@bath.ac.uk}{mpr47@bath.ac.uk}}
Ethan Taylor,$^{3}$ Emily I. Gray,$^{3}$ Stacy Y. Kim,$^{4}$ Eric P. Andersson,$^{5}$ Andrew Pontzen,$^{6}$ \newauthor Oscar Agertz,$^{7}$ Justin I. Read,$^{3}$ Corentin Cadiou,$^{8, 7}$ Robert M. Yates,$^{9}$ Matthew D. A. Orkney,$^{10, 11}$ Dirk \newauthor Scholte,$^{12}$ Am\'elie Saintonge,$^{13, 14}$ Joseph Breneman,$^{15}$ Kristen B. W. McQuinn,$^{15, 16}$ Claudia Muni,$^{13}$ \newauthor and Payel Das$^{3}$
\vspace{0.8mm}
\\
$^{1}$ Department of Physics, University of Bath,  Claverton Down, Bath, BA2 7AY, UK \\ 
$^{2}$ Sub-department of Astrophysics, University of Oxford, DWB, Keble Road, Oxford OX1 3RH, UK \\ 
$^{3}$ Department of Physics, University of Surrey, Guildford GU2 7XH, UK \\ 
$^{4}$ Carnegie Theoretical Astrophysics Center, Carnegie Observatories, 813 Santa Barbara St, Pasadena, CA 91106, USA \\
$^{5}$ Department of Astrophysics, American Museum of Natural History, 200 Central Park West, New York, NY 10024, USA \\
$^{6}$ Institute for Computational Cosmology, Department of Physics, Durham University, South Road, Durham, DH1 3LE, UK \\
$^{7}$ Lund Observatory, Division of Astrophysics, Department of Physics, Lund University, Box 43, SE-221 00 Lund, Sweden \\
$^{8}$ Institut d’Astrophysique de Paris, Sorbonne Universit´es, CNRS, UMR 7095, 98 bis bd Arago, 75014 Paris, France \\
$^{9}$ Centre for Astrophysics Research, University of Hertfordshire, Hatfield, AL10 9AB, UK \\
$^{10}$ Institut de Ciencies del Cosmos (ICCUB), Universitat de Barcelona (IEEC-UB), Martí i Franquès 1, E08028 Barcelona, Spain \\
$^{11}$ Institut d'Estudis Espacials de Catalunya (IEEC), E-08034 Barcelona, Spain \\
$^{12}$ Institute for Astronomy, University of Edinburgh, Royal Observatory, Edinburgh EH9 3HJ, UK \\
$^{13}$ Department of Physics and Astronomy, University College London, Gower Street, London, WC1E 6BT, UK \\
$^{14}$ Max-Planck-Institut für Radioastronomie (MPIfR), Auf dem Hügel 69, D-53121 Bonn, Germany \\
$^{15}$ Department of Physics and Astronomy, Rutgers, The State University of New Jersey, 136 Frelinghuysen Rd, Piscataway, NJ 08854, USA \\
$^{16}$ Space Telescope Science Institute, 3700 San Martin Drive, Baltimore, MD 21218, USA
}

\date{Submitted to MNRAS}

\pubyear{2025}

\begin{document}
\label{firstpage}
\pagerange{\pageref{firstpage}--\pageref{lastpage}}
\maketitle

\begin{abstract}  
 We present a new suite of \textsc{edge} (`Engineering Dwarfs at Galaxy formation's Edge') cosmological zoom simulations. The suite includes 15 radiation-hydrodynamical dwarf galaxies covering the ultra-faint to the dwarf irregular regime ($10^4 \leq \Mstar(z=0) \leq 10^8 \, \Msol$) to enable comparisons with observed scaling relations. Each object in the suite is evolved at high resolution ($\approx 3 \, \pc$) and includes stellar radiation, winds and supernova feedback channels. We compare with previous \textsc{edge} simulations without radiation, finding that radiative feedback results in significantly weaker galactic outflows. This generalizes our previous findings to a wide mass range, and reveals that the effect is most significant at low $\Mstar$. Despite this difference, stellar masses stay within a factor of two of each other, and key scaling relations of dwarf galaxies (size-mass, neutral gas-stellar mass, gas-phase mass-metallicity) emerge correctly in both simulation suites. Only the stellar mass -- stellar metallicity relation is strongly sensitive to the change in feedback. This highlights how obtaining statistical samples of dwarf galaxy stellar abundances with next-generation spectrographs will be key to probing and constraining the baryon cycle of dwarf galaxies.
\end{abstract}




\section{Introduction} \label{sec:intro}

Small `dwarf' galaxies are sensitive probes of galaxy formation and dark matter physics. Within their shallow gravitational potential wells, energetic stellar processes (e.g. supernovae, stellar winds and radiation, collectively `feedback') efficiently drive galactic outflows (see \citealt{Collins2022} for a review), making dwarf galaxies an ideal laboratory to study how feedback processes regulate the growth of galaxies over cosmic time (e.g. \citealt{Naab2017} for a review). Furthermore, the existence of their low-mass dark matter haloes offer leading constraints on the `coldness' of dark matter (e.g. \citealt{Nadler2020}) while their internal dynamics directly probe if dark matter is self-interacting (see \citealt{Pontzen2014, Bullock2017, Sales2022} for reviews). 

The same low mass that makes dwarf galaxies sensitive to important physical processes has historically also made them challenging to observe and characterize. But the advent of wide-field, deep photometric surveys has now revealed an ever-growing number of classical and ultra-faint dwarf galaxies (\citealt{Simon2019} for a review). This has in turn enabled dedicated programs to study the stellar chemistry and kinematics of these systems (e.g. \citealt{Ji2016b, Ji2016c, Ji2016a, Ji2020, Longeard2018, Buttry2022, Simon2023, Bruce2023, Hansen2024}). Furthermore, the new generation of imaging can now be combined with upgrades in radio capabilities (e.g. MeerKAT) and multi-object spectroscopy (e.g. DESI). This has fundamentally changed our ability to characterize the gas contents and gas-phase metallicities of faint dwarf galaxies, pushing characterization to the very faintest end (e.g. \citealt{McQuinn2021}) and vastly extending statistical samples (e.g. \citealt{Scholte2025}). 

This forward trend in observational capabilities and discoveries will continue in the next 5 years, spearheaded by experiments such as the Vera C. Rubin Observatory which is expected to provide a near-complete census of faint ($\magv \approx -6$) dwarf galaxies within 5 Mpc around the Milky Way (\citealt{Mutlu-Pakdil2021}) and the \textit{Nancy Roman Space Telescope}. These new photometric datasets will be complemented by forthcoming spectroscopic information, with for example the 4MOST-Dwarf survey expected to obtain chemical abundances for > 50,000 dwarf galaxy stars (\citealt{Skuladottir2023a}). 

Interpreting these new datasets requires us to develop detailed models of dwarf galaxy formation that can match the data's new statistical power while retaining enough physical fidelity to make robust predictions. This is a challenging task. The small sizes of dwarf galaxies ($\approx 100\, \pc$) require high ($\approx 10 \, \pc$) numerical resolution to resolve their gas reservoirs and star-forming regions over cosmic time. But, at the same time, modelling the full cosmological history of a dwarf galaxy is also essential, as each specific formation scenario plays a key role in setting the $z=0$ properties and observables (e.g. \citealt{Benitez-Llambay2015, Benitez-Llambay2021, Fitts2017, Rey2019UFDScatter, Rey2020, Wright2019, Katz2020, Tarumi2021StellarHalo, Herzog2023}). Simulating many objects, in a cosmological context, and with a high resolution place conflicting demands on the available computing power.  

None the less, the field has made great progress in addressing these challenges in recent years. Improvements in code efficiency and computational power now allow zoomed cosmological simulations of field faint dwarf galaxies with $\approx \pc$ resolution over the full Hubble time (e.g. \citealt{Wheeler2019, Agertz2020EDGE,Gutcke2021CosmologicalSim, Go2025}). These gains go beyond an incremental improvement in resolution, as they allow us to resolve explicitly the cooling radius of supernovae (SNe) explosions, in turn enabling a robust modelling of the emerging momentum and its coupling to the gas (e.g. \citealt{Kimm2015, Kim2015, Martizzi2015, Ohlin2019}). 

In parallel, model improvements are significantly increasing physical fidelity. For example, radiation-hydrodynamics simulations can now account for stellar radiative heating over the cosmological history of a dwarf galaxy (\citealt{Agertz2020EDGE}). Unlike previous subgrid implementation that were attempted to capture radiative effects locally (e.g. \citealt{Agertz2013, Hopkins2018}), explicit radiative transfer gives a physical account of how photons ionize, heat and inject momentum in the gas both locally, and non-locally across the interstellar medium (ISM) and circumgalactic medium (CGM). Beyond radiative transfer, cosmological simulations of dwarf galaxies have also made strides to explicitly sample the initial mass function (IMF) with individual stars (e.g. \citealt{Gutcke2021CosmologicalSim, Andersson2025}), unlocking direct comparisons with resolved-star observations of dwarf galaxies in the Local Volume. And high-resolution cosmological simulations have also started to incorporate detailed models of cosmic ray (e.g. \citealt{Martin-Alvarez2023}) and black hole feedback (e.g. \citealt{Koudmani2021, Koudmani2022, Arjona-Galvez2024}), starting to quantify the importance of these processes at the faint end of galaxy formation at $z=0$. 

Furthermore, lessons learned from these detailed numerical simulations directly inform the development of more realistic semi-analytical and semi-empirical models of dwarf galaxy formation (e.g. \citealt{Benitez-Llambay2020, Wang2021, Kravtsov2022, OLeary2023, Kim2024, Monzon2024}). The statistical power of these models will be invaluable to interpret the populations of dwarf galaxies from next-generation surveys.

The \textsc{edge} collaboration has been a key contributor to these advances. At the core of our approach is the undertaking of suites of highly-detailed, high-resolution cosmological simulations (e.g. \citealt{Agertz2020EDGE, Rey2020, Prgomet2022, Andersson2025}). These simulations are then leveraged to pinpoint the key factors shaping dwarf galaxy observables (\citealt{Rey2019UFDScatter,Rey2022EDGEHI,Rey2024EDGERCs, Goater2024, Gray2025}), and dark matter properties (\citealt{Orkney2021, Orkney2022, Orkney2023}). Learnings from these simulations are then encapsulated into a semi-analytical model to enable the modelling of statistical populations (\citealt{Kim2024}).

In this paper, we present the new generation of \textsc{edge} cosmological zoomed simulations that will be the cornerstone of our future interpretation efforts. This new suite extends the state of the art in multiple ways. First, in physical fidelity. We maintain the characteristic high resolution of the original \textsc{edge} simulations ($\softening \approx 3 \, \pc$; $\mdm \approx 950\, \Msol$) to resolve individual SNe explosions, and combine it with a systematic use of radiative transfer for an explicit account of radiative feedback from stars. This improves the physical realism of the ISM over many more objects than the single dwarf galaxy presented in \citet{Agertz2020EDGE}.

Second, we extend the scale of the simulation suite. We simulate > 15 different dwarf galaxies from ultra-faint dwarf galaxies ($\Mvir \approx 10^9 \, \Msol$; $\Mstar \approx 10^4 \, \Msol$ at $z=0$) to dwarf irregulars ($\Mvir \approx 10^{10} \, \Msol$; $\Mstar \approx 10^7 \, \Msol$). All simulations share the same numerical resolution and feedback physics. Bridging to higher-mass objects without compromising on the physical fidelity opens new regimes of comparisons with observational data, as well as providing a larger range of data points to calibrate future semi-analytical models. 

This combination of scale and fidelity is unique. Furthermore, the ability to compare between two fundamentally different feedback models (without and with photo-ionization feedback in \textsc{edge1} and \textsc{edge2}, respectively) also allows for a new understanding of the theoretical uncertainties in our results. We describe the \textsc{edge2} model and its updates in Section~\ref{sec:methods} and show how it affects the ISM and outflows of dwarf galaxies in Section~\ref{sec:rtimpact}. Section~\ref{sec:scalingrelations} shows how dwarf galaxy scaling relations naturally emerge from \textsc{edge} modelling and their robustness to a large change in sub-grid physics modelling. Section~\ref{sec:numerics} discusses the new numerical challenges associated with explicitly modelling stellar radiation, and we conclude in Section~\ref{sec:conclusion}.

\section{EDGE2 numerical setup} \label{sec:methods}

In this section, we describe the \textsc{edge2} numerical methods and the suite of zoomed cosmological simulations. We briefly summarize the main differences between \textsc{edge1} and \textsc{edge2} in the bullet points below, and refer the reader to \citet{Rey2020} for a more in-depth description of the \textsc{edge1} model. 
\begin{itemize}[leftmargin=2.5mm]
  \item \textsc{edge2} uses non-equilibrium primordial and molecular chemistry via the implementation of \citet{Rosdahl2013} and \citet{Nickerson2018}, as opposed to \textsc{edge1}'s equilibrium cooling from \citet{Courty2004}. 
  \item \textsc{edge2} accounts for stellar photo-ionization and photo-heating using radiative transfer following the setup described in \citet{Agertz2020EDGE}. 
  \item \textsc{edge2} updates the feedback budget to be normalized to a \citet{Kroupa2001} IMF integrated between $0.1$ and $100\, \Msol$ to be consistent with spectral energy distribution (SED) libraries. \textsc{edge1} used a \citet{Chabrier2003} IMF integrated between $0.5$ and $100 \, \Msol$.
  \item \textsc{edge2} tracks the enrichment of eight individual elements (C, N, O, Mg, Al, Si, Fe and Eu) using the \textsc{NuGRID} yields for core-collapse supernovae (CCSNe; \citealt{Pignatari2016, Ritter2018}). \textsc{edge1} tracked only O and Fe and used the \citet{Woosley2007} CCSNe yields for \textsc{edge1}.
  \item \textsc{edge2} updates the Type-Ia SN (SNeIa) model to use a delay-time distribution from \citet{Maoz2012} and yields from \citet{Seitenzahl2013}. \textsc{edge1} used the binary mass function from \citet{Raiteri1996} and yields originally from \citet{Thielemann1986}.
  \item \textsc{edge2} updates the UV background (UVB) from a modified \citet{Haardt1996} to the \citet{Faucher-Giguere2020} photo-ionization and photo-heating rates.
  \item  \textsc{edge2} uses a new high-cadence infrastructure to allow the dynamical tracking of gas and stars on short timescales with tracer particles (\citealt{Cadiou2019}). 
  \item \textsc{edge2} adds five new objects at higher masses ($\Mvir \approx 10^{10}\, \Msol$) to the original suite of initial conditions to grow our sample size (\citealt{Muni2024EDGECores}). Table~\ref{tab:simulations} in Appendix~\ref{app:table} provides all the properties of each object in the \textsc{edge1} and \textsc{edge2} models. 
\end{itemize}

\begin{figure}
  \centering
    \includegraphics[width=\columnwidth]{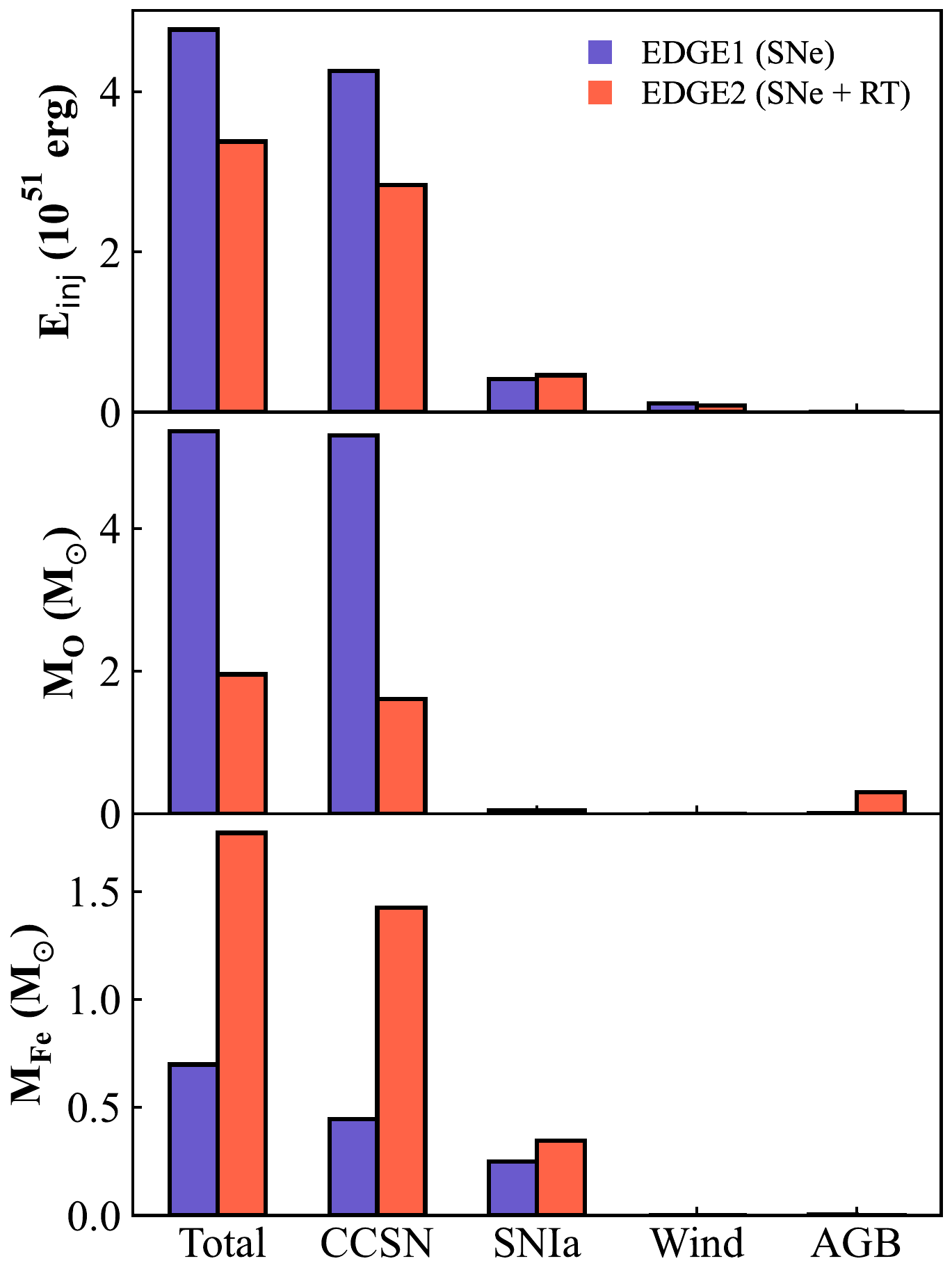}

    \caption{Energy (top), total oxygen (middle), and iron (bottom) injected by a $300 \, \Msol$ stellar particle over its lifetime in the \textsc{edge1} and \textsc{edge2} models. The total is the sum of each individual subcomponent and does not include the radiation budget to enable one-to-one comparison between \textsc{edge2} and \textsc{edge1}. Due to the change in IMF, yields and SNeIa modelling, \textsc{edge2} injects $\approx 50\%$ less CCSN energy, half the oxygen, and twice the iron per stellar population.}
    \label{fig:modelcomparison}
\end{figure}

Figure~\ref{fig:modelcomparison} summarizes how these changes affect the energy injection and metal production budget integrated over 14 billion years in a $300 \, \Msol$, $0.01 \, \Zsol$ stellar population. We describe these changes further in the next sections. 

\subsection{Initial conditions and resolution} \label{sec:sec:ics}
All of our simulated galaxies are evolved to $z=0$ from cosmological zoomed initial conditions constructed with the \textsc{genetic} software (\citealt{Stopyra2021}). All initial conditions use a flat $\Lambda$-cold-dark-matter transfer function generated with \textsc{CAMB} (\citealt{Lewis2000}). Original \textsc{edge1} initial conditions assume a \citet{PlanckCollaboration2014} cosmology with parameters $\Omega_{m} = 0.3086$, $\Omega_{\Lambda} = 0.6914$, $h = 0.6777$, $\sigma_8 = 0.8288$, and $n_s = 0.9611$. The new initial conditions we introduce below use a \citet{PlanckCollaboration2020} cosmology with parameters $\Omega_{m} = 0.3158$, $\Omega_{\Lambda} = 0.6842$, $h = 0.6732$, $\sigma_8 = 0.8117$, and $n_s = 0.9660$. These cosmologies are compatible within 68\% confidence intervals and do not lead to major differences. Initial conditions are evolved analytically using first-order perturbation theory (\citealt{Zeldovich1970}) until $z=99$, at which point numerical integration starts.

To build the original \textsc{edge1} zoomed regions, we first simulate a random cosmological volume (50 Mpc at $512^3$ resolution) only accounting for gravity and dark matter. We then open a cubic zoomed region of 11.5 Mpc (resolution equivalent to $2048^3$, $\mdm = \xMsol{3.8}{6}$) centred on the largest void in the simulation. Having re-simulated this first zoom level, we identify haloes within the $z=0$ void using the \textsc{hop} halo finder (as in \citealt{Eisenstein1998}) and keep only isolated central haloes with no neighbours more massive than them within $5\rvir$. Here, $\rvir$ is the radius encompassing 200 times the critical density of the Universe. We select six dark matter haloes spanning a wide window in present-day halo mass ($\Mvir = \xScientific{1.5}{9}$ to $\xMsol{7}{9}$, where $\Mvir$ is the mass enclosed in a sphere of radius $\rvir$), track them back to the initial conditions again, and generate initial conditions at our final zoomed resolution ($\mdm = 940\, \Msol$, equivalent to $16384^3$). 

New initial conditions presented in this paper follow the same procedure as previously followed with small operational differences (see also \citealt{Muni2024EDGECores}). The original volume is 100 Mpc, first simulated using $256^3$ particles. We centre on the biggest void identified by running a `paired' simulation (see \citealt{Stopyra2021Voids}) and select a $\approx 11.5 \, \Mpc$ cubic subvolume in which the resolution reaches $\mdm = \xMsol{3.9}{6}$. We select three haloes with present-day halo masses from $\Mvir = \xScientific{7}{9}$ to $\xMsol{1}{10}$ and increase the dark matter resolution in their Lagrangian region to $\mdm = 953 \, \Msol$. Resolution between the original and updated initial conditions match to $\approx 1\%$, with slight shift arising from the slight differences in cosmological parameters.  

We also use the `genetic modification' approach to introduce controlled changes in cosmological initial conditions (see \citealt{Roth2016, Rey2018, Stopyra2021, Cadiou2021AngMomGM} for more information about the method and the range of possible modifications). Using this approach, all initial conditions are genetically-modified to ensure that the Lagrangian region of our dwarf galaxies is (almost) at rest with respect to the cosmological box (\citealt{Pontzen2021}). This modification minimizes the streaming of the dwarf galaxy through the simulated volume, mitigating advection errors during integration without affecting its mass growth and local environment.

Furthermore, four of our haloes have genetically-modified initial conditions that craft a targeted change to their mass growth history. One low-mass halo (`Halo 1459') is engineered to form systematically earlier and later, at fixed halo mass today (see \citealt{Rey2019UFDScatter}). Two intermediate mass haloes (`Halo 624', `Halo 600') are modified to respectively be more massive in halo mass overall and to form later at fixed $z=0$ halo mass (see \citealt{Rey2020}). Lastly, a high-mass halo (`Halo 383') is modified to form earlier and later at fixed halo mass today (see \citealt{Gray2025}). Thanks to the nature of the genetic modification algorithm, all other untargeted aspects of each formation scenario is maximally reproduced (e.g. the large-scale environment and cosmic web topology; see e.g. \citealt{Pontzen2017, Rey2019VarianceDMOs} for visuals). All haloes labelled `GM' in Table~\ref{tab:simulations} have had their mass accretion histories genetically modified. Future work will perform controlled modifications to additional haloes in the \textsc{edge2} suite.  

\subsection{Hydrodynamics, radiative transfer and refinement} \label{sec:sec:hydro}

We follow the evolution of dark matter, stars, gas and radiation using the adaptive mesh refinement hydrodynamics code \textsc{ramses-rt} (\citealt{Teyssier2002, Rosdahl2013}). The dynamics of collisionless particles (dark matter and stars) are computed using a multiscale particle-mesh solver estimating densities through a cloud-in-cell approximation (\citealt{Guillet2011}). Fluid dynamics are computed using an HLLC Riemann solver (\citealt{Toro1994}) with the fluid equations closed by assuming an ideal gas equation of state with adiabatic index $\gamma = 5/3$. 

A key addition to this suite is the explicit treatment of the local sources of radiation. We solve the dynamics of the radiation field using the M1 method (\citealt{Rosdahl2013, Rosdahl2015RAMSESRT}) discretizing the light spectrum in six energy bins from the infrared to the UV (same as in \citealt{Agertz2020EDGE}, table 1). Our energy bins are chosen to track radiation that (i) exerts radiation pressure through dust multi-scattering (from 0.1 to 1 eV), (ii) exerts direct radiation pressure (from 1 to 12 eV), (iii) dissociates $\hmol$ (from 12 to 13.6 eV), (iv) ionizes $\hi$ (13.6 to 24.59 eV), (v) ionizes $\hei$ (24.59 to 54.42 eV), and (vi) ionizes $\heii$ (> 54.42 eV). We update the average energies and cross-sections in each band every 10 coarse time-steps by computing the luminosity-weighted average over the spectra of all stellar populations in the simulation volume (see \citealt{Rosdahl2013}). Furthermore, to mitigate computational costs, local radiation around stellar sources is propagated at a reduced speed of light ($c_{\text{reduced}} = c / 100$; see discussions in e.g. \citealt{Gnedin2001, Rosdahl2013}). A time-varying, but spatially uniform, UVB also permeates the simulation box after reionization (see Section~\ref{sec:sec:cooling}). This source is not propagated but contributes to photo-ionization and photo-heating rates.

We use the adaptive nature of \textsc{ramses-rt} to focus computational power in the dense centre of our dwarf galaxies during integration. We split cells when they contain eight dark matter particles, and when their baryonic mass (stars and gas) exceeds $8\, m_{\text{bar}}$, where $m_{\text{bar}} \approx 150 \, \Msol$. Refinement is allowed down to a maximum resolution of $\softening \approx 3$ physical pc. This maximum resolution is achieved throughout the dwarf galaxy's ISM and rapidly degrades as densities decrease (see e.g. \citealt{Pontzen2021}, fig. 3). We maintain an approximately constant resolution in physical units by releasing additional new levels every two-folding of the scale factor (\citealt{Snaith2018}). Refinement is only allowed inside the zoomed region. 

\subsection{Cooling and thermochemistry} \label{sec:sec:cooling}

Compared to the original \textsc{edge} simulations that assumed equilibrium chemistry and cooling, the new simulations of this paper now follow the non-equilibrium thermo-chemistry of the primordial plasma fully coupled to radiative transfer. 

We track the individual ionization fractions of $\hi$, $\hii$, $\hei$, $\heii$, $\heiii$ and $\hmol$. The mass fractions and cooling contributions from atomic species are computed using the semi-implicit solver described in \citet{Rosdahl2013}, accounting for photoionization, collisional ionization and excitation, bremsstrahlung, Compton cooling and heating from the cosmic microwave background, and di-electronic recombination. We assume that ionizing radiation from free-bound ground state recombination radiation is absorbed locally (the on-the-spot approximation; see e.g. \citealt{Nebrin2023}). We follow the formation, advection and destruction of $\hmol$ and its contribution to the cooling rate following the model described in \citet{Nickerson2018}. This model accounts for gas-phase, dust-phase and collisional $\hmol$ formation, destruction through photodissociation, photoionization, and cooling and heating coupled to the local radiation flux. In all runs, we turn off cosmic ray heating and ionization due to the large uncertainties on the cosmic ray ionization rate in dwarf galaxies. We leave to future work an exploration of its impact on the ISM and outflows of dwarf galaxies (see also \citealt{Martin-Alvarez2023}).

Metal cooling uses tabulated cooling rates scaled by metallicity. Below $10^4 \, \K$, we use the \citet{Rosen1995} fine structure cooling rates, while above $10^4 \, \K$, we use \textsc{cloudy} (\citealt{Ferland2017}) tables assuming our updated UVB (\citealt{Faucher-Giguere2020}). These rates are scaled by the total metallicity defined as $Z = (2.09 \, Y_{\rm O} + 1.06 \, Y_{\rm Fe}) / Z_{\odot}$, where $Z_{\odot} = 0.02$. The numerical coefficients are derived from assuming a solar mixture from \citet{Asplund2009} and $Y_{\rm O}$ and $Y_{\rm Fe}$ are the mass fractions of oxygen and iron in the gas cell (see also \citealt{Kim2014}). 

Ionization and heating from the UVB permeates the whole volume in an optically-thin approximation, with high density regions allowed to self-shield from this background. We exponentially damp the UVB photo-ionization and photo-heating rates according to $\exp(- \nh / 10^{-2} \, \cmcube)$ to approximate self-shielding above $\nh \approx 10^{-2} \, \cmcube$ (\citealt{Aubert2010, Rosdahl2012}). Local self-shielding from stellar radiation is treated self-consistently by solving of the coupled equations of radiative transfer and thermo-chemistry. Optically thin UVBs can overheat the early intergalactic medium (\citealt{Onorbe2017}), so we also exponentially damp the UVB as a function of redshift by $\exp(z_{\text{reion}} - z )$. This ensures a continuous ramp up of photo-ionizing and photo-heating rates, rather than the instantaneous switch-on commonly used in UVB tables (e.g. \citealt{Faucher-Giguere2020}). With our choice of $z_{\text{reion}}$, the UVB heating and ionizing rates are at full strength at $z=6$ (see \citealt{Rey2020}).

\subsection{Star formation} \label{sec:sec:sec:sf}

The star formation modelling remains unchanged compared to \citet{Agertz2020EDGE}. We model star formation following a Kennicutt-Schmidt law (\citealt{Schmidt1959, Kennicutt1998}):

\begin{equation}
  \label{eq:schmidt}
  \dot{\rho}_{*}=\epsilonff \frac{\rho_{g}}{\tff}\ \text{for gas cells with} \,\,\rho_{g}>\rho_{\star} \ \text{and} \ T_g < T_{\star} \, ,
\end{equation}
where $\dot{\rho}_{*}$ is the instantaneous star formation rate (SFR) in a gas cell, $\epsilonff$ is the star formation efficiency per free-fall time, $\rho_{g}$ and $T_g$ are the gas cell density and temperature, and $\tff=\sqrt{3\pi/32G\rho}$ is the local free-fall time. We choose $\rho_\star = 300 \, \mppercmcube$, corresponding roughly to the average density of giant molecular clouds and to the density needed to form a $300 \, \Msol$ stellar population at our resolution (see below). We pick a constant $\epsilonff = 10\%$ which, at $\approx3 \, \pc$ resolution, has been shown to reproduce key star formation observables such as the density and energy power spectra of the ISM in Milky Way-like galaxies (\citealt{Grisdale2017}), the properties of giant molecular clouds (\citealt{Grisdale2018}), and the observed star formation efficiency in molecular clouds (\citealt{Grisdale2019}). This value is also compatible with typical predictions from multi-free-fall models and star formation simulations with similar resolution (e.g. \citealt{Federrath2012, Padoan2012}). We also implement a temperature threshold of $T_{\star} = 1000 \, \K$, which is higher than in \textsc{edge1} (100 K) to compensate for the additional heating from radiative feedback. The original motivation for $T_{\star}$ is to ensure that star-forming gas is cold enough to represent (potentially unresolved) molecular gas. As we will see in Section~\ref{sec:numerics:tstar}, however, the choice of this parameter leads to unwanted numerical effects due to the coupling between star formation and radiative heating.

For every gas cell satisfying our threshold conditions, we sample Equation~\eqref{eq:schmidt} stochastically with a Poisson process so that the expectation of the local SFR equals $\dot{\rho}_{*}$ (\citealt{Rasera2006}). Stellar particles have initial masses of $300 \ \Msol$ and assume a \citet{Kroupa2001} IMF. This mass is at the boundary of ensuring that stochastic effects are averaged over at our resolution (see e.g. \citealt{Smith2021IMF}). This is key for radiative feedback which is still modelled as an IMF-averaged process (as opposed to SN feedback for which SNe are individually sampled). We will explore the consequences of these choices in the future with a star-by-star \textsc{edge} implementation (\citealt{Andersson2025}).

\subsection{Stellar feedback} \label{sec:sec:sec:feedback}

A key aspect of the \textsc{edge} model is the detailed account of feedback from massive stars. We account for CCSNe, SNeIa, fast winds from massive stars, slow winds from asymptotic giant branch (AGB) stars, and now radiative feedback. 

Our model tracks the main-sequence lifetime of different progenitors within a stellar particle (\citealt{Agertz2011}), ensuring that stars of different masses inject their feedback on their relevant main-sequence timescale. We model SN explosions as discrete events, computing at each simulation time-step the number of stars exiting the main sequence to turn into CCSNe (equations 6 in \citealt{Agertz2013}). This IMF-averaged number is then randomly sampled through a Poisson process to obtain a discrete number of explosions (\citealt{Agertz2020EDGE}). To maintain consistency with the IMF used for radiative feedback, we normalize this number to the \citet{Kroupa2001} IMF integrated between $0.1$ and $100 \, \Msol$. This differs from \textsc{edge1} which used a \citet{Chabrier2003} IMF integrated between $0.5$ and $100 \, \Msol$, leading to a $\approx 40\%$ decrease in the number of CCSNe per stellar population (Figure~\ref{fig:modelcomparison}, top)

Furthermore, to match the explodability assumptions made by our new chemical enrichment model (Section~\ref{sec:sec:sec:yields}), we now assume that massive stars between $8$ and $30 \, \Msol$ explode as CCSNe, while stars above directly collapse into black holes without releasing energy. Previous \textsc{edge1} modelling assumed that all stars between $8\, \Msol$ and $40\, \Msol$ exploded. When convolved with the IMF, this further reduces the energy budget of CCSNe by $\approx 10$ per cent.

Furthermore, SNeIa now follow a delay time distribution according to \citet{Agertz2020Vintergatan}. We assume that SNeIa start when their parent stellar particle has an age of $38.4\, \Myr$ (the main-sequence lifetime of $8\, \Msol$ star to form the first non-exploding degenerate progenitor) and explode with a rate that initially peaks at $\xScientific{2.6}{-13} \, \Msol^{-1}$ and decays as $t^{-1.12}$ following the empirical determination of \citet{Maoz2012} for field galaxies. As for CCSNe, SNeIa are individually sampled from this rate using a Poisson process. In \textsc{edge1}, we instead modelled the SNeIa rate by integrating over the IMF of secondary, binary companions between $1$ and $8\, \Msol$ (\citealt{Raiteri1996}; see \citealt{Agertz2013} for a detailed description). This modernization leads to a higher number of SNeIa per stellar population (Figure~\ref{fig:modelcomparison}, top), which balances the $40\%$ decrease induced by the change in IMF normalization.

For both SN types, we directly inject thermal energy ($E_{\text{SN}} = 10^{51} \, \text{erg}$ constant at all times) when the cooling radius of the individual SN event is resolved by six resolution elements. This allows us to self-consistently follow the build-up of momentum through the Sedov-Taylor phase by solving the hydrodynamics equations rather than relying on a subgrid implementation. When the cooling radius is unresolved, we switch to a momentum injection. This is a binary switch (see also \citealt{Kim2015} for a similar implementation and \citealt{Kimm2014, Kimm2015} for an interpolated approach). Following \citet{Kim2015}, we inject a Sedov-Taylor momentum of $p_{ST} = 2.95 \times 10^5 \, \Msol \, \kmpers$ scaling with density and gas metallicity according to \citealt{Blondin1998}. We do not account for local velocity dynamics (e.g. \citealt{Hopkins2025SNVelocities}). Appendix~\ref{app:resolvedfeedback} shows that $>90\%$ of CCSNe events and $>60\%$ of SNeIa events are resolved by more than $6^3$ ($> 200$) resolution elements, ensuring that we greatly reduce uncertainties associated to the subgrid modelling of SN feedback in an unresolved ISM.  

We also account for wind feedback from massive O and B stars ($m \geq 8 \, \Msol$) during their main sequence, and from stars of masses $0.5-8.0\,\Msol$ when they reach their AGB phase. OB winds return mass and momentum to the ISM according to the metallicity-dependent budget described in \citet{Agertz2013}. AGB winds continuously release mass and metals over the lifetime of a stellar particle following their IMF-averaged mass-loss (eq. 17 in \citealt{Agertz2013}). No changes are made to these models between \textsc{edge1} and \textsc{edge2} and at our characteristic low metallicities, these wind models contribute little to the feedback energy budget (Figure~\ref{fig:modelcomparison}).

In addition to these changes to the SN budget between \textsc{edge1} and \textsc{edge2}, a large change in stellar feedback comes from the addition of radiative feedback in \textsc{edge2}. In this case, each stellar particle injects radiation according to its age and metallicity following a \citet{Bruzual2003} SED. This SED is a `soft' choice for low-metallicity stellar populations, and we explore in Section~\ref{sec:numerics:seds} the impact of a harder SED accounting for binary stars (\citealt{Stanway2016}). 

\subsection{Metal enrichment} \label{sec:sec:sec:yields}

Another novelty in this suite is the tracking of new chemical elements. In addition to O and Fe tracked by the previous model \citep{Agertz2013}, we now account for the production and advection of C, N, Mg, Al, Si, and Eu. These elements were chosen to sample main $\alpha$ elements commonly observed in stars within dwarf galaxies, as well as the r-process element Eu.

As with previous \textsc{edge} simulations, we do not model the formation, feedback and metal enrichment from primordial metal-free stars. Instead, we initialize all simulations with a floor in oxygen metallicity of $Z_{O} =10^{-3} \, \Zsol$ as an approximation for primordial metal enrichment (e.g. \citealt{Greif2010, Jaacks2018, Visbal2020, Brauer2025}). All other elements have vanishing mass fractions initially. Lowering this floor by an order of magnitude leaves the stellar mass and metallicities of our dwarfs unchanged (\citealt{Agertz2020EDGE}). 

Once star formation starts, winds from O and B stars, winds from AGB stars, CCSNe, and SNeIa all inject chemical elements on the same timescale as their feedback (Section~\ref{sec:sec:sec:feedback}). Once injected, each element is advected passively with the gas.  

Yields for CCSNe, OB winds and AGB winds are interpolated from the tables provided by \textsc{NuGrid} \citep{Pignatari2016, Ritter2018}. Discrete injection events from CCSNe are interpolated across the table in progenitor mass and metallicity. Since the \textsc{NuGrid} tables are only available for CCSN progenitor masses between 12 and 25 $\Msol$, we linearly extrapolate to cover the range assumed in our model (8-30 $\Msol$). Future \textsc{edge} simulations will leverage new yield tables explicitly including low progenitor masses (M. Limongi et al. in preparation) to alleviate this issue, as well as updating our assumptions for the CCSN progenitor mass range to reflect new understanding of the super-AGB phase and electron-capture SNe (\citealt{Gil-Pons2018,Limongi2024ECSNe}).

Discrete SNeIa inject chemical elements according to \citet{Seitenzahl2013} assuming a (constant) solar metallicity (metallicity-dependence of these yields is weak). For AGB winds, we compute the IMF-averaged number of AGB stars in a given timestep (\citealt{Agertz2013}) and inject the corresponding yields from \textsc{NuGrid} for each element. OB winds follow the same procedure, but instead inject according to a time-dependent fitting function that was calibrated in \citet{Agertz2013}. 

Combined, these updates have a significant impact on the metal production budget per stellar population. Figure~\ref{fig:modelcomparison} shows that oxygen production is roughly halved in \textsc{edge2} (middle panel), primarily driven by a strong decrease in CCSNe yield. Part of this decrease is due to the change in IMF normalization reducing overall CCSNe numbers by $\approx 40\%$, but is strongly driven by the change in yield tables. \textsc{edge1} tables (\citealt{Woosley2007}) had an exponential scaling of oxygen production with progenitor mass (see e.g. \citealt{Kim2015}, eq. 5) with high-mass stars producing copious amounts of metals. This was known to significantly over-produce oxygen and alpha elements compared to Milky Way observations (e.g. \citealt{Agertz2020Vintergatan}). The scaling with progenitor mass is much weaker in the \textsc{NuGrid} tables. Importantly, iron production is increased by a factor $\approx2.5$ in \textsc{edge2} (bottom panel), again largely driven by the change in CCSN yield table. This reflects the recent trends in yield computations with newer models all producing significantly more iron than previous iterations (e.g. \citealt{Limongi2018}). 

Finally, for r-process production of Eu, we include an effective model for neutron-star mergers (NSNS) inspired by \citet{Naiman2018}. We assume a constant relative fraction $\xScientific{4.6}{-2}$ between NSNS and SNeIa consistent with the NSNS rates observed in the local Universe \citep{LIGO2017BNS}. NSNS events are sampled discretely using the same mechanism as SNeIa with each NSNS injecting a europium yield of $10^{-5}\,\Msol$ \citep{Cote2018} but no other elements. NSNS events do not inject energy, mass or momentum.

\subsection{High-cadence outputs} \label{sec:sec:sec:tracers}

The Eulerian nature of our code prevents us from efficiently tracking the Lagrangian history of gas flows. This is particularly problematic in dwarf galaxies where dynamical times and cooling times in the centre are much shorter than the cadence at which we can save simulation outputs (see e.g. \citealt{Rey2022EDGEHI, Rey2024EDGERCs} for examples of these limitations). 

To remedy to this, \textsc{edge2} uses a Monte-Carlo particle tracer algorithm (\citealt{Cadiou2019}). Tracers are designed to statistically track gas flows and exchange mass with stellar tracers to track the full baryon cycle of gas as it is accreted and recycled through star formation. Tracer dynamics is solved using the same physical solvers as the rest of the simulation, but they do not source or contribute towards hydrodynamical, gravity or radiative forces. We spawn five tracers per high-resolution gas cell, leading to a tracer mass of $223\, \Msol$ comparable to the stellar particle mass. Our smallest dwarf galaxies have $\approx2$ million tracers, while our most massive objects can have up to $20$ million tracers. 

The position and velocity of tracers is stored on disc every $4.5 \, \Myr$ (compared to every $100 \, \Myr$ for full simulation output), along with the density, pressure and gravitational potential of the gas at the location of the tracer. Additionally, the same information with the same cadence is dumped for a selection of 10,000 dark matter particles, selected randomly from the main progenitor of the dwarf at $z=2$. 

\subsection{Data processing and analysis} \label{sec:sec:sec:analysis}

We process each \textsc{edge2} simulation with the \textsc{adaptahop} halo and subhalo finder (\citealt{Aubert2004, Tweed2009}) retaining dark matter structures with more than 100 particles. \textsc{edge1} simulations were processed with the \textsc{hop} halo finder (\citealt{Eisenstein1998}). We match haloes and subhaloes between simulation snapshots to build merger trees using the \textsc{pynbody} (\citealt{Pontzen2013}) and \textsc{tangos} (\citealt{Pontzen2018}) libraries. Halo centres are identified using the shrinking-sphere algorithm (\citealt{Power2003}).

\begin{figure*}
  \centering
    \includegraphics[width=\textwidth]{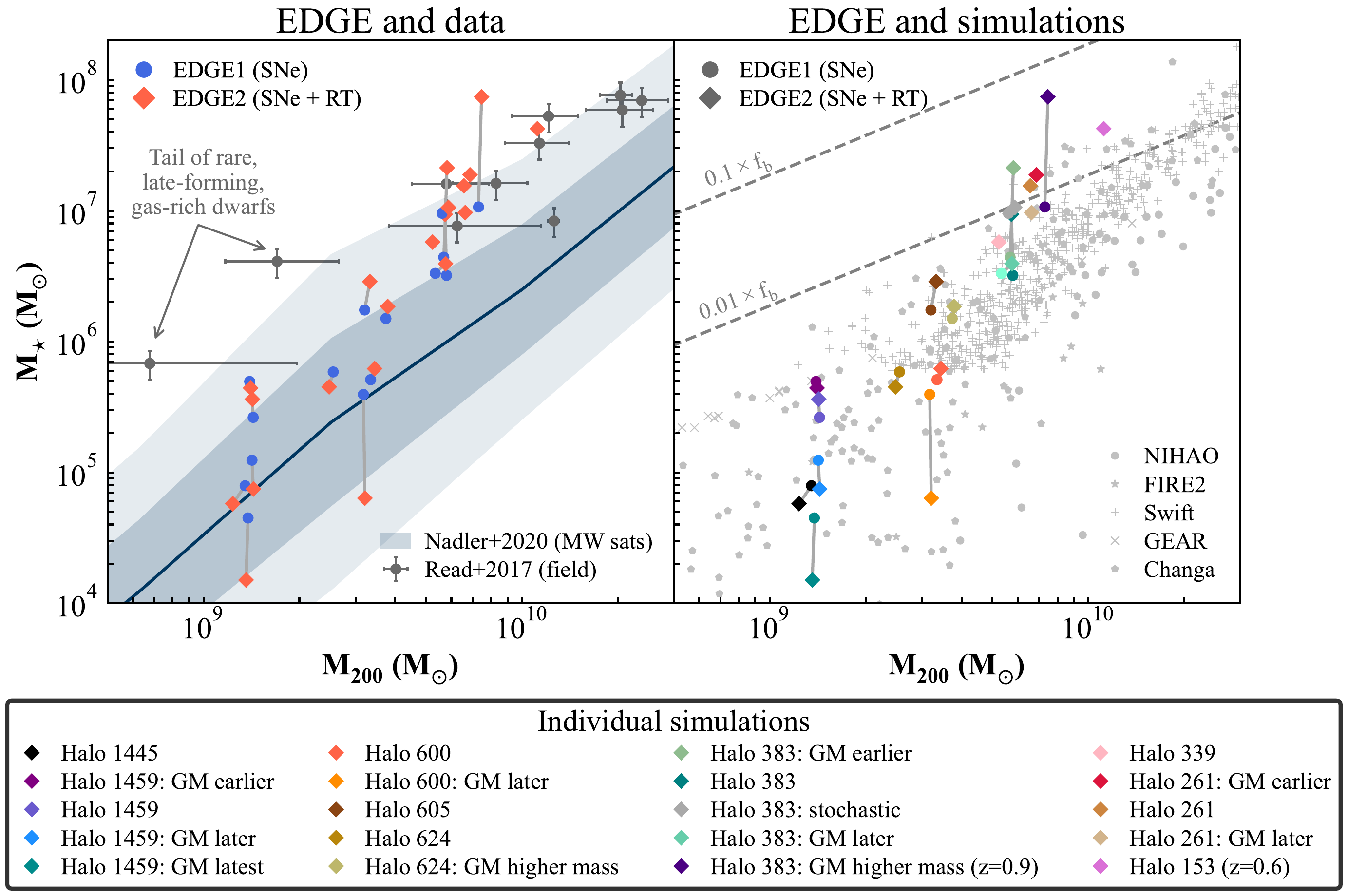}

    \caption{Stellar and halo masses of dwarf galaxies simulated with the \textsc{edge1} and \textsc{edge2} models (circles and diamonds, respectively). Both suites are broadly compatible with the stellar-mass-halo-mass relation inferred from Milky Way satellites at the low-mass end (left-hand panel, blue contours; \citealt{Nadler2020}). Higher-mass objects ($\Mstar \approx \, 10^7 \, \Msol$) in \textsc{edge2} (red) closely match the empirical data measured in isolated dwarf irregulars (left-hand panel, grey points; \citealt{Read2017}). This match to observational data further extends to stellar sizes, gas contents, and stellar and gas-phase metallicities (Figure~\ref{fig:mvsize}~--\ref{fig:mstaroxygen}). The average difference in $\Mstar$ between \textsc{edge1} and \textsc{edge2} is $\approx 20\%$, i.e. well converged compared to the scatter in $\Mstar$ at given $\Mvir$ between different galaxy formation models (right-hand panel, grey points). This is primarily driven by a change in ISM structure (Figure~\ref{fig:ismstacks}) and galactic outflow strength (Figure~\ref{fig:outflows}). Section~\ref{sec:sec:mstarmhalo} discusses the response of individual formation histories (symbols linked with a line and caption). 
    }
    \label{fig:mstarmhalo}
\end{figure*}

\section{The impact of radiative feedback on dwarf galaxies} \label{sec:rtimpact}

\subsection{The stellar mass-halo mass relation} \label{sec:sec:mstarmhalo}

Figure~\ref{fig:mstarmhalo} shows how the integrated stellar masses, $\Mstar$, respond to the change from the original \textsc{edge1} (blue circles) to the updated \textsc{edge2} (red diamonds) model. $\Mstar$ is computed by summing all stellar particles within $\rvir$ and lines connect dwarf galaxies sharing the same cosmological initial conditions and formation scenarios (see the legend for individual names).

Comparing our results with empirical determinations of the $\MstarMhalo$ relation from observed dwarf galaxies (left-hand and middle panels) shows that \textsc{edge} dwarf galaxies are within observational uncertainties. In both the \textsc{edge1} and \textsc{edge2} models, low-mass dwarf galaxies ($\Mstar\approx 10^4 - 10^6 \, \Msol$; $\Mvir \approx \, 10^9 \Msol$) scatter around the median inferred from Milky Way satellites (blue line showing the peak halo mass from \citealt{Nadler2020}) and span the breadth of the 16-84 confidence interval at fixed $\Mvir$ (blue contours). In our suite, this scattering around the median is largely driven by our systematic exploration of different formation histories at fixed halo mass (\citealt{Rey2019UFDScatter, Gray2025}). 

An agreement between Milky-Way satellites and simulated field dwarfs might appear surprising at first. But in the low mass regime, the shutdown of gas accretion in the smallest haloes due to cosmic reionization (`reionization feedback') is the dominant mechanism regulating the stellar mass and gas content (e.g. \citealt{Efstathiou1992, Gnedin2000, Hoeft2006, Okamoto2008, Noh2014, Benitez-Llambay2020}). Furthermore, few faint and ultra-faint observed dwarf galaxies have orbital parameters that lead to strong tidal interactions with the Milky Way (\citealt{Simon2018, McConnachie2020, McConnachie2021}). As a result, the faint-end of the $\MstarMhalo$ relation is likely to be shaped by their pre-reionization ($z\geq6$) evolution, rather than limited by the specific environment of our Galaxy.

As dwarf galaxies grow in mass, ($\Mstar\geq 10^6 \, \Msol$; $\Mvir\geq 5 \times 10^9 \, \Msol$), reionization feedback becomes less dominant. This enables field dwarf galaxies to accrete late-time gas and enable dark matter halo mass measurements through rotation curves (\citealt{Read2017}; grey points in left-hand panel). At high $\Mvir$ all the way to $\Mvir\approx 10^{10} \, \Msol$, our simulated dwarf galaxies closely match the individual measurements of field, gas-rich dwarf irregulars. But, importantly, the two lowest mass objects in the \citet{Read2017} sample lack a simulated counterpart. Rather than a failure of the model, this more likely reflect the rarity of finding such low-mass, gas-rich galaxies on which to perform rotation curve measurements. At halo masses $\Mvir \approx 2 \times 10^{9} \, \Msol$, only very specific formation histories that form especially late will allow gas accretion and star formation (\citealt{Benitez-Llambay2021}). Such objects are not present either in the \textsc{edge1} and \textsc{edge2} suites, but calibrating a semi-analytical on \textsc{edge1} results to create a large population naturally recovers them (\citealt{Kim2024}).

\begin{figure*}
  \centering
    \includegraphics[width=\textwidth]{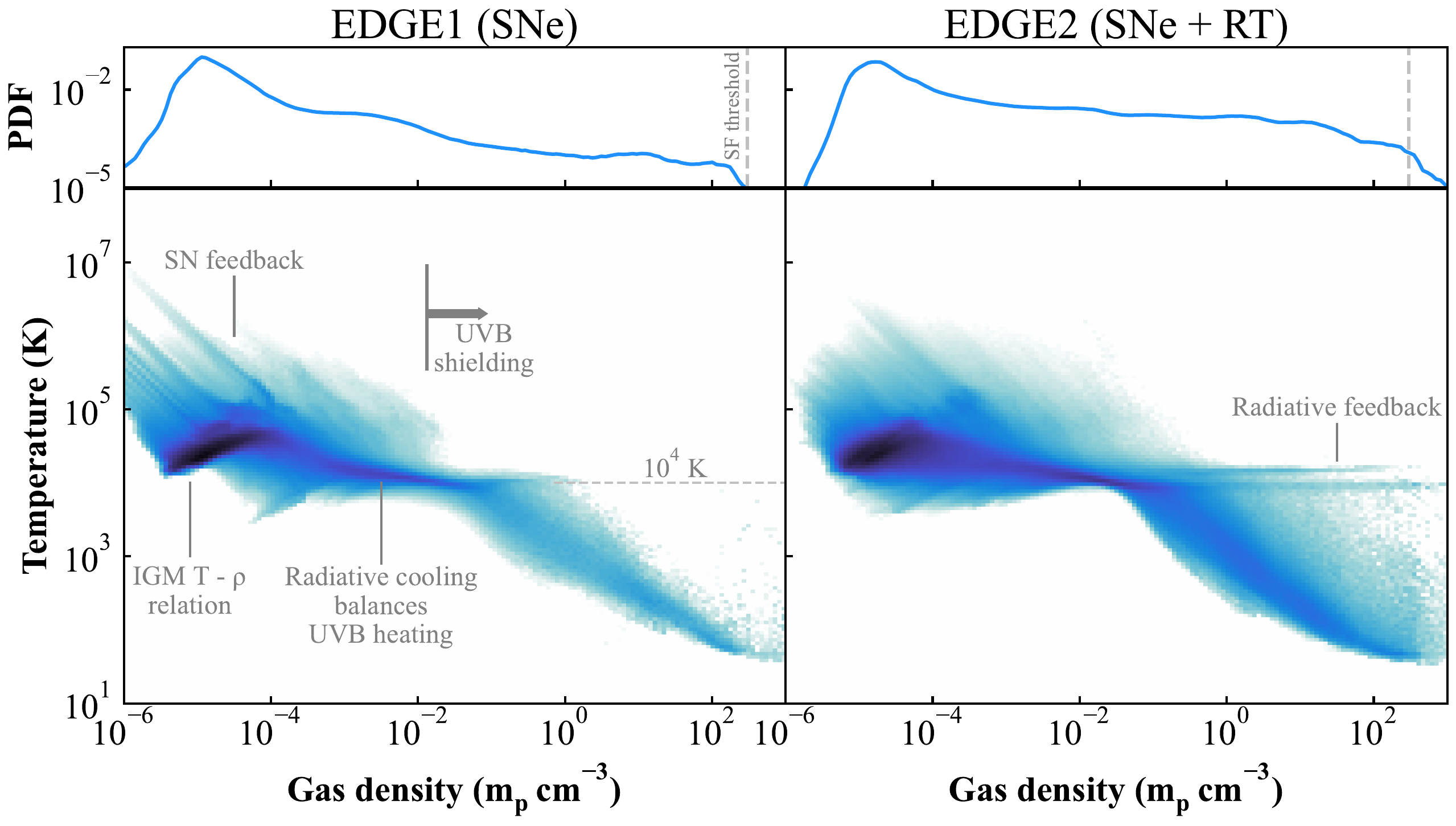}

    \caption{Temperature-density diagrams of the ISM of the dwarf galaxy (`Halo 383 (early)'; $\Mstar \approx 10^7 \, \Msol$) with the \textsc{edge1} (left) and \textsc{edge2} model (right). The diagrams are 2D mass-weighted PDF averaged over the last 4 billion years of evolution. Many features are shared between models (annotated). The addition of radiative feedback leads to warm ($T\approx10^4 \, \K$) and dense ($\nh \geq 10 \, \cmcube$) ISM gas. With this change, the ISM is also significantly denser overall (top panels) and the strength of galactic outflows is reduced (Figure~\ref{fig:outflows}).}
    \label{fig:ismstacks}
\end{figure*}

On average, $\Mstar$ in \textsc{edge1} and \textsc{edge2} differ by 34 per cent across the suite, and are within a factor of two of each other for all galaxies except for three specific formation histories that we discuss further below. The magnitude of these shifts is encouragingly tight given the large changes in cooling and heating physics between models. To emphasize that such shifts are well within theoretical uncertainties, the right-hand panel of Figure~\ref{fig:mstarmhalo} shows a compilation of simulated field dwarf galaxies (\citealt{Benitez-Llambay2021, Herzog2023}, plusses, `Swift'; \citealt{Revaz2018}, crosses, `Gear'; \citealt{Wang2015, Tollet2016}, dots, `Nihao'; \citet{Fitts2017, Wheeler2019}, stars, `FIRE-2'; \citet{Munshi2019, Munshi2021}, pentagons, `Changa'). At given $\Mvir$, predictions from different simulation groups can differ by over an order of magnitude in $\Mstar$ (e.g. pentagons against stars around $\Mvir \approx 10^{10} \, \Msol$), highlighting the small-in-comparison shifts between \textsc{edge1} and \textsc{edge2}. 

Before turning to comparing our dwarf galaxies to observations (Section~\ref{sec:scalingrelations}), we highlight several trends in Figure~\ref{fig:mstarmhalo} that will help us establish differences between our two numerical models:

\begin{itemize}[leftmargin=2.5mm]
  \item \textsc{edge2} systematically suppresses stellar masses in low-mass systems ($\Mstar \leq 10^6 \, \Msol$). This trend is reversed in higher-mass dwarfs ($\Mstar \geq 10^6 \, \Msol$), for which $\Mstar$ is systematically increased. Sections~\ref{sec:sec:ism} and~\ref{sec:sec:outflows} show that the inclusion of radiative feedback leads to a fundamentally different structure of the ISM and reduces the efficiency of galactic outflows. This suppression in outflow loading factors is more pronounced at lower masses, driving $\Mstar$ down, and less at higher masses, driving $\Mstar$ up. 
  \item One of our simulated dwarfs (`Halo 383: GM higher mass', purple in the right-hand panel) significantly over-shoots the stellar-mass-halo-mass relation, with data for this simulation shown at $z=0.9$ when we stopped the simulation due to the numerical cost incurred by the high $\Mstar$. Little $\Mvir$ growth is expected after this time. Section~\ref{sec:numerics} shows that, at high $\Mstar$, the reduced strength of galactic outflows stems from numerical issues leading to increasingly difficult regulation. We stopped `Halo 153' (pink) at $z=0.6$ for the same reason.
  \item Two low-mass galaxies show a large $\Mstar$ suppression with \textsc{edge2} (right-hand panel, `Halo 600: GM later' in orange and `Halo1459: GM latest' in turquoise). These two objects share the same characteristic of assembling late for their halo masses, from multiple small building blocks during the reionization era each with $\Mvir(z=6)\leq \, 10^8 \, \Msol$ that later come together in dry mergers (\citealt{Rey2019UFDScatter, Rey2020}). Each of these small building blocks sees their $\Mstar$ suppressed by radiative feedback, compounding the effect on the total $\Mstar$ compared to a comparatively massive object at the time of reionization.
\end{itemize}

\subsection{The structure of the ISM} \label{sec:sec:ism}

Figure~\ref{fig:ismstacks} shows the difference in ISM structure between models using SN-only feedback (left) and including radiative feedback (right). For illustrative purposes, we plot the 2D temperature-density distributions averaged over the $\approx 100$ snapshots along the formation history of `Halo 383 (early)' ($\Mstar\approx10^7 \, \Msol$). We pick this galaxy as it has a stable and gas-rich ISM at all times, in both models. But all points highlighted below generalize across the suite.      

In both models, we observe similar features in the temperature-density diagram (labelled on plot). Namely, (i) the upper tail of the IGM temperature-density relation (diagonal track at low densities; see \citealt{McQuinn2016} for a review); (ii) the distinctive thermal equilibrium curve around $10^4\, \K$ where radiative cooling balances photo-heating from the external UVB (see e.g. \citealt{Smith2017}, fig. 8); (iii) the break of this thermal equilibrium when gas starts self-shielding against the UVB; and (iv) high-temperature ($T\geq 10^5\, \K$) gas at low densities resulting from SN-driven outflows. 

The first notable difference is the warm ($T\geq 10^4\, \K$) and dense ($\rho \geq 10 \, \mppercmcube$) gas present in the \textsc{edge2} model (right-hand panel) and absent in \textsc{edge1} (left-hand panel). This is the direct result of the explicit modelling of radiative feedback in $\hii$ regions around stars, ensuring that \textsc{edge2} galaxies now capture a key gas phase for the overall structure of the ISM and its emission lines. The `doubled' horizontal track results from the mixing of photo-ionized gas with the surrounding ISM -- the upper track is fully ionized, while the lower track has significant neutral fraction, driving a difference in molecular weight and thus a $\approx 0.5$ shift in temperature. Part of this mixing is physical, at the edges of $\hii$ regions, and part is numerical (Section~\ref{sec:numerics:sspheres}). 

The second difference is that the ISM is overall denser with non-equilibrium cooling and photo-ionization feedback (top panels show marginalized density PDFs). Surprisingly, this trend extends beyond the density threshold for star formation (dashed lines), with the \textsc{edge2} model maintaining a significant amount of gas at densities above the star-formation threshold ($\geq 300 \, \mppercmcube$; grey dashed) that contrasts with the plummeting PDF in \textsc{edge1}. This reflects our numerical choices in the star formation algorithm, which we explore further in Section~\ref{sec:numerics}. As we show next, the much denser ISM in \textsc{edge2} correlates with strongly reduced galactic outflows. 

\subsection{The strength of galactic outflows} \label{sec:sec:outflows}

\begin{figure}
  \centering
    \includegraphics[width=\columnwidth]{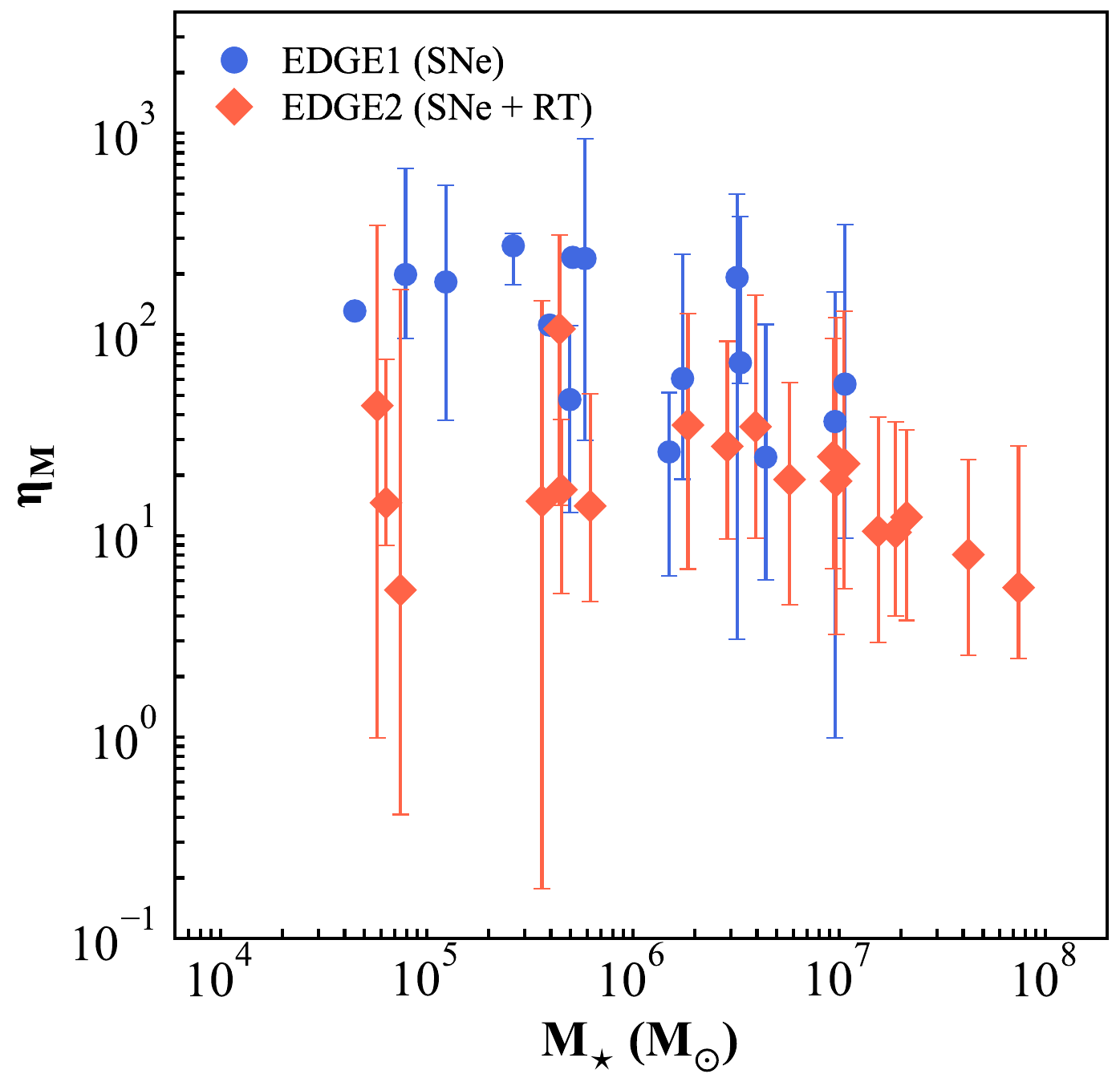}

    \caption{Mass loading factors measured through a spherical shell at $0.25\, \rvir$ as a function of galaxy stellar mass. Error bars show the 16-84 confidence interval around the median over the cosmological history of each individual dwarf galaxy. Radiative feedback (red) reduces outflow loading factors for every galaxy compared to SN-only feedback (blue). This reduction is up to 1.5 dex for the least massive dwarf galaxies.}
    \label{fig:outflows}
\end{figure}

\begin{figure*}
  \centering
    \includegraphics[width=\textwidth]{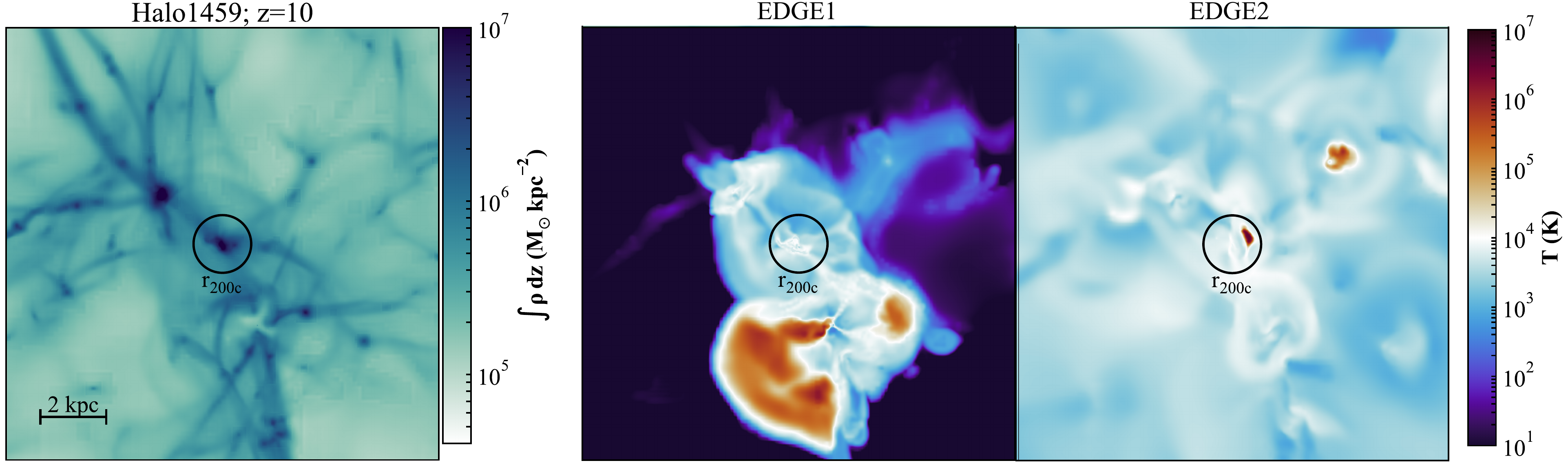}

    \caption{Gas density integrated along the line of sight (left) and thin slices of the temperature around the same low-mass dwarf galaxy ($\Mstar \approx 10^5\, \Msol$) at $z=10$, in \textsc{edge1} (middle) and \textsc{edge2} (right). Accounting for radiative feedback has a dramatic impact beyond $\rvir$ (right-hand panel). Radiation efficiently escapes from the dwarf's small building blocks, heating the volume to $T \geq 10^3 \, \K$ and suppressing inflows and correlated star formation across a large volume. Hot, SN-driven outflows eventually achieve the same effect (middle), but on a timescale longer than radiative feedback.  
    }
    \label{fig:tempmaps}
\end{figure*}

Figure~\ref{fig:outflows} shows the gas mass loading factor, $\massloading$, over the cosmological history of each galaxy in the suite as a function of their $\Mstar$. We define $\massloading = \dot{M}_{\text{out}} / \SFR_{\text{10 Myr}}$, where the mass outflow rate $\dot{M}_{\text{out}}$ is measured through a spherical shell centred on the galaxy that spans a radial range between $0.2\, \rvir$ and $0.3\, \rvir$ and only includes outflowing gas (see also \citet{Rey2024Outflows} for further details). $\SFR_{\text{10 Myr}}$ is the star formation rate averaged over 10 Myr. Error-bars show the 16-84 confidence interval of the distribution of $\massloading$ values over time, with the median shown as a symbol. Some galaxies for which no saved snapshots have $\SFR_{\text{10 Myr}} > 0$ are missing from the plot.

Figure~\ref{fig:outflows} demonstrates that radiative feedback systematically decreases the strength of galactic outflows, reducing their ability to remove mass from the central galaxy. Scatter as a function of time is large, reflecting the stochastic nature of the star formation-feedback cycle, but the suppression is systematic and close to an order of magnitude. Since $\Mstar$ remains close between \textsc{edge1} and \textsc{edge2} (Figure~\ref{fig:mstarmhalo}), this highlights a fundamental change in the way star formation is regulated. With radiative feedback, gas is prevented from forming stars by gentle heating, rather than being mechanically removed from the centre in blastwaves when considering only SN feedback. We further show in Appendix~\ref{app:sfhs} that the star formation histories (SFHs) are on average less bursty as a result, aligning with this picture.

A reduction in the strength of galactic outflows and of the burstiness of star formation due to radiative feedback is well established in isolated non-cosmological dwarf galaxies (e.g. \citealt{Emerick2018, Smith2020PhotoRT, Deng2024Rigel, Andersson2024ClusterMF}) and was previously noted for a single low-mass cosmological dwarf in \citet{Agertz2020EDGE}. Our simulations extend these findings to a much wider range of masses and highlight trends with host $\Mstar$. 

In particular, the median $\massloading$ reduces with increasing $\Mstar$ in the \textsc{edge1} model, as expected from the scaling of SN-driven outflows with $\Mstar$ (e.g. \citealt{Muratov2015, Christensen2016, Nelson2019TNGOutflows, Mitchell2020, Pandya2021}). In contrast, \textsc{edge2} sees $\massloading$ stay relatively flat over a wide range of $\Mstar$, even slightly increasing with mass. This is a direct consequence of the different feedback mechanisms at play in the two models.

In \textsc{edge1}, there is no rapid feedback mechanism to regulate star formation, the clustering of SNe across all $\Mstar$ is high, and the efficiency of mechanical outflows decreases as the gravitational potential wells get deeper. 

In contrast, \textsc{edge2} sees some of the weakest outflows and the strongest outflow suppression compared to \textsc{edge1} in the lowest-mass objects. This reflects the unique regime of very low-mass galaxies that formed all their stars before reionization.

In this case, stellar radiation efficiently escapes the ISM of the small building blocks that will make the $z=0$ galaxy but remain spatially distinct at $z>10$ (Figure~\ref{fig:tempmaps}, left-hand panel). The radiation propagates outwards much more rapidly than a mechanical outflow, heating up gas to $T \geq 10^3 \, \K$ well beyond the main progenitor's $\rvir$ (right-hand panel)\footnote{Note that the temperatures in Figure~\ref{fig:tempmaps} are such that most of the hydrogen in the volume remains neutral, thereby not impacting the cosmological timing of hydrogen reionization.}. This in turn suppresses gas inflows and correlated star formation across a much larger volume. This regulation mechanism is fundamentally different to the mechanical outflows driven by SNe in \textsc{edge1} (middle panel). As the galaxies get larger in mass, this effect occurs at ever higher redshift and contributes less to shaping the overall $\Mstar$ of the galaxy. 

Lastly, we note that observational values for $\massloading$ at $\Mstar \approx 10^7 \, \Msol$ have varied from $\approx 10$ (e.g. \citealt{Chisholm2017}), to $\approx 0.1-1$ (e.g. \citealt{McQuinn2019Outflows,Xu2022EmpressOutflows}), to $10^{-2}$ (e.g. \citealt{Marasco2023}), bracketing the values reported here. There are clear limitations in directly comparing the values in Figure~\ref{fig:outflows} to observations. Here, $\massloading$ is computed including all gas phases rather those that are bright in emission lines used to make the measurement. We also use a radius well outside that probed by observations which would likely affect the measured $\massloading$ (e.g. \citealt{Muratov2015}). Finally, the values in Figure~\ref{fig:outflows} are averaged over the full cosmological history, rather than at a time of star formation when the dwarf galaxy's gas is observable. A more careful comparison is thus warranted to establish the realism of either \textsc{edge1} and \textsc{edge2} outflows. This will be the focus of a future paper.

\section{The emergence of dwarf galaxy scaling relations and their differentiating power} \label{sec:scalingrelations}

\subsection{ The stellar size-luminosity relation}
\begin{figure}
  \centering
    \includegraphics[width=\columnwidth]{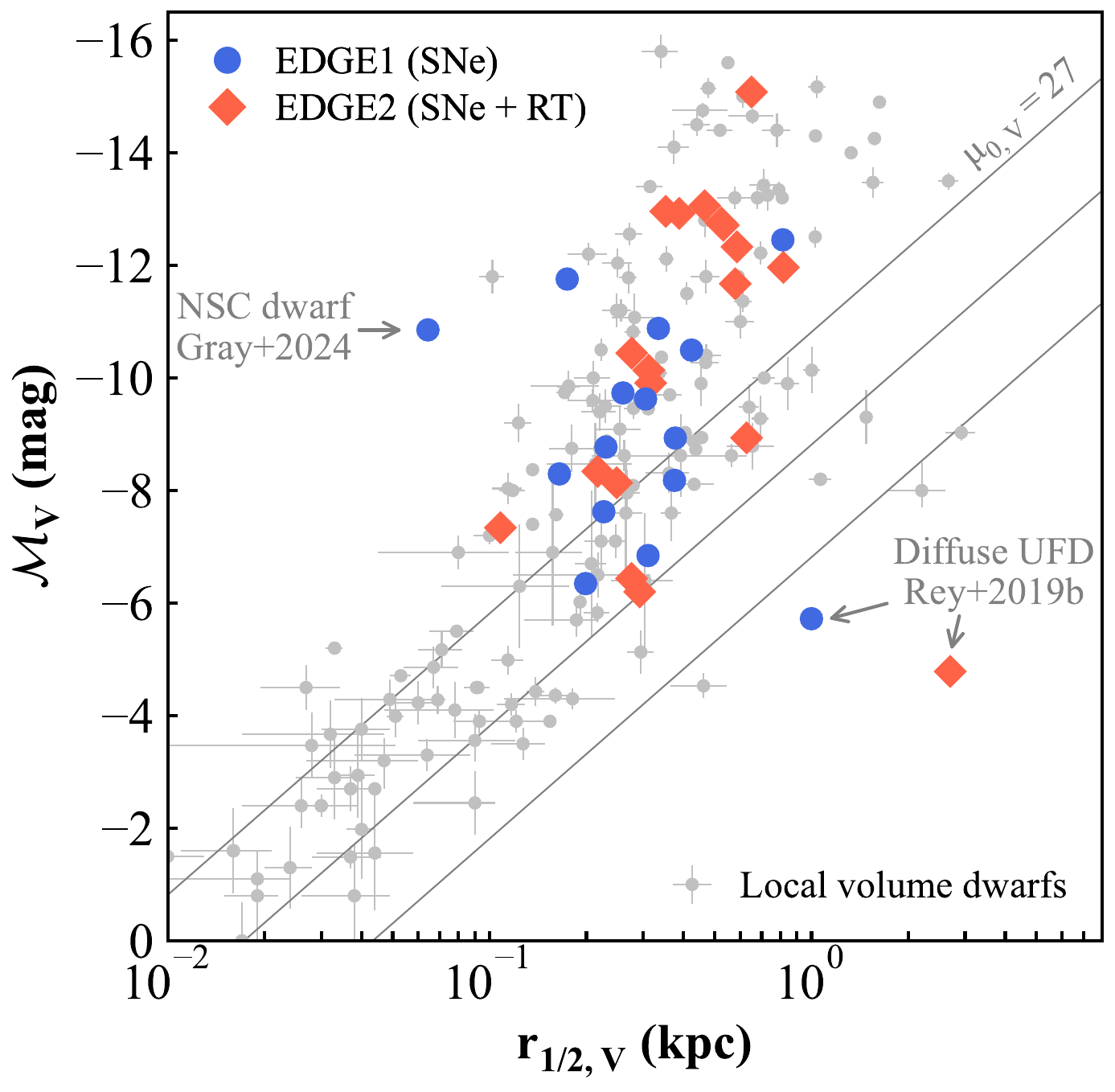}

    \caption{Absolute V-band total magnitude of simulated dwarf galaxies against their V-band half-light radius. With both models (red and blue), simulated dwarf galaxies populate the observational scatter observed around the Local Volume (grey points; error-bars showing the dispersion around the median) showcasing the limited power of stellar sizes in discriminating between galaxy formation models. Three exceptions lack an observational counterpart (labelled), but do not pose significant challenges to either model (see the text discussion).}
    \label{fig:mvsize}
\end{figure}

Figure~\ref{fig:mvsize} shows the absolute V-band magnitude, $\magv$, and projected half-light radius, $\rhalflight$, of simulated dwarf galaxies. To obtain these quantities, we compute the luminosity of all star particles within $\rvir$ as a function of their mass, age and metallicity according to the single stellar population model derived from \textsc{parsec} isochrones (\citealt{Bressan2012, Nguyen2022}). $\rhalflight$ is then obtained along a random line of sight. Simulated properties are compared to observational data from the compilations of \citet{McConnachie2012, Kirby2013, Kirby2014, Simon2019} augmented with individual candidates and detections from \citet{Torrealba2016, Torrealba2018, Torrealba2019, Homma2019, Homma2024, Mau2020, Bennet2022, Richstein2022, Sand2022, Cerny2023DELVE, Cerny2023DELVE6, Cerny2023PegIV, Jones2023PAVODwarf, Jones2024Corvus, McQuinn2023, Collins2024, Li2024Hedgehog, Martinez-Delgado2024, McNanna2024, Smith2023BootesV, Smith2024Faintest, Tan2025}. We curate the observational data to use the most recent reference when multiple observations of the same system are available. 

Irrespective of the model, nearly all dwarf galaxies are within the observational scatter of $\magv$-$\rhalflight$ and there is no clear trend or systematic offset between models. On average, \textsc{edge2} dwarfs are marginally larger than their \textsc{edge1} counterparts, and high-mass \textsc{edge2} dwarfs are systematically brighter in line with their increased $\Mstar$ (Figure~\ref{fig:mstarmhalo}).  

Two simulated dwarfs lack direct observed analogues (labelled on plot). In the bottom right, the same dwarf galaxy in both models is an ultra-faint ($\magv \geq -6$) that is also extremely extended ($\rhalflight \geq 1 \, \kpc$). This large $\rhalflight$ is driven by the specific assembly history of this galaxy that sees its stellar component built from the mergers of small building blocks that deposit stars on wide orbits (see \citealt{Rey2019UFDScatter} for details). As discussed in Section~\ref{sec:sec:mstarmhalo}, photo-ionization feedback is particularly efficient in these small building blocks, and this object becomes even more diffuse in the \textsc{edge2} model. In both cases, the lack of observational counterpart can be explained by their central surface brightnesses being beyond currently-observable capabilities ($\mu_{0, V} > 31 \, \text{mag arcsec}^{-2}$). Such dwarfs should be revealed by forthcoming surveys such as on the Vera Rubin C. Observatory. 

In the left-hand corner, one \textsc{edge1} dwarf (blue point) has a particular formation history that leads to a compact nuclear star cluster that dominates the light of the galaxy. This galaxy shares observational characteristics with ultra-compact dwarfs and nuclear star clusters not included in Figure~\ref{fig:mvsize} (see \citealt{Gray2025} for a discussion). Due to the reduced outflow strength and less clustered star formation, this galaxy does not qualify as a nuclear star cluster in \textsc{edge2}. 

Our results highlight the natural emergence of the $\magv$-$\rhalflight$ scaling relation, in both median and scatter, from our cosmological modelling. This emergence is robust to a large change in galaxy formation  modelling from \textsc{edge1} to \textsc{edge2}. Such convergence hints that, if stellar masses are reasonably predicted at a given $\Mvir$, the $\magv$-$\rhalflight$ relation and its scatter follow from the diversity of cosmological assemblies. This in turn brings confidence that this relation can be well predicted with physics-informed semi-analytical arguments applied to dark-matter-only simulations (S. Nigudkar et al. in preparation).

\subsection{The stellar mass-metallicity relation} \label{sec:sec:massmetallicity}

\begin{figure}
  \centering
    \includegraphics[width=\columnwidth]{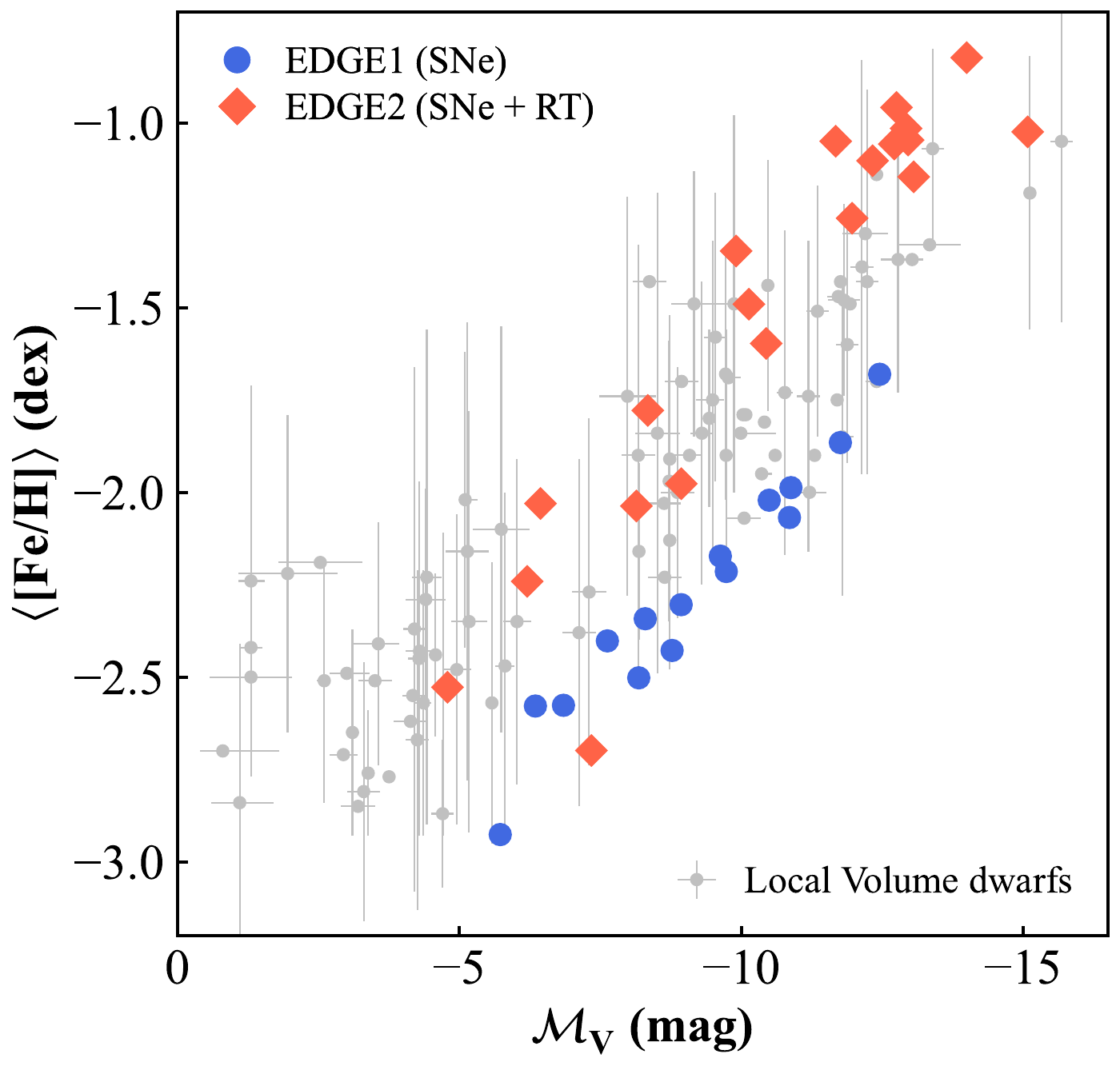}

    \caption{Absolute total V-band magnitude of our simulated galaxies against the average iron content of their stars. Both models track the slope of the mass-metallicity relation, but are offset with one another due to the changes in metal production tables and feedback physics.  }
    \label{fig:mstarmetallicity}
\end{figure}

Figure~\ref{fig:mstarmetallicity} shows the $\magv$-$\averagefeh$ relation where $\averagefeh$ is the average stellar iron abundance. We derive $\feh$ for each stellar particle within $\rvir$ from their iron mass fractions and compute $\averagefeh$ as the mass-weighted average (see \citealt{Escala2018}, eq. 3 and 4). Observed data are compiled from \citet{Kirby2013, Kirby2014, Simon2019} augmented and updated with data from \citet{Kirby2017, Kirby2020, Li2017, Li2018, Longeard2018, Fritz2019, Ji2019, Ji2021, Collins2020, Collins2021, Pace2020, Taibi2020, Wojno2020, Jenkins2021, Chiti2021, Chiti2022GrusI, Bruce2023, Charles2023, Cerny2023PegIV, Smith2023BootesV,Hansen2024, Heiger2024, Kvasova2024, Tan2025}. Data error-bars show the dispersion around $\averagefeh$ (when available) rather than measurement errors on the mean. 

Both \textsc{edge1} and \textsc{edge2} models accurately track the slope of the $\magv$-$\averagefeh$ relation and are within the observational scatter. However, \textsc{edge1} dwarfs populate the lower end of observed points over the range of $\magv$, while \textsc{edge2} dwarfs populate the upper end. The origin of the $\approx 0.5$ dex offset between models is two-fold. First, the \textsc{edge2} model produces $\approx2.5$ more iron per stellar mass formed than \textsc{edge1} (Figure~\ref{fig:modelcomparison}). And second, the inclusion of radiative feedback leads to a more gentle regulation of star formation with weaker galactic outflows (Section~\ref{sec:rtimpact}), and thus metal retention in the ISM. These two effects compound one another, with both iron production and retention increased in the \textsc{edge2} ISM, in turn leading to higher $\averagefeh$.

These results cement the $\magv$-$\averagefeh$ relation as a sensitive probe of star formation and stellar evolution physics at low metallicities (see also \citealt{Agertz2020EDGE, Prgomet2022, Sanati2023}). The normalization at a given $\magv$ is in particular directly related to the strength of galactic outflows in dwarf galaxies, while the slope emerges from the cosmological relation between stellar and halo masses. A way to more strongly discriminate between feedback models is thus to increase the precision of $\averagefeh$ measurements in dwarf galaxies. This will occur in the forthcoming years as more stellar abundances per dwarf galaxy become available (e.g. \citealt{Skuladottir2023a}). 

Another promising route is to combine $\magv$-$\averagefeh$ with more observables of the star formation and feedback cycle. Given the sensitivity of iron abundances, further ratios beyond chemical elements are likely to provide complementary constraints on metal production and retention. We will use the extended number of chemical elements tracked in \textsc{edge2} to explore this in future work. Next, we focus on the gas contents and gas-phase metallicity of our simulated dwarf galaxies. 

\subsection{The gas content of dwarf galaxies}

\begin{figure}
  \centering
    \includegraphics[width=\columnwidth]{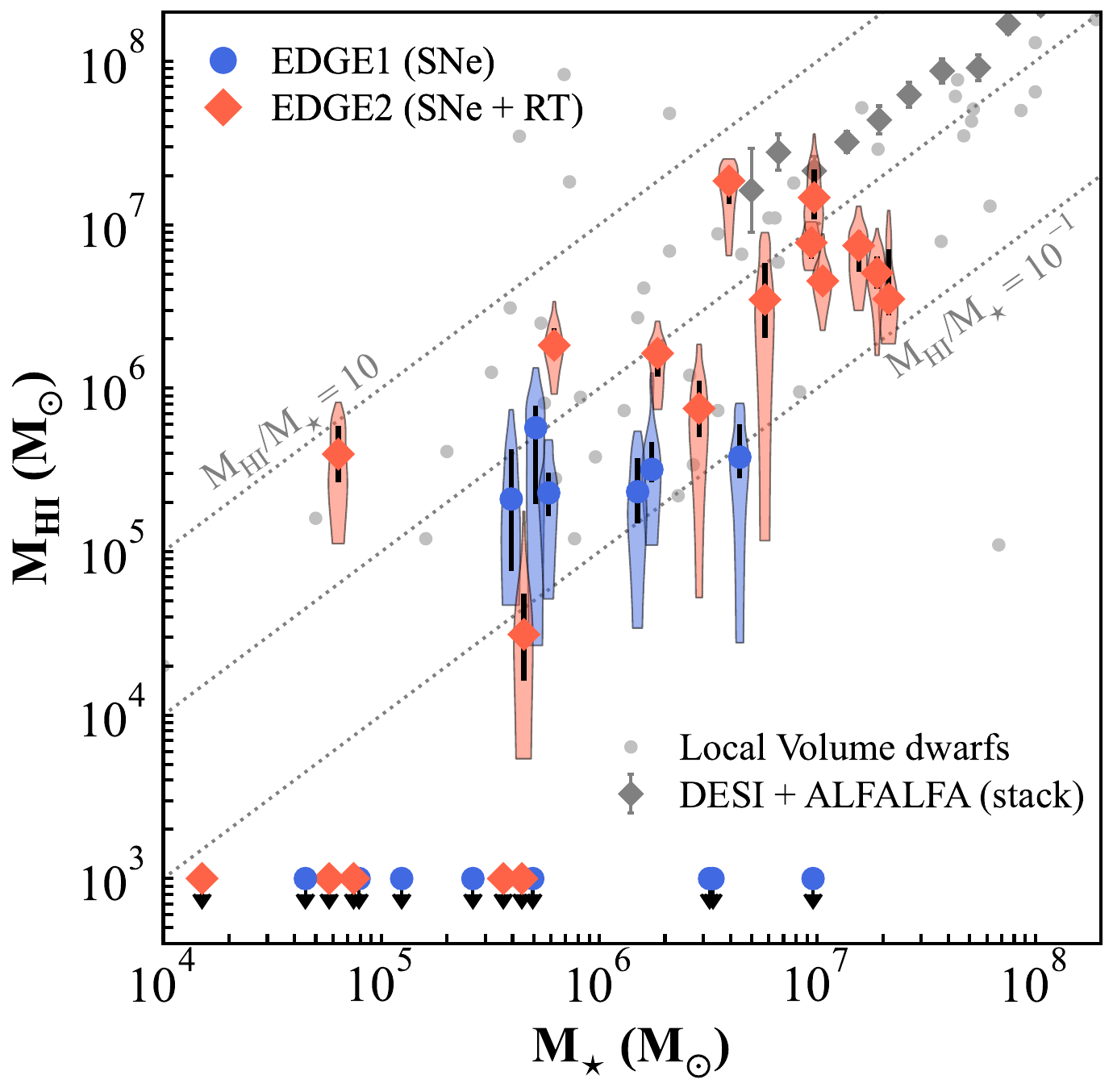}

    \caption{$\hi$ and stellar masses of our simulated dwarf galaxies in each model. The structure of the $\MstarMhi$ is robust to our change of model, with a bimodal structure at the low-mass end and significant time variability in $\hi$ content due to stellar feedback (violins show the distribution over the last 4 Gyr). Both models are within the observed scatter at low masses (grey circles; see the text for the compilation of individual detections), with \textsc{edge2} approaching the $\MstarMhi$ relation of a stacked mass-complete sample at higher masses (grey diamonds; \citealt{Scholte2025}).}
    \label{fig:mstarmhi}
\end{figure}

Figure~\ref{fig:mstarmhi} shows the $\MstarMhi$ relation for our simulated dwarf galaxies, where $\Mhi$ is the total $\hi$ mass within $\rvir$ (see \citealt{Rey2022EDGEHI} for how we derive $\Mhi$ in \textsc{edge1}, while natively tracks out-of-equilibrium $\hi$ fractions). Violins in Figure~\ref{fig:mstarmhi} show the distribution of $\Mhi$ over the last 4 billion years (symbols and lines showing the medians and 16-84 confidence intervals, respectively). This is long enough to average over the time variability due to stellar feedback, but short enough that $\Mstar$ does not vary significantly (e.g. \citealt{Rey2022EDGEHI}). 

Both models predict a similar structure for the $\MstarMhi$ relation, with the most distinct feature being a bimodality between gas-deficient dwarfs with vanishing $\hi$ contents (upper limits at the bottom) overlapping in $\Mstar$ with gas-rich systems. This bimodality arises from the interplay between the UVB and stellar feedback which regulate the gas and $\hi$ contents, and the diversity of possible mass-growth histories in a $\Lambda$CDM universe (\citealt{Rey2022EDGEHI}; see also e.g. \citealt{Benitez-Llambay2020, Benitez-Llambay2021, Kim2024}).

For dwarf galaxies with measurable $\hi$ contents, both models produce dwarf galaxies within the observational scatter from a compilation of individual  gas-rich field dwarfs (\citealt{McConnachie2012, Cole2014, McQuinn2015, McQuinn2020, McQuinn2021, Sand2015, Adams2018, Brunker2019, Janesh2019, Hargis2020, Bennet2022, Xu2023FASTGasRichDwarf}; grey circles). In the case of the \textsc{edge2} model, this agreement is striking for $\Mstar \approx 10^7\, \Msol$ where simulated dwarfs cluster around the $\MstarMhi$ relation from \citet{Scholte2025} that combines \textsc{desi} and \textsc{alfalfa} (grey diamonds) to stack a mass-complete sample. Note that the error-bars on the stack show the error on the median, not the expected population scatter which can be read from individual measurements. 

Across the suites, \textsc{edge2} galaxies are on average more $\hi$-rich than \textsc{edge1} galaxies at similar $\Mstar$, as expected since radiative feedback dampens galactic outflows (Figure~\ref{fig:outflows}). Similarly, both models predict substantial variability (violins), but $\Mhi$ is more stable over time in the \textsc{edge2} model (the extent of the 16-84 confidence interval is 53\% smaller on average). And an even more notable difference is the lack of $\hi$-deficient, higher-mass dwarf galaxies ($\Mstar \geq 10^6\, \Msol$) in \textsc{edge2}. Runaway star formation and highly clustered SN feedback during mergers can vacate the whole $\hi$ reservoir and self-quench a dwarf galaxy, while a more gentle, radiative regulation of star formation allows them to retain more gas. 

We conclude that the exact shape of the bimodality at low $\Mstar$, the median $\Mhi$ at a given $\Mstar$ and the scatter around this median due to time variability are all sensitive to the efficiency with which dwarf galaxies drive galactic outflows. This is promising for future comparisons with forthcoming radio surveys (e.g. Wallaby, \citealt{Koribalski2020}; Apertif-Medium deep, \citealt{vanCappellen2022}). But the fact that the structure and the key features of the $\MstarMhi$ relation robustly emerge from our modelling, despite large changes in the cooling and heating physics of the \textsc{edge} model, highlight that constraining power will only be unlocked by leveraging large populations of dwarf galaxies. This requires developing a new generation of semi-analytical models building on the insights presented here to generate large statistical samples of $\hi$ dwarfs, which we will present in future work (S. Hutton et al. in preparation). 

\subsection{The gas-phase mass-metallicity relation}

\begin{figure}
  \centering
    \includegraphics[width=\columnwidth]{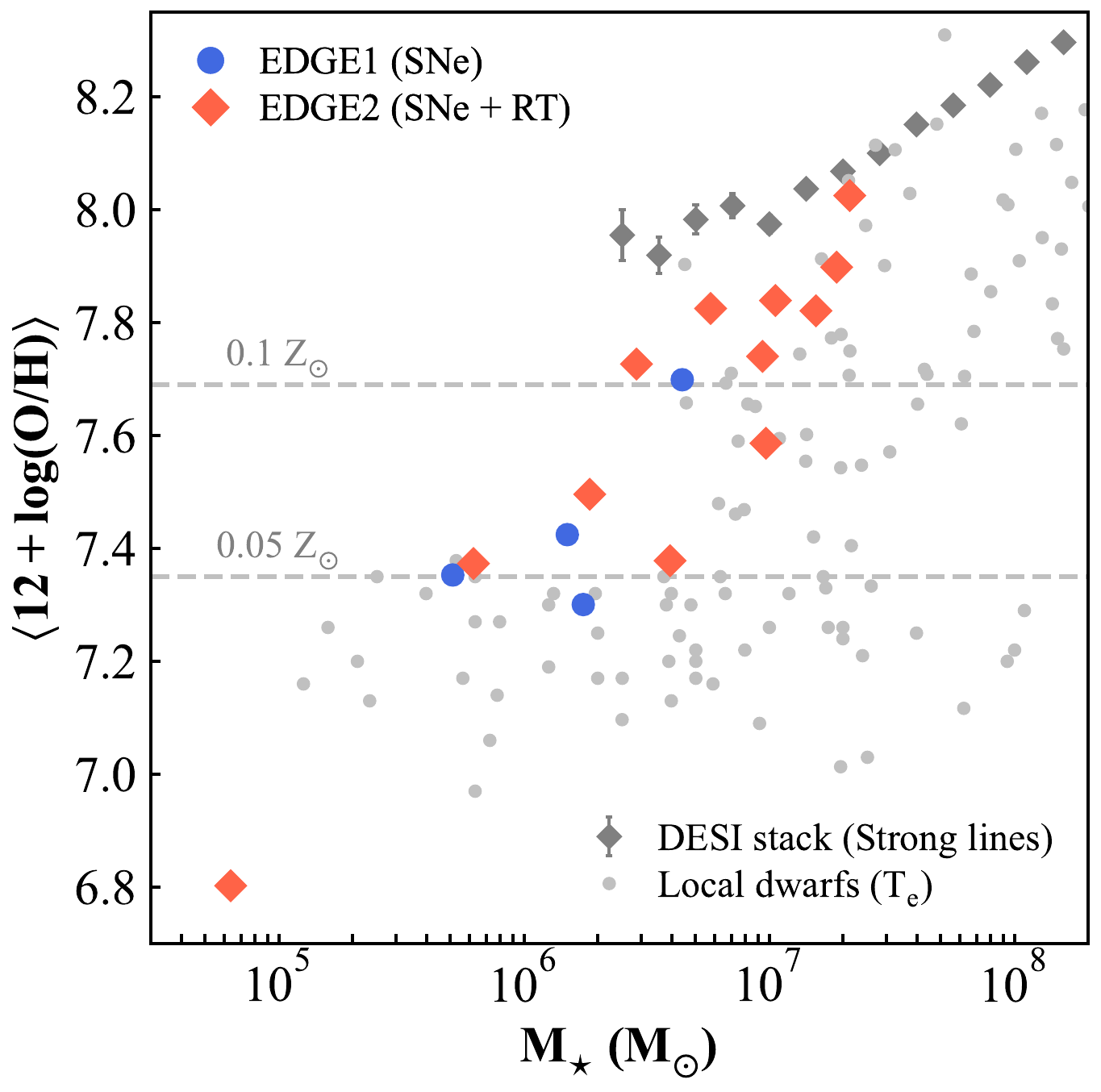}

    \caption{Stellar mass versus gas-phase oxygen metallicity for simulated dwarf galaxies that have formed new stars within the last 4 Myr. Both models track the slope of the mass-metallicity relation and are broadly compatible with individual measurements of metal-poor, low-mass dwarfs (grey). \textsc{edge2} dwarfs (red) are slightly more metal-rich at a given stellar mass and populate the upper end of the observational scatter. But these effects are smaller than uncertainties arising from measuring gas-phase metallicities with different calibration methods (strong-line calibration as diamonds, direct-method as circles).   
    }
    \label{fig:mstaroxygen}
\end{figure}

Figure~\ref{fig:mstaroxygen} shows the $\Mstar$-$\averageOmetallicity$ relation at $z=0$, where $\averageOmetallicity$ is the mass-weighted average oxygen metallicity in the gas phase. To compute $\averageOmetallicity$, we select gas within $2\, \rhalflight$ of dwarf galaxies that have formed new stars in the last 4 billion years, as measured data of oxygen metallicity almost invariably rely on ionized emission lines in $\hii$ regions. We do not account for dust depletion which should be small at our low metallicities.

Observed data (grey circles) shows measurements of individual $z=0$ dwarf galaxies compiled by \citealt{Yates2020} and J. Breneman et al. in preparation. All these points measure the electron temperature $T_e$ from auroral lines to derive $\averageOmetallicity$. We omit their error-bars for visual clarity. We also show the stack of emission-line-selected dwarf galaxies from DESI (\citealt{Scholte2025}; pentagons) with metallicities derived from strong-line calibration (\citealt{Nakajima2022}).  

Both \textsc{edge1} and \textsc{edge2} are broadly within the scatter of individual measurements of gas-phase metallicities and track the slope of the relation. On average, \textsc{edge2} dwarfs are slightly more oxygen rich ($\approx 0.1$ dex) at similar stellar masses than \textsc{edge1} dwarfs. This much smaller upshift than for stellar $\averagefeh$ ($\approx 0.5$ dex) is readily explained by the difference in chemical elements used for the observable. While the weaker galactic outflows in \textsc{edge2} increase oxygen retention in the ISM (driving $\averageOmetallicity$ higher), oxygen production per stellar mass formed is more than halved with our updated CCSNe yields (Figure~\ref{fig:modelcomparison}), driving $\averageOmetallicity$ lower. When combined, this leads to a small change in the overall $\averageOmetallicity$ of our galaxies.

The \textsc{edge2} model predicts one example of a very low-mass $\Mstar\leq 10^5\, \Msol$, very metal-poor, star-forming dwarf galaxy (bottom left corner). This object (`Halo 600: GM Later') is quenched by cosmic reionization early on, and achieves a high-enough dynamical mass to re-ignite star formation particularly late ($z=0.2$, \citealt{Rey2020}). The delayed growth drives down $\Mstar$ and $\averageOmetallicity$ along the slope of the mass-metallicity relation. This highlights the possibility for low-mass dwarf galaxies even more oxygen-deficient than currently known forming through such rare assembly histories.

We conclude that, like for the stellar-phase mass metallicity relation (Figure~\ref{fig:mstarmetallicity}), the normalization of the gas-phase mass metallicity is sensitive to the galaxy formation assumptions made. However, given the implementation choices between \textsc{edge1} and \textsc{edge2}, the shifts observed ($\approx 0.1$ dex) are smaller than systematic uncertainties in metallicity measurements. This is illustrated by the $\approx 0.3$ dex offset between individual and stacked measurements (grey circles and diamonds). Such offset is of the magnitude expected given their distinct metallicity calibration (strong line method validated with `semi-direct' observations, versus `semi-direct' and `direct' auroral measurements; see e.g. \citealt{Kewley2008, Yates2020} for a discussion). And shifts in $\averageOmetallicity$ of $\approx 0.1$ dex are also well within the uncertainties of comparing direct-method measurements to simulation values (e.g. \citealt{Cameron2023ISMTe}). A more detailed exploration is thus warranted to establish the precise constraining power of the gas-phase mass-metallicity relation on feedback models in the dwarf galaxy regime. 

\section{Numerical limitations and remaining uncertainties in the \textsc{edge2} model} \label{sec:numerics}

\begin{figure*}
  \centering
    \includegraphics[width=\textwidth]{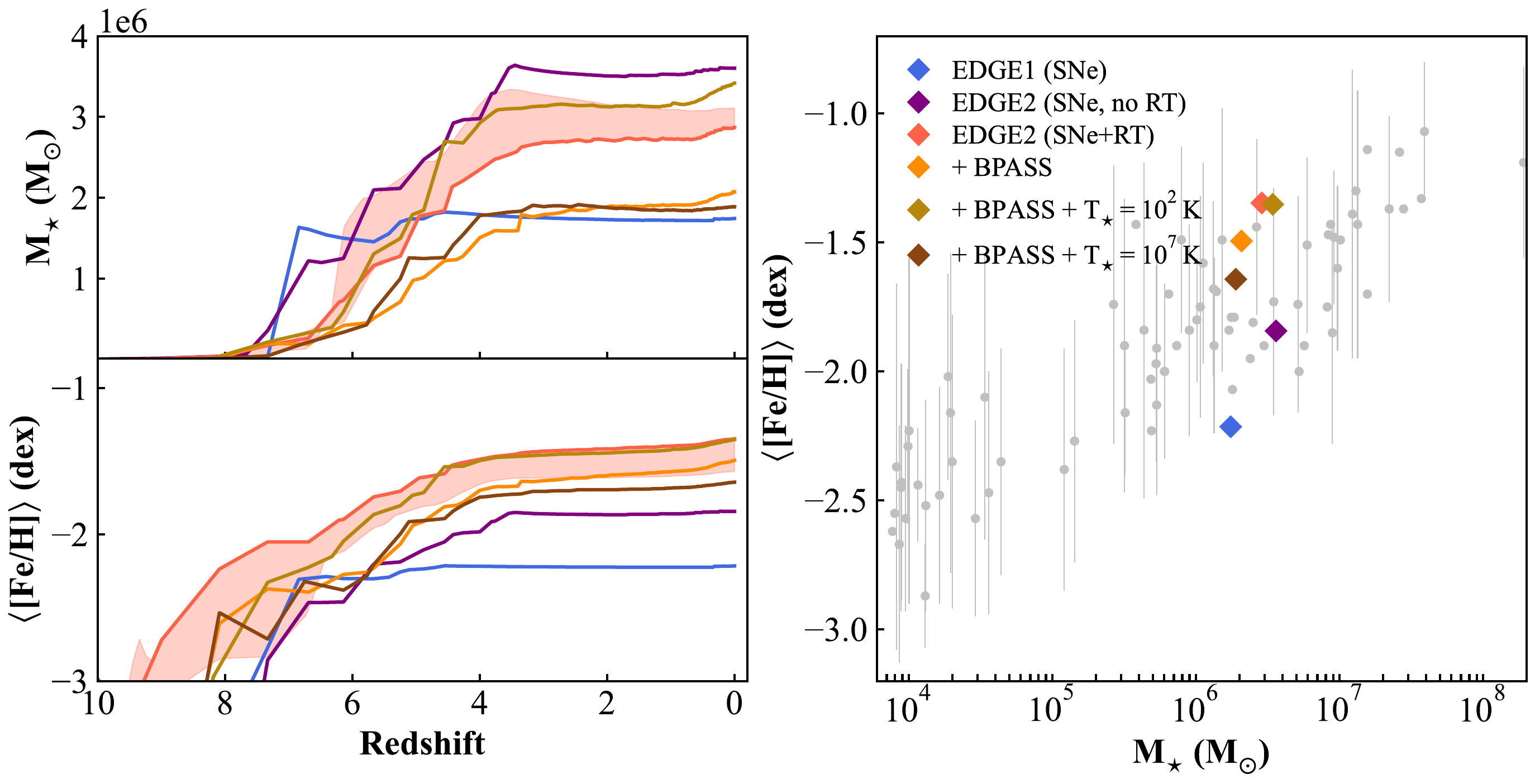}

    \caption{Stellar mass and metallicity growth over cosmic time (left) and at $z=0$ for multiple versions of the same dwarf galaxy (`Halo 605') varying input parameters. The increased iron production between \textsc{edge1} and \textsc{edge2} (see Figure~\ref{fig:modelcomparison}) drives a $\approx 0.4$ dex increase in $\averagefeh$ at $z=0$ (blue versus purple). Further adding radiative feedback (purple versus red) increases $\averagefeh$ by another $\approx 0.5$ dex by enhancing iron retention in the ISM. Varying the SED (orange versus red) and parameters of the star formation algorithm (gold and brown versus orange) has a smaller overall effect, but modulates the clustering of star formation and the ability to drive more or less powerful outflows. Intrinsic stochasticity due to chaotic noise in the simulations is small compared to these shifts (red envelopes).
    }
    \label{fig:halo605discussion}
\end{figure*}

The comparison between new and previous generation of \textsc{edge} simulations has revealed impressive convergence in observables (Section~\ref{sec:scalingrelations}), but fundamentally different ISM and outflow structures (Section~\ref{sec:rtimpact}). In this section, we discuss the sensitivity of these comparisons to key uncertainties, discuss numerical limitations of our radiative simulations, and highlight key systematics yet to be quantified when modelling small dwarf galaxies.

\subsection{Direct numerical tests} \label{sec:numerics:inputvariations}

To explore uncertainties in our modelling, we re-simulate multiple times a dwarf galaxy midway through the mass-range of our sample (`Halo 605'; $\Mstar\approx 10^6 \, \Msol$). This dwarf survives reionization and maintains star formation for a Hubble time, providing a long baseline to compare between model inputs. Figure~\ref{fig:halo605discussion} shows how $\Mstar$ and $\averagefeh$ respond to variations in key numerical and physical parameters that we detail next.

\subsubsection{Chaotic noise and stochasticity} \label{sec:numerics:stochasticity}

We start by quantifying the intrinsic noise level and stochasticity in our model. The chaotic nature of galaxy formation combined with finite numerical precision can lead to divergent evolution from the same initial conditions (see e.g. discussion in \citealt{Keller2019, Genel2019}). 

To quantify the magnitude of this noise term, we re-simulate our test galaxy twice, each time seeding a different truncation error. We achieve this by re-simulating the same initial condition on a different number of cores, which changes the order of arithmetic operations and their truncation errors when communicating across a supercomputer's network. 

The envelopes (left-hand panels) around the fiducial \textsc{edge2} run (red) in Figure~\ref{fig:halo605discussion} show the extent of the differences in $\Mstar$ and $\averagefeh$ spanned by these two additional stochastic re-simulations. At early times, when few stellar particles are present, stochasticity effects can be substantial, but the different simulations have converged with one another by $z\approx 6$. 

Averaging over the three simulations at $z=0$, we find $\Mstar = \xScientificErrorBar{2.97}{0.07}{0.09}{6} \, \Msol$ and $\averagefeh = -1.41_{-0.11}^{+0.04}$. Such differences are small compared to the factor-of-a-few changes going from the \textsc{edge1} to the \textsc{edge2} model. They also remain small compared to other input variations that we explore next. 

Admittedly, three simulations is a limited sample size to make statistical statements. The numerical costs of these simulations make a wider exploration impractical, but we note that the magnitude of the scatter in $\Mstar$ is comparable to previous findings in the \textsc{edge1} context (\citealt{Pontzen2021}). We conclude that chaotic stochasticity, while present, plays a small role in the trends and shifts in observables discussed in this paper.

\subsubsection{Radiative feedback versus other inputs} \label{sec:numerics:yields}
In addition to radiative feedback, the \textsc{edge2} model makes significant updates to the chemical enrichment modelling and SN feedback budget (Figure~\ref{fig:modelcomparison}). To isolate the effect of these changes on $\averagefeh$, we re-simulate our test dwarf galaxy in the \textsc{edge2} model turning off non-equilibrium cooling and radiative feedback. 
 
Comparing this new run to the original \textsc{edge1} version in Figure~\ref{fig:halo605discussion} (blue against purple, respectively), we find a $\approx 2\times$ increase in $\Mstar$ and $\approx 0.4$ dex increase in $\averagefeh$. Conversely, contrasting runs with and without radiative feedback (purple versus red, respectively), $\Mstar$ is decreased by $\approx 20\%$ while $\averagefeh$ further increases by another $\approx 0.5$ dex. All of these shifts are large compared to chaotic stochasticity (Section~\ref{sec:numerics:stochasticity}).

To summarize, our updates to the IMF and to the chemical enrichment modelling between \textsc{edge1} and \textsc{edge2} drives roughly half of the total increase in $\averagefeh$ at $z=0$. The other half is due to radiative feedback promoting metal retention through weaker outflows. It is thus clear that both of these inputs play an important role in setting the normalization of the stellar mass-metallicity relation. In the future, we will provide a much more detailed exploration of CCSNe yield models and further variations of the SNeIa inputs (E. Andersson et al. in preparation)

\subsubsection{The spectral energy distribution} \label{sec:numerics:seds}
Another key input to radiative feedback is the choice of the SED for stellar populations. Figure~\ref{fig:halo605discussion} shows the response of our test dwarf galaxy when swapping from our fiducial SED (\citealt{Bruzual2003}, red) to a harder and more ionizing \textsc{bpass2.2} SED (\citealt{Stanway2016}; orange) that takes into account binary populations of massive stars.

As expected with increasing the hardness and number of ionizing photons per stellar population, the dwarf galaxy is $\approx30\%$ fainter overall and $\approx 0.25$ dex more metal poor. These shifts are more modest than the other numerical choices we have explored, but remain larger than the intrinsic stochasticity. We conclude that the choice of SED is a subdominant parameter in setting the stellar masses and metallicities. At present, we therefore settle on the fiducial SED for the \textsc{edge2} suite as a whole.
 
\subsubsection{The coupling between photo-ionization feedback and the star formation algorithm} \label{sec:numerics:tstar}

Figure~\ref{fig:ismstacks} shows that the \textsc{edge2} ISM has a longer high-density tail, with significant gas staying above our density threshold for star formation. This stems from the combined conditions that star-forming gas has to be dense and cold, with our fiducial setup restricting star-forming gas to have $T < T_{\star} = 1000 \, \K$ (Section~\ref{sec:sec:sec:sf}). 

Figure~\ref{fig:halo605discussion} shows two more re-simulations of our test dwarf galaxy lowering and raising this temperature threshold, while keeping the density threshold the same at $300 \, \mppercmcube$. The input SED in all cases is taken to be \textsc{bpass2.2} (Section~\ref{sec:numerics:seds}).

Lowering $T_{\star}$ to $100\, \K$ (gold) increases the stellar mass by $\approx60\%$ compared to the fiducial value with the same SED (orange), despite having made the conditions to form stars more restrictive. This is combined with an increase in $\averagefeh$ by $\approx0.2$ dex, indicating increased metal retention and weaker outflows. Effectively removing $T_{\star}$ by increasing it to $10^7 \, \K$ (brown) lowers $\Mstar$ and $\averagefeh$. 

We show in Appendix~\ref{app:sfstats} that $T_{\star}$ in fact controls the clustering of star-formation and SNe events. Following a star formation event, the dense gas surrounding the newborn stellar particle is immediately heated to $T \geq T_{\star} = 10^3\, \K$ by radiative feedback. Any gas in the immediate vicinity of a star-forming event is thus prevented from forming new stars, introducing an effective limit on the clustering of star formation. Figure~\ref{fig:sfsnstats605} shows how lowering $T_{\star}$ compounds this effect, leading to star formation and SN events forming at ever-increasing densities and in turn reducing the efficiency of galactic outflows at clearing this high-density gas. 

This behaviour reflects the difficulties of modelling stellar photo-heating, even at the high resolution of these simulations. If all processes around $\hii$ regions are captured, a star formation site would remain a mixture of cold gas below $T_{\star}$ allowed to form stars, and of warm photo-heated gas above $T_{\star}$ prevented from star formation. If, however, resolution is limited, this multiphase star formation site is numerically mixed into a partially-neutral, lukewarm phase. Restricting star formation to cold gas then unphysically shuts down correlated star formation. As shown in Appendix~\ref{app:resolvedfeedback}, our simulations are in this regime of partially-resolved radiative feedback, for which removing $T_{\star}$ provides a way to rebuild correlation between SF events that have been spuriously erased. 

\subsubsection{Convergence with dark matter resolution} \label{sec:numerics:resolution}
Using a similar \textsc{edge} setup with non-equilibrium cooling and radiative transfer, \citet{Agertz2020EDGE} demonstrated that our fiducial dark matter resolution ($\mdm \approx 950\, \Msol$) resolves well the cooling and ISM of small dark matter haloes. Re-simulating a low-mass object with eight times more dark matter particles while keeping the spatial resolution fixed, they found well converged stellar masses, sizes and iron metallicities (compare `Fiducial+RT' to `Hires + RT' in their table~2). 

We repeat this experiment with the updated \textsc{edge2} model, re-simulating one of our low-mass object, `Halo 1459 (earlier)', improving $\mdm$ from $940\, \Msol$ to $118\, \Msol$ and keeping $\softening = 3\, \pc$. At $z=0$, $\Mstar$ changes from $\xMsol{7.4}{4}$ to $\xMsol{7.9}{4}$, $\rhalflight$ from 230 to 190 pc, and $\averagefeh$ from -2.03 to -2.11. These shifts are comparable in magnitude to those induced by chaotic stochasticity (Section~\ref{sec:numerics:stochasticity}) and we conclude that the observables presented in this work are robust to the choice of dark matter resolution. 

\subsection{The impact of partially-resolved Str\"{o}mgren spheres} \label{sec:numerics:sspheres}

In Appendix~\ref{app:resolvedfeedback}, we show that at our resolution ($\Delta x = 3\, \pc$) the Str\"{o}mgren radii, $R_{S}$, of photo-ionized $\hii$ regions from radiative feedback events are only partially resolved. This is a key numerical limitation of our simulations, which has a number of consequences we discuss here.

First, an unresolved Str\"{o}mgren radius around a stellar cluster will mix two gas regions, one warm and ionized, one cold and neutral, into a single element that is lukewarm and partially ionized. Such out-of-equilibrium dense, warm and mostly-neutral gas cools very efficiently through collisional excitation, generating a source of numerically-induced cooling that is yet to be fully understood (e.g. \citealt{Deng2024RadFeedback}). 

Furthermore, this numerically-mixed gas from unresolved Str\"{o}mgren radii interacts with the temperature threshold for star formation (Section~\ref{sec:numerics:tstar}), preventing star formation in larger regions of space than should physically be allowed. This then suppresses the clustering of star formation and subsequent CCSNe feedback. In turn, this may mean that \textsc{edge2} outflows are somewhat too weak, but we stress that the impressive agreement with observed scaling relations (Section~\ref{sec:scalingrelations}) makes it unclear whether this is a significant issue. In future work, we will quantitatively compare the characteristics of our outflows to observations to obtain an orthogonal constraint (e.g. \citealt{McQuinn2019Outflows, Xu2022EmpressOutflows, Marasco2023}). 

How to best address the numerical losses from under-resolved Str\"{o}mgren spheres is a key open question for the next-generation of radiation-hydrodynamics dwarf galaxy simulations. 

An obvious solution is to simply increase numerical resolution until $R_{S}$ is resolved, providing a robust test of the physical-versus-numerical reduction in the strength of galactic outflows. Such an endeavour will be computationally demanding, but could already be achieved using super-Lagrangian strategies around star formation events similar to those implemented for black hole accretion (e.g. \citealt{Curtis2015, Beckmann2018, Angles-Alcazar2021}). However, given that numerical resolution is already at $\approx 3\, \pc$ across the ISM, going further will likely require algorithmic improvements to, for example explicitly sample the IMF in star formation events and inject radiative feedback star by star rather than as a population average as is done here. 

An alternative approach could be to keep the resolution unchanged, but revise subgrid algorithms. For example, one can probabilistically draw the ionization fraction of $\hii$ regions when they are under-resolved (\citealt{Hopkins2022FIRE3}), or allow stars to form at either lower densities or in more massive particles to boost their ionizing power and thus $R_{S}$ (e.g. \citealt{Katz2024MegPilot}). Future studies quantifying the effects of these assumptions on dwarf galaxy properties will be crucial.

Lastly, under-resolved Str\"{o}mgren spheres also have an important impact when comparing simulated and observed dwarf galaxies. Since cooling is efficient in the spurious warm-dense, partially-neutral ISM phase, it leads to much brighter emission lines luminosities compared to what should be the true luminosity of the dwarf galaxy. This requires corrections to obtain accurate estimates (e.g. \citealt{Katz2023, Ejdetjarn2024}) and will be a key problem to solve when comparing \textsc{edge2} data with observed emission lines.

\subsection{Untested theoretical uncertainties} \label{sec:numerics:systematics}

Beyond the numerical limitations and parameter exploration discussed above, there remain several astrophysical inputs needed for cosmological zoom simulations whose impact we are yet to explore.  

For example, the choice of the IMF, of the UVB, and of the chemical yields all have consequences for the cooling, heating and metallicity balance of the ISM. We have not yet been able to test all of these independently within the new \textsc{edge2} model, but many of these uncertainties have been explored in previous iterations of the \textsc{edge} model. In particular, \citet{Rey2020} discuss the consequences of varying the redshift at which cosmic reionization happens (see also \citealt{Benitez-Llambay2020, Katz2020}), \citet{Prgomet2022} investigate the impact of a metallicity-dependent IMF, while \citet{Andersson2025} show convergence tests against a new feedback budget sourced by individual stars. Future work will also present a thorough exploration of the impact of yield inputs on \textsc{edge} galaxy chemistry (E. Andersson et al. in preperation)  

And despite the milestone advance of including non-local stellar radiative input, other galaxy formation processes remain missing from this breed of \textsc{edge} dwarf galaxies. Exotic stellar evolution tracks (e.g. hypernovae, pair-instability SNe, variable-energy SNe, metal-free primordial stars) can have dramatic effects on such small dwarf galaxies (e.g. \citealt{Jeon2017, Jeon2021FaintPopIIISNe, Gutcke2021CosmologicalSim, Sanati2023}) and leave distinct traces in their chemistry. The growing abundance of evidence of active galactic nucleii in dwarf galaxies (see e.g. \citealt{Reines2013, Reines2020, Burke2022, Davis2022, Mezcua2024} and summary in \citealt{Wasleske2024}) and their associated outflows (e.g. \citealt{Liu2020, Liu2024}) calls for the inclusion of black hole processes at this mass scale, although different implementations do not converge on their importance over a Hubble time (e.g. \citealt{Koudmani2021, Koudmani2022, Sharma2023, Arjona-Galvez2024, DeAlmeida2024}). Similarly, cosmic ray feedback provides a different avenue to regulate star formation, with its efficiency in faint dwarf galaxies now starting to be quantified (e.g. \citealt{Martin-Alvarez2023}). Future extensions of the \textsc{edge} project will explore these avenues.

\section{Summary and conclusion} \label{sec:conclusion}
We present new \textsc{edge} radiation-hydrodynamics cosmological zoomed simulations covering dwarf galaxy formation from the ultra-faint to the dwarf irregular regime ($10^9 \leq \Mvir \leq 10^{10} \, \Msol$; $10^4 \leq \Mstar \leq 10^8 \, \Msol$ at $z=0$). Our combination of a uniformly high resolution ($\approx 3\, \pc$, $\mdm\approx 950 \, \Msol$), a large sample size (15 galaxies) and a detailed stellar feedback modelling (resolved SN feedback and explicit radiative feedback) is unprecedented. Leveraging previous-generation \textsc{edge} simulations with similar resolution but without radiative feedback, we systematically compare the response of dwarf galaxy observables to this change in physical modelling. 

The addition of radiative feedback leads to a fundamentally different ISM structure (Figure~\ref{fig:ismstacks}) in which star formation and gas accretion are increasingly regulated by radiative heating from massive stars, rather than mechanical removal from the halo centre. This change in regulation mode leads to a strong suppression of the mass outflow rates in dwarf galaxies over their cosmological history (Figure~\ref{fig:outflows}). These findings reaffirm that radiative feedback reduces galactic outflows in dwarf galaxies, extending previous results (\citealt{Agertz2020EDGE} for a single cosmological dwarf and e.g. \citealt{Emerick2018, Smith2020PhotoRT, Deng2024Rigel, Andersson2024ClusterMF} for isolated non-cosmological examples) to a 3-dex range in $\Mstar$. The extended mass range also shows that the suppression is the most pronounced in low-mass objects ($\Mstar \leq 10^{6} \, \Msol$) for which shallow gravitational potential wells increases sensitivity to radiative feedback (Figure~\ref{fig:tempmaps}).

Despite the significant change in outflow behaviours, stellar masses at $z=0$ are converged to $\approx 40\%$ on average across the suite, a shift well within uncertainties (Figure~\ref{fig:mstarmhalo}). In fact, most scaling relations of dwarf galaxy observables cannot distinguish between these two models. The $\magv$-$\rhalflight$ relation (Figure~\ref{fig:mvsize}), the $\MstarMhi$ relation (Figure~\ref{fig:mstarmhi}), and the $\Mstar$-$\averageOmetallicity$ relation (Figure~\ref{fig:mstaroxygen}) all showcase small shifts compared to the width of the observational scatter and its error bars. 

Only the $\magv$-$\averagefeh$ presents a strong response to our model changes, with both models tracking the observed slope of the mass-metallicity relation but respectively scattering at the lower and upper end of the observed scatter (Figure~\ref{fig:mstarmetallicity}). Higher $\averagefeh$ in the new simulations are driven by the combined update to the chemical modelling -- yield updates increasing iron production -- and the weaker galactic outflows -- that increase iron retention in the ISM (Figure~\ref{fig:halo605discussion}).

These results confirm the expectation from \citet{Agertz2020EDGE} that the $\magv$-$\averagefeh$ relation is the most sensitive probe of stellar and ISM physics in dwarf galaxies. There, we used a single dwarf galaxy ran with and without radiative transfer. This paper extends these results to a much wider range of stellar and halo masses, showing that the slope of the $\magv$-$\averagefeh$ readily emerges in both models but that the normalization of the relation directly reflects the combination of modelling choices that control iron production (e.g. CCSNe yields) and retention (e.g. outflow strength). 

A promising way to distinguish between these two effects will be to leverage the wealth of new chemical data from spectroscopic surveys (e.g. $> 50,000$ chemical abundances of dwarf galaxy stars with 4MOST; \citealt{Skuladottir2023a}). Unpicking the relative abundances of each element and their scatter will provide stronger constraints on the chemical yields of stars at low metallicities. In parallel, direct observations of dwarf galaxies' outflows, their velocities, and their mass-loading factors is now possible thanks to the advancements in IFU technology (e.g. \citealt{McQuinn2019Outflows, Xu2022EmpressOutflows, Marasco2023}). A careful comparison of these emission-line derived observations to simulations like the \textsc{edge2} suite should allow us to constrain the physical mechanisms driving galactic outflows in dwarf galaxies. Furthermore, additional dynamical scaling relations (e.g. \citealt{VandenBroucke2016, DiPaolo2019, Romeo2020}) and new observables such as the distribution of ionized metal absorbers in the CGM of dwarf irregulars (e.g. \citealt{Zheng2019, Zheng2020, Zheng2024}) can inform us on the baryon cycle of dwarf galaxies and offer orthogonal constraints on the accretion-outflow cycle to be explored in the coming years.

The robust emergence of several dwarf galaxy scaling relations and the relative convergence of our results signal that cosmological simulations of dwarf galaxies have entered an era of precision. Even with a committed theoretical variation (the inclusion of a new feedback channel) which changes by order of magnitudes the outflow behaviour of simulated dwarf galaxies, much of their general characteristics remain broadly compatible with observational data. None the less, there remains clear numerical limitations to our simulations. While SN feedback is now well resolved and converged, radiative feedback remains less accurately captured (Section~\ref{sec:numerics}). Moreover, other theoretical uncertainties that impact the strength of galactic outflows in dwarf galaxies remain to be quantified, for example the importance of active galactic nucleus or cosmic rays in this regime. Addressing these challenges will be the focus of future works from the \textsc{edge} collaboration. But ultimately, the convergence of basic scaling relations makes for an exciting moment, providing us with a solid foundation to differentiate between these mechanisms and interpret new dwarf galaxy data coming in the next decade.  

\section*{Acknowledgements}
We thank the referee for constructive criticism that improved the quality of the paper. MR thanks Astha Agarwal, Harley Katz, Taysun Kimm, Joakim Rosdahl and Mahsa Sanati for useful discussions and comments during the preparation of this manuscript. MR acknowledges support from the Beecroft Fellowship funded by Adrian Beecroft. ET acknowledges the UKRI Science and Technology Facilities Council (STFC) for support (grant ST/V50712X/1). O.A. acknowledges support from the Knut and Alice Wallenberg Foundation, the Swedish Research Council (grant 2019-04659), the Swedish National Space Agency (SNSA Dnr 2023-00164), and the LMK foundation. JIR would like to thank the STFC for support from grants ST/Y002865/1 and ST/Y002857/1. This project has received funding from the European Union’s Horizon 2020 research and innovation programme under grant agreement No. 818085 GMGalaxies. 
This work was performed using the DiRAC Data Intensive service at Leicester, operated by the University of Leicester IT Services, which forms part of the STFC DiRAC HPC Facility (www.dirac.ac.uk). The equipment was funded by BEIS capital funding via STFC capital grants ST/K000373/1 and ST/R002363/1 and STFC DiRAC Operations grant ST/R001014/1. DiRAC is part of the National e-Infrastructure. The authors acknowledge the use of the UCL Grace High Performance Computing Facility, the Surrey Eureka supercomputer facility, and their associated support services. This work was partially supported by the UCL Cosmoparticle Initiative.

We thank the developers and maintainers of \textsc{pynbody} (\citealt{Pontzen2013}), \textsc{tangos} (\citealt{Pontzen2018}), \textsc{NumPy} (\citealt{vanderWalt2011, Harris2020}), \textsc{SciPy} (\citealt{Virtanen2020}), \textsc{jupyter} (\citealt{Ragan-Kelley2014}), \textsc{matplotlib} (\citealt{Hunter2007}), the Astrophysics Data Service and the arXiv preprint repository for providing open-source softwares that were used extensively in this work.

\section*{Data Availability}
The data underlying this article will be shared upon reasonable request to the corresponding author.

\section*{Author contributions}
The main roles of the authors were, using the CRediT (Contribution Roles Taxonomy) system\footnote{\url{https://authorservices.wiley.com/author-resources/Journal-Authors/open-access/credit.html}}: 

MR: Conceptualization; Data curation; Formal analysis; Funding acquisition; Investigation; Writing – original draft. 
ET: Data curation; Investigation; Writing – review and editing.
EG: Data curation; Investigation; Writing – review and editing.
SK: Data curation; Investigation; Writing – review and editing.
EA: Investigation; Methodology; Software; Writing – review and editing.
AP: Conceptualization; Funding acquisition; Software; Writing – review and editing. 
OA: Conceptualization; Methodology; Software; Writing – review and editing.
JR: Conceptualization; Project administration; Resources; Writing – review and editing. 
CC: Methodology; Software; Writing – review and editing.
RY: Writing – review and editing.
MO: Data Curation; Investigation; Writing – review and editing. 
DS: Data curation; Writing – review and editing.
AS: Conceptualization; Writing – review and editing.
JB: Data curation; Writing – review and editing.
KQ: Writing – review and editing.
CM: Writing – review and editing.
PD: Writing – review and editing.




\bibliographystyle{mnras}
\bibliography{EDGE2-launch} 

\begin{thebibliography}{}
\makeatletter
\relax
\def\mn@urlcharsother{\let\do\@makeother \do\$\do\&\do\#\do\^\do\_\do\%\do\~}
\def\mn@doi{\begingroup\mn@urlcharsother \@ifnextchar [ {\mn@doi@} {\mn@doi@[]}}
\def\mn@doi@[#1]#2{\def\@tempa{#1}\ifx\@tempa\@empty \href {http://dx.doi.org/#2} {doi:#2}\else \href {http://dx.doi.org/#2} {#1}\fi \endgroup}
\def\mn@eprint#1#2{\mn@eprint@#1:#2::\@nil}
\def\mn@eprint@arXiv#1{\href {http://arxiv.org/abs/#1} {{\tt arXiv:#1}}}
\def\mn@eprint@dblp#1{\href {http://dblp.uni-trier.de/rec/bibtex/#1.xml} {dblp:#1}}
\def\mn@eprint@#1:#2:#3:#4\@nil{\def\@tempa {#1}\def\@tempb {#2}\def\@tempc {#3}\ifx \@tempc \@empty \let \@tempc \@tempb \let \@tempb \@tempa \fi \ifx \@tempb \@empty \def\@tempb {arXiv}\fi \@ifundefined {mn@eprint@\@tempb}{\@tempb:\@tempc}{\expandafter \expandafter \csname mn@eprint@\@tempb\endcsname \expandafter{\@tempc}}}

\bibitem[\protect\citeauthoryear{Adams \& Oosterloo}{Adams \& Oosterloo}{2018}]{Adams2018}
Adams E. A.~K.,  Oosterloo T.~A.,  2018, \mn@doi [A\&A] {10.1051/0004-6361/201732017}, 612, A26

\bibitem[\protect\citeauthoryear{Agertz, Teyssier  \& Moore}{Agertz et~al.}{2011}]{Agertz2011}
Agertz O.,  Teyssier R.,   Moore B.,  2011, \mn@doi [MNRAS] {10.1111/j.1365-2966.2010.17530.x}, 410, 1391

\bibitem[\protect\citeauthoryear{Agertz, Kravtsov, Leitner  \& Gnedin}{Agertz et~al.}{2013}]{Agertz2013}
Agertz O.,  Kravtsov A.~V.,  Leitner S.~N.,   Gnedin N.~Y.,  2013, \mn@doi [ApJ] {10.1088/0004-637X/770/1/25}, 770, 25

\bibitem[\protect\citeauthoryear{Agertz et~al.,}{Agertz et~al.}{2020}]{Agertz2020EDGE}
Agertz O.,  et~al., 2020, \mn@doi [MNRAS] {10.1093/mnras/stz3053}, 491, 1656

\bibitem[\protect\citeauthoryear{Agertz et~al.,}{Agertz et~al.}{2021}]{Agertz2020Vintergatan}
Agertz O.,  et~al., 2021, \mn@doi [MNRAS] {10.1093/mnras/stab322}, 503, 5826

\bibitem[\protect\citeauthoryear{Andersson, Mac~Low, Agertz, Renaud  \& Li}{Andersson et~al.}{2024}]{Andersson2024ClusterMF}
Andersson E.~P.,  Mac~Low M.-M.,  Agertz O.,  Renaud F.,   Li H.,  2024, \mn@doi [A\&A] {10.1051/0004-6361/202347792}, 681, 13

\bibitem[\protect\citeauthoryear{Andersson, Rey, Pontzen, Cadiou, Agertz, Read  \& Martin}{Andersson et~al.}{2025}]{Andersson2025}
Andersson E.~P.,  Rey M.~P.,  Pontzen A.,  Cadiou C.,  Agertz O.,  Read J.~I.,   Martin N.~F.,  2025, \mn@doi [ApJ] {10.3847/1538-4357/ad99d6}, 978, 129

\bibitem[\protect\citeauthoryear{{Angl{\'e}s-Alc{\'a}zar} et~al.,}{{Angl{\'e}s-Alc{\'a}zar} et~al.}{2021}]{Angles-Alcazar2021}
{Angl{\'e}s-Alc{\'a}zar} D.,  et~al., 2021, \mn@doi [ApJ] {10.3847/1538-4357/ac09e8}, 917, 53

\bibitem[\protect\citeauthoryear{{Arjona-Galvez}, Di~Cintio  \& Grand}{{Arjona-Galvez} et~al.}{2024}]{Arjona-Galvez2024}
{Arjona-Galvez} E.,  Di~Cintio A.,   Grand R. J.~J.,  2024, \mn@doi [A\&A] {10.1051/0004-6361/202449439}, 690, A286

\bibitem[\protect\citeauthoryear{Asplund, Grevesse, Sauval  \& Scott}{Asplund et~al.}{2009}]{Asplund2009}
Asplund M.,  Grevesse N.,  Sauval A.~J.,   Scott P.,  2009, \mn@doi [ARA\&A] {10.1146/annurev.astro.46.060407.145222}, 47, 481

\bibitem[\protect\citeauthoryear{Aubert \& Teyssier}{Aubert \& Teyssier}{2010}]{Aubert2010}
Aubert D.,  Teyssier R.,  2010, \mn@doi [ApJ] {10.1088/0004-637X/724/1/244}, 724, 244

\bibitem[\protect\citeauthoryear{Aubert, Pichon  \& Colombi}{Aubert et~al.}{2004}]{Aubert2004}
Aubert D.,  Pichon C.,   Colombi S.,  2004, \mn@doi [MNRAS] {10.1111/j.1365-2966.2004.07883.x}, 352, 376

\bibitem[\protect\citeauthoryear{Beckmann, Slyz  \& Devriendt}{Beckmann et~al.}{2018}]{Beckmann2018}
Beckmann R.~S.,  Slyz A.,   Devriendt J.,  2018, \mn@doi [MNRAS] {10.1093/mnras/sty931}, 478, 995

\bibitem[\protect\citeauthoryear{{Benitez-Llambay} \& Frenk}{{Benitez-Llambay} \& Frenk}{2020}]{Benitez-Llambay2020}
{Benitez-Llambay} A.,  Frenk C.,  2020, \mn@doi [MNRAS] {10.1093/mnras/staa2698}, 498, 4887

\bibitem[\protect\citeauthoryear{{Benitez-Llambay} \& Fumagalli}{{Benitez-Llambay} \& Fumagalli}{2021}]{Benitez-Llambay2021}
{Benitez-Llambay} A.,  Fumagalli M.,  2021, \mn@doi [ApJ] {10.3847/2041-8213/ac3006}, 921, L9

\bibitem[\protect\citeauthoryear{{Ben{\'i}tez-Llambay}, Navarro, Abadi, Gottl{\"o}ber, Yepes, Hoffman  \& Steinmetz}{{Ben{\'i}tez-Llambay} et~al.}{2015}]{Benitez-Llambay2015}
{Ben{\'i}tez-Llambay} A.,  Navarro J.~F.,  Abadi M.~G.,  Gottl{\"o}ber S.,  Yepes G.,  Hoffman Y.,   Steinmetz M.,  2015, \mn@doi [MNRAS] {10.1093/mnras/stv925}, 450, 4207

\bibitem[\protect\citeauthoryear{Bennet et~al.,}{Bennet et~al.}{2022}]{Bennet2022}
Bennet P.,  et~al., 2022, \mn@doi [ApJ] {10.3847/1538-4357/ac356c}, 924, 98

\bibitem[\protect\citeauthoryear{Blondin, Wright, Borkowski  \& Reynolds}{Blondin et~al.}{1998}]{Blondin1998}
Blondin J.~M.,  Wright E.~B.,  Borkowski K.~J.,   Reynolds S.~P.,  1998, \mn@doi [ApJ] {10.1086/305708}, 500, 342

\bibitem[\protect\citeauthoryear{Brauer et~al.,}{Brauer et~al.}{2025}]{Brauer2025}
Brauer K.,  et~al., 2025, \mn@doi [ApJ] {10.3847/1538-4357/ada4a1}, 980, 41

\bibitem[\protect\citeauthoryear{Bressan, Marigo, Girardi, Salasnich, Dal~Cero, Rubele  \& Nanni}{Bressan et~al.}{2012}]{Bressan2012}
Bressan A.,  Marigo P.,  Girardi {\relax L{\'e}o}.,  Salasnich B.,  Dal~Cero C.,  Rubele S.,   Nanni A.,  2012, \mn@doi [MNRAS] {10.1111/j.1365-2966.2012.21948.x}, 427, 127

\bibitem[\protect\citeauthoryear{Bruce, Li, Pace, Heiger, Song  \& Simon}{Bruce et~al.}{2023}]{Bruce2023}
Bruce J.,  Li T.~S.,  Pace A.~B.,  Heiger M.,  Song Y.-Y.,   Simon J.~D.,  2023, \mn@doi [ApJ] {10.3847/1538-4357/acc943}, 950, 167

\bibitem[\protect\citeauthoryear{Brunker et~al.,}{Brunker et~al.}{2019}]{Brunker2019}
Brunker S.~W.,  et~al., 2019, \mn@doi [AJ] {10.3847/1538-3881/aafb39}, 157, 76

\bibitem[\protect\citeauthoryear{Bruzual \& Charlot}{Bruzual \& Charlot}{2003}]{Bruzual2003}
Bruzual G.,  Charlot S.,  2003, \mn@doi [MNRAS] {10.1046/j.1365-8711.2003.06897.x}, 344, 1000

\bibitem[\protect\citeauthoryear{Bullock \& {Boylan-Kolchin}}{Bullock \& {Boylan-Kolchin}}{2017}]{Bullock2017}
Bullock J.~S.,  {Boylan-Kolchin} M.,  2017, \mn@doi [ARA\&A] {10.1146/annurev-astro-091916-055313}, 55, 343

\bibitem[\protect\citeauthoryear{Burke et~al.,}{Burke et~al.}{2022}]{Burke2022}
Burke C.~J.,  et~al., 2022, \mn@doi [MNRAS] {10.1093/mnras/stac2262}, 516, 2736

\bibitem[\protect\citeauthoryear{Buttry et~al.,}{Buttry et~al.}{2022}]{Buttry2022}
Buttry R.,  et~al., 2022, \mn@doi [MNRAS] {10.1093/mnras/stac1441}, 514, 1706

\bibitem[\protect\citeauthoryear{Cadiou, Dubois  \& Pichon}{Cadiou et~al.}{2019}]{Cadiou2019}
Cadiou C.,  Dubois Y.,   Pichon C.,  2019, \mn@doi [A\&A] {10.1051/0004-6361/201834496}, 621, A96

\bibitem[\protect\citeauthoryear{Cadiou, Pontzen  \& Peiris}{Cadiou et~al.}{2021}]{Cadiou2021AngMomGM}
Cadiou C.,  Pontzen A.,   Peiris H.~V.,  2021, \mn@doi [MNRAS] {10.1093/mnras/stab440}, 502, 5480

\bibitem[\protect\citeauthoryear{Cameron, Katz  \& Rey}{Cameron et~al.}{2023}]{Cameron2023ISMTe}
Cameron A.~J.,  Katz H.,   Rey M.~P.,  2023, \mn@doi [MNRAS] {10.1093/mnrasl/slad046}, 552, L89

\bibitem[\protect\citeauthoryear{Cerny et~al.,}{Cerny et~al.}{2023a}]{Cerny2023PegIV}
Cerny W.,  et~al., 2023a, \mn@doi [ApJ] {10.3847/1538-4357/aca1c3}, 942, 111

\bibitem[\protect\citeauthoryear{Cerny et~al.,}{Cerny et~al.}{2023b}]{Cerny2023DELVE}
Cerny W.,  et~al., 2023b, \mn@doi [ApJ] {10.3847/1538-4357/acdd78}, 953, 1

\bibitem[\protect\citeauthoryear{Cerny et~al.,}{Cerny et~al.}{2023c}]{Cerny2023DELVE6}
Cerny W.,  et~al., 2023c, \mn@doi [ApJL] {10.3847/2041-8213/aced84}, 953, L21

\bibitem[\protect\citeauthoryear{Chabrier}{Chabrier}{2003}]{Chabrier2003}
Chabrier G.,  2003, \mn@doi [PASP] {10.1086/376392}, 115, 763

\bibitem[\protect\citeauthoryear{Charles et~al.,}{Charles et~al.}{2023}]{Charles2023}
Charles E. J.~E.,  et~al., 2023, \mn@doi [MNRAS] {10.1093/mnras/stad752}, 521, 3527

\bibitem[\protect\citeauthoryear{Chisholm, Tremonti, Leitherer  \& Chen}{Chisholm et~al.}{2017}]{Chisholm2017}
Chisholm J.,  Tremonti C.~A.,  Leitherer C.,   Chen Y.,  2017, \mn@doi [MNRAS] {10.1093/mnras/stx1164}, 469, 4831

\bibitem[\protect\citeauthoryear{Chiti et~al.,}{Chiti et~al.}{2021}]{Chiti2021}
Chiti A.,  et~al., 2021, \mn@doi [Nat Ast] {10.1038/s41550-020-01285-w}, 5, 392

\bibitem[\protect\citeauthoryear{Chiti, Simon, Frebel, Pace, Ji  \& Li}{Chiti et~al.}{2022}]{Chiti2022GrusI}
Chiti A.,  Simon J.~D.,  Frebel A.,  Pace A.~B.,  Ji A.~P.,   Li T.~S.,  2022, \mn@doi [ApJ] {10.3847/1538-4357/ac96ed}, 939, 41

\bibitem[\protect\citeauthoryear{Christensen, Dav{\'e}, Governato, Pontzen, Brooks, Munshi, Quinn  \& Wadsley}{Christensen et~al.}{2016}]{Christensen2016}
Christensen C.~R.,  Dav{\'e} R.,  Governato F.,  Pontzen A.,  Brooks A.,  Munshi F.,  Quinn T.,   Wadsley J.,  2016, \mn@doi [ApJ] {10.3847/0004-637X/824/1/57}, 824, 57

\bibitem[\protect\citeauthoryear{Cioffi, McKee  \& Bertschinger}{Cioffi et~al.}{1988}]{Cioffi1988}
Cioffi D.~F.,  McKee C.~F.,   Bertschinger E.,  1988, \mn@doi [ApJ] {10.1086/166834}, 334, 252

\bibitem[\protect\citeauthoryear{Cole, Weisz, Dolphin, Skillman, McConnachie, Brooks  \& Leaman}{Cole et~al.}{2014}]{Cole2014}
Cole A.~A.,  Weisz D.~R.,  Dolphin A.~E.,  Skillman E.~D.,  McConnachie A.~W.,  Brooks A.~M.,   Leaman R.,  2014, \mn@doi [ApJ] {10.1088/0004-637X/795/1/54}, 795, 54

\bibitem[\protect\citeauthoryear{Collins \& Read}{Collins \& Read}{2022}]{Collins2022}
Collins M. L.~M.,  Read J.~I.,  2022, \mn@doi [Nat Ast] {10.1038/s41550-022-01657-4}, 6, 647

\bibitem[\protect\citeauthoryear{Collins, Tollerud, Rich, Ibata, Martin, Chapman, Gilbert  \& Preston}{Collins et~al.}{2020}]{Collins2020}
Collins M. L.~M.,  Tollerud E.~J.,  Rich R.~M.,  Ibata R.~A.,  Martin N.~F.,  Chapman S.~C.,  Gilbert K.~M.,   Preston J.,  2020, \mn@doi [MNRAS] {10.1093/mnras/stz3252}, 491, 3496

\bibitem[\protect\citeauthoryear{Collins et~al.,}{Collins et~al.}{2021}]{Collins2021}
Collins M. L.~M.,  et~al., 2021, \mn@doi [MNRAS] {10.1093/mnras/stab1624}, 505, 5686

\bibitem[\protect\citeauthoryear{Collins et~al.,}{Collins et~al.}{2024}]{Collins2024}
Collins M. L.~M.,  et~al., 2024, \mn@doi [MNRAS] {10.1093/mnras/stae199}, 528, 2614

\bibitem[\protect\citeauthoryear{C{\^o}t{\'e} et~al.,}{C{\^o}t{\'e} et~al.}{2018}]{Cote2018}
C{\^o}t{\'e} B.,  et~al., 2018, \mn@doi [ApJ] {10.3847/1538-4357/aaad67}, 855, 99

\bibitem[\protect\citeauthoryear{Courty \& Alimi}{Courty \& Alimi}{2004}]{Courty2004}
Courty S.,  Alimi J.~M.,  2004, \mn@doi [A\&A] {10.1051/0004-6361:20031736}, 416, 875

\bibitem[\protect\citeauthoryear{Curtis \& Sijacki}{Curtis \& Sijacki}{2015}]{Curtis2015}
Curtis M.,  Sijacki D.,  2015, \mn@doi [MNRAS] {10.1093/mnras/stv2246}, 454, 3445

\bibitem[\protect\citeauthoryear{Davis et~al.,}{Davis et~al.}{2022}]{Davis2022}
Davis F.,  et~al., 2022, \mn@doi [MNRAS] {10.1093/mnras/stac068}, 511, 4109

\bibitem[\protect\citeauthoryear{De~Almeida, Mamon, Dekel  \& Lima~Neto}{De~Almeida et~al.}{2024}]{DeAlmeida2024}
De~Almeida A.~P.,  Mamon G.~A.,  Dekel A.,   Lima~Neto G.~B.,  2024, \mn@doi [A\&A] {10.1051/0004-6361/202449939}, 687, A131

\bibitem[\protect\citeauthoryear{Deng, Li, Kannan, Smith, Vogelsberger  \& Bryan}{Deng et~al.}{2024a}]{Deng2024RadFeedback}
Deng Y.,  Li H.,  Kannan R.,  Smith A.,  Vogelsberger M.,   Bryan G.~L.,  2024a, \mn@doi [MNRAS] {10.1093/mnras/stad3202}, 527, 478

\bibitem[\protect\citeauthoryear{Deng, Li, Liu, Kannan, Smith  \& Bryan}{Deng et~al.}{2024b}]{Deng2024Rigel}
Deng Y.,  Li H.,  Liu B.,  Kannan R.,  Smith A.,   Bryan G.~L.,  2024b, \mn@doi [A\&A] {10.1051/0004-6361/202450699}, 691, A231

\bibitem[\protect\citeauthoryear{Di~Paolo, Salucci  \& Fontaine}{Di~Paolo et~al.}{2019}]{DiPaolo2019}
Di~Paolo C.,  Salucci P.,   Fontaine J.~P.,  2019, \mn@doi [ApJ] {10.3847/1538-4357/aaffd6}, 873, 106

\bibitem[\protect\citeauthoryear{Efstathiou}{Efstathiou}{1992}]{Efstathiou1992}
Efstathiou G.,  1992, \mn@doi [MNRAS] {10.1093/mnras/256.1.43P}, 256, 43P

\bibitem[\protect\citeauthoryear{Eisenstein \& Hut}{Eisenstein \& Hut}{1998}]{Eisenstein1998}
Eisenstein D.~J.,  Hut P.,  1998, \mn@doi [ApJ] {10.1086/305535}, 498, 137

\bibitem[\protect\citeauthoryear{Ejdetj{\"a}rn, Agertz, {\"O}stlin, Rey  \& Renaud}{Ejdetj{\"a}rn et~al.}{2024}]{Ejdetjarn2024}
Ejdetj{\"a}rn T.,  Agertz O.,  {\"O}stlin G.,  Rey M.~P.,   Renaud F.,  2024, \mn@doi [MNRAS] {10.1093/mnras/stae2099}, 534, 135

\bibitem[\protect\citeauthoryear{Emerick, Bryan  \& Mac~Low}{Emerick et~al.}{2018}]{Emerick2018}
Emerick A.,  Bryan G.~L.,   Mac~Low M.-M.,  2018, \mn@doi [ApJ] {10.3847/2041-8213/aae315}, 865, L22

\bibitem[\protect\citeauthoryear{Escala et~al.,}{Escala et~al.}{2018}]{Escala2018}
Escala I.,  et~al., 2018, \mn@doi [MNRAS] {10.1093/mnras/stx2858}, 474, 2194

\bibitem[\protect\citeauthoryear{{Faucher-Gigu{\`e}re}}{{Faucher-Gigu{\`e}re}}{2020}]{Faucher-Giguere2020}
{Faucher-Gigu{\`e}re} C.-A.,  2020, \mn@doi [MNRAS] {10.1093/mnras/staa302}, 493, 1614

\bibitem[\protect\citeauthoryear{Federrath \& Klessen}{Federrath \& Klessen}{2012}]{Federrath2012}
Federrath C.,  Klessen R.~S.,  2012, \mn@doi [ApJ] {10.1088/0004-637X/761/2/156}, 761, 156

\bibitem[\protect\citeauthoryear{Ferland et~al.,}{Ferland et~al.}{2017}]{Ferland2017}
Ferland G.~J.,  et~al., 2017, Revista Mexicana de Astronomia y Astrofisica, 53, 385

\bibitem[\protect\citeauthoryear{Fitts et~al.,}{Fitts et~al.}{2017}]{Fitts2017}
Fitts A.,  et~al., 2017, \mn@doi [MNRAS] {10.1093/mnras/stx1757}, 471, 3547

\bibitem[\protect\citeauthoryear{Fritz, Carrera, Battaglia  \& Taibi}{Fritz et~al.}{2019}]{Fritz2019}
Fritz T.~K.,  Carrera R.,  Battaglia G.,   Taibi S.,  2019, \mn@doi [A\&A] {10.1051/0004-6361/201833458}, 623, A129

\bibitem[\protect\citeauthoryear{Genel et~al.,}{Genel et~al.}{2019}]{Genel2019}
Genel S.,  et~al., 2019, \mn@doi [ApJ] {10.3847/1538-4357/aaf4bb}, 871, 21

\bibitem[\protect\citeauthoryear{{Gil-Pons}, Doherty, Guti{\'e}rrez, Siess, Campbell, Lau  \& Lattanzio}{{Gil-Pons} et~al.}{2018}]{Gil-Pons2018}
{Gil-Pons} P.,  Doherty C.~L.,  Guti{\'e}rrez J.~L.,  Siess L.,  Campbell S.~W.,  Lau H.~B.,   Lattanzio J.~C.,  2018, \mn@doi [Publ. Astron. Soc. Aust.] {10.1017/pasa.2018.42}, 35, e038

\bibitem[\protect\citeauthoryear{Gnedin}{Gnedin}{2000}]{Gnedin2000}
Gnedin N.~Y.,  2000, \mn@doi [ApJ] {10.1086/317042}, 542, 535

\bibitem[\protect\citeauthoryear{Gnedin \& Abel}{Gnedin \& Abel}{2001}]{Gnedin2001}
Gnedin N.~Y.,  Abel T.,  2001, \mn@doi [New Astronomy] {10.1016/S1384-1076(01)00068-9}, 6, 437

\bibitem[\protect\citeauthoryear{Go et~al.,}{Go et~al.}{2025}]{Go2025}
Go M.,  et~al., 2025, \mn@doi [ApJ] {10.3847/1538-4357/add2fa}, 986, 214

\bibitem[\protect\citeauthoryear{Goater et~al.,}{Goater et~al.}{2024}]{Goater2024}
Goater A.,  et~al., 2024, \mn@doi [MNRAS] {10.1093/mnras/stad3354}, 527, 2403

\bibitem[\protect\citeauthoryear{Gray et~al.,}{Gray et~al.}{2025}]{Gray2025}
Gray E.~I.,  et~al., 2025, MNRAS, 539, 1167

\bibitem[\protect\citeauthoryear{Greif, Glover, Bromm  \& Klessen}{Greif et~al.}{2010}]{Greif2010}
Greif T.~H.,  Glover S. C.~O.,  Bromm V.,   Klessen R.~S.,  2010, \mn@doi [ApJ] {10.1088/0004-637X/716/1/510}, 716, 510

\bibitem[\protect\citeauthoryear{Grisdale, Agertz, Romeo, Renaud  \& Read}{Grisdale et~al.}{2017}]{Grisdale2017}
Grisdale K.,  Agertz O.,  Romeo A.~B.,  Renaud F.,   Read J.~I.,  2017, \mn@doi [MNRAS] {10.1093/mnras/stw3133}, 466, 1093

\bibitem[\protect\citeauthoryear{Grisdale, Agertz, Renaud  \& Romeo}{Grisdale et~al.}{2018}]{Grisdale2018}
Grisdale K.,  Agertz O.,  Renaud F.,   Romeo A.~B.,  2018, \mn@doi [MNRAS] {10.1093/mnras/sty1595}, 479, 3167

\bibitem[\protect\citeauthoryear{Grisdale, Agertz, Renaud, Romeo, Devriendt  \& Slyz}{Grisdale et~al.}{2019}]{Grisdale2019}
Grisdale K.,  Agertz O.,  Renaud F.,  Romeo A.~B.,  Devriendt J.,   Slyz A.,  2019, \mn@doi [MNRAS] {10.1093/mnras/stz1201}, 486, 5482

\bibitem[\protect\citeauthoryear{Guillet \& Teyssier}{Guillet \& Teyssier}{2011}]{Guillet2011}
Guillet T.,  Teyssier R.,  2011, \mn@doi [J. Comput. Phys.] {10.1016/j.jcp.2011.02.044}, 230, 4756

\bibitem[\protect\citeauthoryear{Gutcke, Pakmor, Naab  \& Springel}{Gutcke et~al.}{2022}]{Gutcke2021CosmologicalSim}
Gutcke T.~A.,  Pakmor R.,  Naab T.,   Springel V.,  2022, \mn@doi [MNRAS] {10.1093/mnras/stac867}, 513, 1372

\bibitem[\protect\citeauthoryear{Haardt \& Madau}{Haardt \& Madau}{1996}]{Haardt1996}
Haardt F.,  Madau P.,  1996, \mn@doi [ApJ] {10.1086/177035}, 461, 20

\bibitem[\protect\citeauthoryear{Hansen, Simon, Li, Sharkey, Ji, Thompson, Reggiani  \& Galarza}{Hansen et~al.}{2024}]{Hansen2024}
Hansen T.~T.,  Simon J.~D.,  Li T.~S.,  Sharkey D.,  Ji A.~P.,  Thompson I.~B.,  Reggiani H.~M.,   Galarza J.~Y.,  2024, \mn@doi [ApJ] {10.3847/1538-4357/ad3a52}, 968, 21

\bibitem[\protect\citeauthoryear{Hargis et~al.,}{Hargis et~al.}{2020}]{Hargis2020}
Hargis J.~R.,  et~al., 2020, \mn@doi [ApJ] {10.3847/1538-4357/ab58d2}, 888, 31

\bibitem[\protect\citeauthoryear{Harris et~al.,}{Harris et~al.}{2020}]{Harris2020}
Harris C.~R.,  et~al., 2020, \mn@doi [Nature] {10.1038/s41586-020-2649-2}, 585, 357

\bibitem[\protect\citeauthoryear{Heiger et~al.,}{Heiger et~al.}{2024}]{Heiger2024}
Heiger M.~E.,  et~al., 2024, \mn@doi [ApJ] {10.3847/1538-4357/ad0cf7}, 961, 234

\bibitem[\protect\citeauthoryear{Herzog, {Ben{\'i}tez-Llambay}  \& Fumagalli}{Herzog et~al.}{2023}]{Herzog2023}
Herzog G.,  {Ben{\'i}tez-Llambay} A.,   Fumagalli M.,  2023, \mn@doi [MNRAS] {10.1093/mnras/stac3282}, 518, 6305

\bibitem[\protect\citeauthoryear{Hoeft, Yepes, Gottl{\"o}ber  \& Springel}{Hoeft et~al.}{2006}]{Hoeft2006}
Hoeft M.,  Yepes G.,  Gottl{\"o}ber S.,   Springel V.,  2006, \mn@doi [MNRAS] {10.1111/j.1365-2966.2006.10678.x}, 371, 401

\bibitem[\protect\citeauthoryear{Homma et~al.,}{Homma et~al.}{2019}]{Homma2019}
Homma D.,  et~al., 2019, \mn@doi [PASJ] {10.1093/pasj/psz076}, 71, 94

\bibitem[\protect\citeauthoryear{Homma et~al.,}{Homma et~al.}{2024}]{Homma2024}
Homma D.,  et~al., 2024, \mn@doi [PASJ] {10.1093/pasj/psae044}, 76, 733

\bibitem[\protect\citeauthoryear{Hopkins}{Hopkins}{2025}]{Hopkins2025SNVelocities}
Hopkins P.~F.,  2025, OpJA, 8, 44

\bibitem[\protect\citeauthoryear{Hopkins et~al.,}{Hopkins et~al.}{2018}]{Hopkins2018}
Hopkins P.~F.,  et~al., 2018, \mn@doi [MNRAS] {10.1093/mnras/sty1690}, 480, 800

\bibitem[\protect\citeauthoryear{Hopkins et~al.,}{Hopkins et~al.}{2022}]{Hopkins2022FIRE3}
Hopkins P.~F.,  et~al., 2022, \mn@doi [MNRAS] {10.1093/mnras/stac3489}, 519, 3154

\bibitem[\protect\citeauthoryear{Hui \& Gnedin}{Hui \& Gnedin}{1997}]{Hui1997}
Hui L.,  Gnedin N.~Y.,  1997, \mn@doi [MNRAS] {10.1093/mnras/292.1.27}, 292, 27

\bibitem[\protect\citeauthoryear{Hunter}{Hunter}{2007}]{Hunter2007}
Hunter J.~D.,  2007, \mn@doi [CiSE] {10.1109/MCSE.2007.55}, 9, 90

\bibitem[\protect\citeauthoryear{Jaacks, Thompson, Finkelstein  \& Bromm}{Jaacks et~al.}{2018}]{Jaacks2018}
Jaacks J.,  Thompson R.,  Finkelstein S.~L.,   Bromm V.,  2018, \mn@doi [MNRAS] {10.1093/mnras/sty062}, 475, 4396

\bibitem[\protect\citeauthoryear{Janesh, Rhode, Salzer, Janowiecki, Adams, Haynes, Giovanelli  \& Cannon}{Janesh et~al.}{2019}]{Janesh2019}
Janesh W.,  Rhode K.~L.,  Salzer J.~J.,  Janowiecki S.,  Adams E. A.~K.,  Haynes M.~P.,  Giovanelli R.,   Cannon J.~M.,  2019, \mn@doi [AJ] {10.3847/1538-3881/ab12d3}, 157, 183

\bibitem[\protect\citeauthoryear{Jenkins, Li, Pace, Ji, Koposov  \& {Mutlu-Pakdil}}{Jenkins et~al.}{2021}]{Jenkins2021}
Jenkins S.~A.,  Li T.~S.,  Pace A.~B.,  Ji A.~P.,  Koposov S.~E.,   {Mutlu-Pakdil} B.,  2021, \mn@doi [ApJ] {10.3847/1538-4357/ac1353}, 920, 92

\bibitem[\protect\citeauthoryear{Jeon, Besla  \& Bromm}{Jeon et~al.}{2017}]{Jeon2017}
Jeon M.,  Besla G.,   Bromm V.,  2017, \mn@doi [ApJ] {10.3847/1538-4357/aa8c80}, 848, 85

\bibitem[\protect\citeauthoryear{Jeon, Bromm, Besla, Yoon  \& Choi}{Jeon et~al.}{2021}]{Jeon2021FaintPopIIISNe}
Jeon M.,  Bromm V.,  Besla G.,  Yoon J.,   Choi Y.,  2021, \mn@doi [MNRAS] {10.1093/mnras/staa4017}, 502, 1

\bibitem[\protect\citeauthoryear{Ji, Frebel, Simon  \& Geha}{Ji et~al.}{2016a}]{Ji2016b}
Ji A.~P.,  Frebel A.,  Simon J.~D.,   Geha M.,  2016a, \mn@doi [ApJ] {10.3847/0004-637X/817/1/41}, 817, 41

\bibitem[\protect\citeauthoryear{Ji, Frebel, Simon  \& Chiti}{Ji et~al.}{2016b}]{Ji2016c}
Ji A.~P.,  Frebel A.,  Simon J.~D.,   Chiti A.,  2016b, \mn@doi [ApJ] {10.3847/0004-637X/830/2/93}, 830, 93

\bibitem[\protect\citeauthoryear{Ji, Frebel, Ezzeddine  \& Casey}{Ji et~al.}{2016c}]{Ji2016a}
Ji A.~P.,  Frebel A.,  Ezzeddine R.,   Casey A.~R.,  2016c, \mn@doi [ApJ] {10.3847/2041-8205/832/1/L3}, 832, L3

\bibitem[\protect\citeauthoryear{Ji, Simon, Frebel, Venn  \& Hansen}{Ji et~al.}{2019}]{Ji2019}
Ji A.~P.,  Simon J.~D.,  Frebel A.,  Venn K.~A.,   Hansen T.~T.,  2019, \mn@doi [ApJ] {10.3847/1538-4357/aaf3bb}, 870, 83

\bibitem[\protect\citeauthoryear{Ji et~al.,}{Ji et~al.}{2020}]{Ji2020}
Ji A.~P.,  et~al., 2020, \mn@doi [ApJ] {10.3847/1538-4357/ab6213}, 889, 27

\bibitem[\protect\citeauthoryear{Ji et~al.,}{Ji et~al.}{2021}]{Ji2021}
Ji A.~P.,  et~al., 2021, \mn@doi [ApJ] {10.3847/1538-4357/ac1869}, 921, 32

\bibitem[\protect\citeauthoryear{Jones et~al.,}{Jones et~al.}{2023}]{Jones2023PAVODwarf}
Jones M.~G.,  et~al., 2023, \mn@doi [ApJL] {10.3847/2041-8213/ad0130}, 957, L5

\bibitem[\protect\citeauthoryear{Jones et~al.,}{Jones et~al.}{2024}]{Jones2024Corvus}
Jones M.~G.,  et~al., 2024, \mn@doi [ApJL] {10.3847/2041-8213/ad676e}, 971, L37

\bibitem[\protect\citeauthoryear{Katz et~al.,}{Katz et~al.}{2020}]{Katz2020}
Katz H.,  et~al., 2020, \mn@doi [MNRAS] {10.1093/mnras/staa639}, 494, 2200

\bibitem[\protect\citeauthoryear{Katz et~al.,}{Katz et~al.}{2023}]{Katz2023}
Katz H.,  et~al., 2023, \mn@doi [Open J. of Astrophysics] {10.21105/astro.2309.03269}, 6, 44

\bibitem[\protect\citeauthoryear{Katz, Rey, Cadiou, Kimm  \& Agertz}{Katz et~al.}{2024}]{Katz2024MegPilot}
Katz H.,  Rey M.~P.,  Cadiou C.,  Kimm T.,   Agertz O.,  2024, The {{Impact}} of {{Star Formation}} and {{Feedback Recipes}} on the {{Stellar Mass}} and {{Interstellar Medium}} of {{High-Redshift Galaxies}} (\mn@eprint {arXiv} {2411.07282})

\bibitem[\protect\citeauthoryear{Keller, Wadsley, Wang  \& Kruijssen}{Keller et~al.}{2019}]{Keller2019}
Keller B.~W.,  Wadsley J.~W.,  Wang L.,   Kruijssen J. M.~D.,  2019, \mn@doi [MNRAS] {10.1093/mnras/sty2859}, 482, 2244

\bibitem[\protect\citeauthoryear{Kennicutt}{Kennicutt}{1998}]{Kennicutt1998}
Kennicutt Jr. R.~C.,  1998, \mn@doi [ApJ] {10.1086/305588}, 498, 541

\bibitem[\protect\citeauthoryear{Kewley \& Ellison}{Kewley \& Ellison}{2008}]{Kewley2008}
Kewley L.~J.,  Ellison S.~L.,  2008, \mn@doi [ApJ] {10.1086/587500}, 681, 1183

\bibitem[\protect\citeauthoryear{Kim \& Ostriker}{Kim \& Ostriker}{2015}]{Kim2015}
Kim C.-G.,  Ostriker E.~C.,  2015, \mn@doi [ApJ] {10.1088/0004-637X/802/2/99}, 802, 99

\bibitem[\protect\citeauthoryear{Kim et~al.,}{Kim et~al.}{2014}]{Kim2014}
Kim J.-h.,  et~al., 2014, \mn@doi [ApJS] {10.1088/0067-0049/210/1/14}, 210, 14

\bibitem[\protect\citeauthoryear{Kim et~al.,}{Kim et~al.}{2024}]{Kim2024}
Kim S.~Y.,  et~al., 2024, {{EDGE}}: {{Predictable Scatter}} in the {{Stellar Mass--Halo Mass Relation}} of {{Dwarf Galaxies}} (\mn@eprint {arXiv} {2408.15214})

\bibitem[\protect\citeauthoryear{Kimm \& Cen}{Kimm \& Cen}{2014}]{Kimm2014}
Kimm T.,  Cen R.,  2014, \mn@doi [ApJ] {10.1088/0004-637X/788/2/121}, 788, 121

\bibitem[\protect\citeauthoryear{Kimm, Cen, Devriendt, Dubois  \& Slyz}{Kimm et~al.}{2015}]{Kimm2015}
Kimm T.,  Cen R.,  Devriendt J.,  Dubois Y.,   Slyz A.,  2015, \mn@doi [MNRAS] {10.1093/mnras/stv1211}, 451, 2900

\bibitem[\protect\citeauthoryear{Kirby, Cohen, Guhathakurta, Cheng, Bullock  \& Gallazzi}{Kirby et~al.}{2013}]{Kirby2013}
Kirby E.~N.,  Cohen J.~G.,  Guhathakurta P.,  Cheng L.,  Bullock J.~S.,   Gallazzi A.,  2013, \mn@doi [ApJ] {10.1088/0004-637X/779/2/102}, 779, 102

\bibitem[\protect\citeauthoryear{Kirby, Bullock, {Boylan-Kolchin}, Kaplinghat  \& Cohen}{Kirby et~al.}{2014}]{Kirby2014}
Kirby E.~N.,  Bullock J.~S.,  {Boylan-Kolchin} M.,  Kaplinghat M.,   Cohen J.~G.,  2014, \mn@doi [MNRAS] {10.1093/mnras/stu025}, 439, 1015

\bibitem[\protect\citeauthoryear{Kirby, Rizzi, Held, Cohen, Cole, Manning, Skillman  \& Weisz}{Kirby et~al.}{2017}]{Kirby2017}
Kirby E.~N.,  Rizzi L.,  Held E.~V.,  Cohen J.~G.,  Cole A.~A.,  Manning E.~M.,  Skillman E.~D.,   Weisz D.~R.,  2017, \mn@doi [ApJ] {10.3847/1538-4357/834/1/9}, 834, 9

\bibitem[\protect\citeauthoryear{Kirby, Gilbert, Escala, Wojno, Guhathakurta, Majewski  \& Beaton}{Kirby et~al.}{2020}]{Kirby2020}
Kirby E.~N.,  Gilbert K.~M.,  Escala I.,  Wojno J.,  Guhathakurta P.,  Majewski S.~R.,   Beaton R.~L.,  2020, \mn@doi [AJ] {10.3847/1538-3881/ab5f0f}, 159, 46

\bibitem[\protect\citeauthoryear{Koribalski et~al.,}{Koribalski et~al.}{2020}]{Koribalski2020}
Koribalski B.~S.,  et~al., 2020, \mn@doi [ASS] {10.1007/s10509-020-03831-4}, 365, 118

\bibitem[\protect\citeauthoryear{Koudmani, Henden  \& Sijacki}{Koudmani et~al.}{2021}]{Koudmani2021}
Koudmani S.,  Henden N.~A.,   Sijacki D.,  2021, \mn@doi [MNRAS] {10.1093/mnras/stab677}, 503, 3568

\bibitem[\protect\citeauthoryear{Koudmani, Sijacki  \& Smith}{Koudmani et~al.}{2022}]{Koudmani2022}
Koudmani S.,  Sijacki D.,   Smith M.~C.,  2022, \mn@doi [MNRAS] {10.1093/mnras/stac2252}, 516, 2112

\bibitem[\protect\citeauthoryear{Kravtsov \& Manwadkar}{Kravtsov \& Manwadkar}{2022}]{Kravtsov2022}
Kravtsov A.,  Manwadkar V.,  2022, \mn@doi [MNRAS] {10.1093/mnras/stac1439}, 514, 2667

\bibitem[\protect\citeauthoryear{Kroupa}{Kroupa}{2001}]{Kroupa2001}
Kroupa P.,  2001, \mn@doi [MNRAS] {10.1046/j.1365-8711.2001.04022.x}, 322, 231

\bibitem[\protect\citeauthoryear{Kvasova, Kirby  \& Beaton}{Kvasova et~al.}{2024}]{Kvasova2024}
Kvasova K.,  Kirby E.~N.,   Beaton R.~L.,  2024, \mn@doi [ApJ] {10.3847/1538-4357/ad55f0}, 972, 180

\bibitem[\protect\citeauthoryear{{LIGO and Virgo Scientific Collaboration} et~al.,}{{LIGO and Virgo Scientific Collaboration} et~al.}{2017}]{LIGO2017BNS}
{LIGO and Virgo Scientific Collaboration} et~al., 2017, \mn@doi [PRL] {10.1103/PhysRevLett.119.161101}, 119, 161101

\bibitem[\protect\citeauthoryear{Lewis, Challinor  \& Lasenby}{Lewis et~al.}{2000}]{Lewis2000}
Lewis A.,  Challinor A.,   Lasenby A.,  2000, \mn@doi [ApJ] {10.1086/309179}, 538, 473

\bibitem[\protect\citeauthoryear{Li et~al.,}{Li et~al.}{2017}]{Li2017}
Li T.~S.,  et~al., 2017, \mn@doi [ApJ] {10.3847/1538-4357/aa6113}, 838, 8

\bibitem[\protect\citeauthoryear{Li et~al.,}{Li et~al.}{2018}]{Li2018}
Li T.~S.,  et~al., 2018, \mn@doi [ApJ] {10.3847/1538-4357/aab666}, 857, 145

\bibitem[\protect\citeauthoryear{Li, Greene, Carlsten  \& Danieli}{Li et~al.}{2024}]{Li2024Hedgehog}
Li J.,  Greene J.~E.,  Carlsten S.~G.,   Danieli S.,  2024, \mn@doi [ApJL] {10.3847/2041-8213/ad5b59}, 975, L23

\bibitem[\protect\citeauthoryear{Limongi \& Chieffi}{Limongi \& Chieffi}{2018}]{Limongi2018}
Limongi M.,  Chieffi A.,  2018, \mn@doi [ApJS] {10.3847/1538-4365/aacb24}, 237, 13

\bibitem[\protect\citeauthoryear{Limongi, Roberti, Chieffi  \& Nomoto}{Limongi et~al.}{2024}]{Limongi2024ECSNe}
Limongi M.,  Roberti L.,  Chieffi A.,   Nomoto K.,  2024, \mn@doi [ApJS] {10.3847/1538-4365/ad12c1}, 270, 28

\bibitem[\protect\citeauthoryear{Liu, Veilleux, Canalizo, Rupke, {Manzano-King}, Bohn  \& U}{Liu et~al.}{2020}]{Liu2020}
Liu W.,  Veilleux S.,  Canalizo G.,  Rupke D. S.~N.,  {Manzano-King} C.~M.,  Bohn T.,   U V.,  2020, \mn@doi [ApJ] {10.3847/1538-4357/abc269}, 905, 166

\bibitem[\protect\citeauthoryear{Liu et~al.,}{Liu et~al.}{2024}]{Liu2024}
Liu W.,  et~al., 2024, \mn@doi [ApJ] {10.3847/1538-4357/ad2b63}, 965, 152

\bibitem[\protect\citeauthoryear{Longeard et~al.,}{Longeard et~al.}{2018}]{Longeard2018}
Longeard N.,  et~al., 2018, \mn@doi [MNRAS] {10.1093/mnras/sty1986}, 480, 2609

\bibitem[\protect\citeauthoryear{Maoz, Mannucci  \& Brandt}{Maoz et~al.}{2012}]{Maoz2012}
Maoz D.,  Mannucci F.,   Brandt T.~D.,  2012, \mn@doi [MNRAS] {10.1111/j.1365-2966.2012.21871.x}, 426, 3282

\bibitem[\protect\citeauthoryear{Marasco et~al.,}{Marasco et~al.}{2023}]{Marasco2023}
Marasco A.,  et~al., 2023, \mn@doi [A\&A] {10.1051/0004-6361/202244895}, 670, A92

\bibitem[\protect\citeauthoryear{{Martin-Alvarez}, Sijacki, Haehnelt, Farcy, Dubois, Belokurov, Rosdahl  \& {Lopez-Rodriguez}}{{Martin-Alvarez} et~al.}{2023}]{Martin-Alvarez2023}
{Martin-Alvarez} S.,  Sijacki D.,  Haehnelt M.~G.,  Farcy M.,  Dubois Y.,  Belokurov V.,  Rosdahl J.,   {Lopez-Rodriguez} E.,  2023, \mn@doi [MNRAS] {10.1093/mnras/stad2559}, 525, 3806

\bibitem[\protect\citeauthoryear{{Martinez-Delgado}, Stein, Pawlowski, Makarov, Makarova, Donatiello  \& Lang}{{Martinez-Delgado} et~al.}{2024}]{Martinez-Delgado2024}
{Martinez-Delgado} D.,  Stein M.,  Pawlowski M.~S.,  Makarov D.,  Makarova L.,  Donatiello G.,   Lang D.,  2024, Tracing Satellite Planes in the {{Sculptor}} Group: {{II}}. {{Discovery}} of Five Faint Dwarf Galaxies in the {{DESI Legacy Survey}} (\mn@eprint {arXiv} {2405.03769})

\bibitem[\protect\citeauthoryear{Martizzi, {Faucher-Gigu{\`e}re}  \& Quataert}{Martizzi et~al.}{2015}]{Martizzi2015}
Martizzi D.,  {Faucher-Gigu{\`e}re} C.-A.,   Quataert E.,  2015, \mn@doi [MNRAS] {10.1093/mnras/stv562}, 450, 504

\bibitem[\protect\citeauthoryear{Mau et~al.,}{Mau et~al.}{2020}]{Mau2020}
Mau S.,  et~al., 2020, \mn@doi [ApJ] {10.3847/1538-4357/ab6c67}, 890, 136

\bibitem[\protect\citeauthoryear{McConnachie}{McConnachie}{2012}]{McConnachie2012}
McConnachie A.~W.,  2012, \mn@doi [AJ] {10.1088/0004-6256/144/1/4}, 144, 4

\bibitem[\protect\citeauthoryear{McConnachie \& Venn}{McConnachie \& Venn}{2020}]{McConnachie2020}
McConnachie A.~W.,  Venn K.~A.,  2020, \mn@doi [AJ] {10.3847/1538-3881/aba4ab}, 160, 124

\bibitem[\protect\citeauthoryear{McConnachie, Higgs, Thomas, Venn, C{\^o}t{\'e}, Battaglia  \& Lewis}{McConnachie et~al.}{2021}]{McConnachie2021}
McConnachie A.~W.,  Higgs C.~R.,  Thomas G.~F.,  Venn K.~A.,  C{\^o}t{\'e} P.,  Battaglia G.,   Lewis G.~F.,  2021, \mn@doi [MNRAS] {10.1093/mnras/staa3740}, 501, 2363

\bibitem[\protect\citeauthoryear{McNanna et~al.,}{McNanna et~al.}{2024}]{McNanna2024}
McNanna M.,  et~al., 2024, \mn@doi [ApJ] {10.3847/1538-4357/ad07d0}, 961, 126

\bibitem[\protect\citeauthoryear{McQuinn}{McQuinn}{2016}]{McQuinn2016}
McQuinn M.,  2016, \mn@doi [ARA\&A] {10.1146/annurev-astro-082214-122355}, 54, 313

\bibitem[\protect\citeauthoryear{McQuinn et~al.,}{McQuinn et~al.}{2015}]{McQuinn2015}
McQuinn K. B.~W.,  et~al., 2015, \mn@doi [ApJ] {10.1088/0004-637X/812/2/158}, 812, 158

\bibitem[\protect\citeauthoryear{McQuinn, {van Zee}  \& Skillman}{McQuinn et~al.}{2019}]{McQuinn2019Outflows}
McQuinn {\relax Kristen}. B.~W.,  {van Zee} L.,   Skillman E.~D.,  2019, \mn@doi [ApJ] {10.3847/1538-4357/ab4c37}, 886, 74

\bibitem[\protect\citeauthoryear{McQuinn et~al.,}{McQuinn et~al.}{2020}]{McQuinn2020}
McQuinn {\relax Kristen}. B.~W.,  et~al., 2020, \mn@doi [ApJ] {10.3847/1538-4357/ab7447}, 891, 181

\bibitem[\protect\citeauthoryear{McQuinn et~al.,}{McQuinn et~al.}{2021}]{McQuinn2021}
McQuinn K. B.~W.,  et~al., 2021, \mn@doi [ApJ] {10.3847/1538-4357/ac03ae}, 918, 23

\bibitem[\protect\citeauthoryear{McQuinn, Mao, Buckley, Shih, Cohen  \& Dolphin}{McQuinn et~al.}{2023}]{McQuinn2023}
McQuinn K. B.~W.,  Mao Y.-Y.,  Buckley M.~R.,  Shih D.,  Cohen R.~E.,   Dolphin A.~E.,  2023, \mn@doi [ApJ] {10.3847/1538-4357/acaec9}, 944, 14

\bibitem[\protect\citeauthoryear{Mezcua \& S{\'a}nchez}{Mezcua \& S{\'a}nchez}{2024}]{Mezcua2024}
Mezcua M.,  S{\'a}nchez H.~D.,  2024, \mn@doi [MNRAS] {10.1093/mnras/stae292}, 528, 5252

\bibitem[\protect\citeauthoryear{Mitchell, Schaye, Bower  \& Crain}{Mitchell et~al.}{2020}]{Mitchell2020}
Mitchell P.~D.,  Schaye J.,  Bower R.~G.,   Crain R.~A.,  2020, \mn@doi [MNRAS] {10.1093/mnras/staa938}, 494, 3971

\bibitem[\protect\citeauthoryear{Monzon, van~den Bosch  \& Mitra}{Monzon et~al.}{2024}]{Monzon2024}
Monzon J.~S.,  van~den Bosch F.~C.,   Mitra K.,  2024, \mn@doi [ApJ] {10.3847/1538-4357/ad834e}, 976, 197

\bibitem[\protect\citeauthoryear{Muni, Pontzen, Read, Agertz, Rey  \& Taylor}{Muni et~al.}{2025}]{Muni2024EDGECores}
Muni C.,  Pontzen A.,  Read J.~I.,  Agertz O.,  Rey M.~P.,   Taylor E.,  2025, \mn@doi [MNRAS] {10.1093/mnras/stae2748}, 536, 314

\bibitem[\protect\citeauthoryear{Munshi, Brooks, Christensen, Applebaum, {Holley-Bockelmann}, Quinn  \& Wadsley}{Munshi et~al.}{2019}]{Munshi2019}
Munshi F.,  Brooks A.~M.,  Christensen C.,  Applebaum E.,  {Holley-Bockelmann} K.,  Quinn T.~R.,   Wadsley J.,  2019, \mn@doi [ApJ] {10.3847/1538-4357/ab0085}, 874, 40

\bibitem[\protect\citeauthoryear{Munshi, Brooks, Applebaum, Christensen, Sligh  \& Quinn}{Munshi et~al.}{2021}]{Munshi2021}
Munshi F.,  Brooks A.,  Applebaum E.,  Christensen C.,  Sligh J.~P.,   Quinn T.,  2021, \mn@doi [ApJ] {10.3847/1538-4357/ac0db6}, 923, 35

\bibitem[\protect\citeauthoryear{Muratov, Kere{\v s}, {Faucher-Gigu{\`e}re}, Hopkins, Quataert  \& Murray}{Muratov et~al.}{2015}]{Muratov2015}
Muratov A.~L.,  Kere{\v s} D.,  {Faucher-Gigu{\`e}re} C.-A.,  Hopkins P.~F.,  Quataert E.,   Murray N.,  2015, \mn@doi [MNRAS] {10.1093/mnras/stv2126}, 454, 2691

\bibitem[\protect\citeauthoryear{{Mutlu-Pakdil} et~al.,}{{Mutlu-Pakdil} et~al.}{2021}]{Mutlu-Pakdil2021}
{Mutlu-Pakdil} B.,  et~al., 2021, \mn@doi [ApJ] {10.3847/1538-4357/ac0db8}, 918, 88

\bibitem[\protect\citeauthoryear{Naab \& Ostriker}{Naab \& Ostriker}{2017}]{Naab2017}
Naab T.,  Ostriker J.~P.,  2017, \mn@doi [ARA\&A] {10.1146/annurev-astro-081913-040019}, 55, 59

\bibitem[\protect\citeauthoryear{Nadler et~al.,}{Nadler et~al.}{2020}]{Nadler2020}
Nadler E.~O.,  et~al., 2020, \mn@doi [ApJ] {10.3847/1538-4357/ab846a}, 893, 48

\bibitem[\protect\citeauthoryear{Naiman et~al.,}{Naiman et~al.}{2018}]{Naiman2018}
Naiman J.~P.,  et~al., 2018, \mn@doi [MNRAS] {10.1093/mnras/sty618}, 477, 1206

\bibitem[\protect\citeauthoryear{Nakajima et~al.,}{Nakajima et~al.}{2022}]{Nakajima2022}
Nakajima K.,  et~al., 2022, \mn@doi [ApJS] {10.3847/1538-4365/ac7710}, 262, 3

\bibitem[\protect\citeauthoryear{Nebrin}{Nebrin}{2023}]{Nebrin2023}
Nebrin O.,  2023, \mn@doi [RAA] {10.3847/2515-5172/acd37a}, 7, 90

\bibitem[\protect\citeauthoryear{Nelson et~al.,}{Nelson et~al.}{2019}]{Nelson2019TNGOutflows}
Nelson D.,  et~al., 2019, \mn@doi [MNRAS] {10.1093/mnras/stz2306}, 490, 3234

\bibitem[\protect\citeauthoryear{Nguyen et~al.,}{Nguyen et~al.}{2022}]{Nguyen2022}
Nguyen C.~T.,  et~al., 2022, \mn@doi [A\&A] {10.1051/0004-6361/202244166}, 665, A126

\bibitem[\protect\citeauthoryear{Nickerson, Teyssier  \& Rosdahl}{Nickerson et~al.}{2018}]{Nickerson2018}
Nickerson S.,  Teyssier R.,   Rosdahl J.,  2018, \mn@doi [MNRAS] {10.1093/mnras/sty1556}, 479, 3206

\bibitem[\protect\citeauthoryear{Noh \& McQuinn}{Noh \& McQuinn}{2014}]{Noh2014}
Noh Y.,  McQuinn M.,  2014, \mn@doi [MNRAS] {10.1093/mnras/stu1412}, 444, 503

\bibitem[\protect\citeauthoryear{O'Leary, Steinwandel, Moster, Martin  \& Naab}{O'Leary et~al.}{2023}]{OLeary2023}
O'Leary J.~A.,  Steinwandel U.~P.,  Moster B.~P.,  Martin N.,   Naab T.,  2023, \mn@doi [MNRAS] {10.1093/mnras/stad166}, 520, 897

\bibitem[\protect\citeauthoryear{Ohlin, Renaud  \& Agertz}{Ohlin et~al.}{2019}]{Ohlin2019}
Ohlin L.,  Renaud F.,   Agertz O.,  2019, \mn@doi [MNRAS] {10.1093/mnras/stz705}, 485, 3887

\bibitem[\protect\citeauthoryear{Okamoto, Gao  \& Theuns}{Okamoto et~al.}{2008}]{Okamoto2008}
Okamoto T.,  Gao L.,   Theuns T.,  2008, \mn@doi [MNRAS] {10.1111/j.1365-2966.2008.13830.x}, 390, 920

\bibitem[\protect\citeauthoryear{O{\~n}orbe, Hennawi  \& Luki{\'c}}{O{\~n}orbe et~al.}{2017}]{Onorbe2017}
O{\~n}orbe J.,  Hennawi J.~F.,   Luki{\'c} Z.,  2017, \mn@doi [ApJ] {10.3847/1538-4357/aa6031}, 837, 106

\bibitem[\protect\citeauthoryear{Orkney et~al.,}{Orkney et~al.}{2021}]{Orkney2021}
Orkney M. D.~A.,  et~al., 2021, \mn@doi [MNRAS] {10.1093/mnras/stab1066}, 504, 3509

\bibitem[\protect\citeauthoryear{Orkney et~al.,}{Orkney et~al.}{2022}]{Orkney2022}
Orkney M. D.~A.,  et~al., 2022, \mn@doi [MNRAS] {10.1093/mnras/stac1755}, 515, 185

\bibitem[\protect\citeauthoryear{Orkney, Taylor, Read, Rey, Pontzen, Agertz, Kim  \& Delorme}{Orkney et~al.}{2023}]{Orkney2023}
Orkney M. D.~A.,  Taylor E.,  Read J.~I.,  Rey M.~P.,  Pontzen A.,  Agertz O.,  Kim S.~Y.,   Delorme M.,  2023, \mn@doi [MNRAS] {10.1093/mnras/stad2516}, 525, 3516

\bibitem[\protect\citeauthoryear{Pace et~al.,}{Pace et~al.}{2020}]{Pace2020}
Pace A.~B.,  et~al., 2020, \mn@doi [MNRAS] {10.1093/mnras/staa1419}, 495, 3022

\bibitem[\protect\citeauthoryear{Padoan, Haugb{\o}lle  \& Nordlund}{Padoan et~al.}{2012}]{Padoan2012}
Padoan P.,  Haugb{\o}lle T.,   Nordlund {\AA}.,  2012, \mn@doi [ApJL] {10.1088/2041-8205/759/2/L27}, 759, L27

\bibitem[\protect\citeauthoryear{Pandya et~al.,}{Pandya et~al.}{2021}]{Pandya2021}
Pandya V.,  et~al., 2021, \mn@doi [MNRAS] {10.1093/mnras/stab2714}, 508, 2979

\bibitem[\protect\citeauthoryear{Pignatari et~al.,}{Pignatari et~al.}{2016}]{Pignatari2016}
Pignatari M.,  et~al., 2016, \mn@doi [ApJS] {10.3847/0067-0049/225/2/24}, 225, 24

\bibitem[\protect\citeauthoryear{{Planck Collaboration} et~al.,}{{Planck Collaboration} et~al.}{2014}]{PlanckCollaboration2014}
{Planck Collaboration} et~al., 2014, \mn@doi [A\&A] {10.1051/0004-6361/201321591}, 571, A16

\bibitem[\protect\citeauthoryear{{Planck Collaboration} et~al.,}{{Planck Collaboration} et~al.}{2020}]{PlanckCollaboration2020}
{Planck Collaboration} et~al., 2020, \mn@doi [A\&A] {10.1051/0004-6361/201833910}, 641, A6

\bibitem[\protect\citeauthoryear{Pontzen \& Governato}{Pontzen \& Governato}{2014}]{Pontzen2014}
Pontzen A.,  Governato F.,  2014, \mn@doi [Nature] {10.1038/nature12953}, 506, 171

\bibitem[\protect\citeauthoryear{Pontzen \& Tremmel}{Pontzen \& Tremmel}{2018}]{Pontzen2018}
Pontzen A.,  Tremmel M.,  2018, \mn@doi [ApJS] {10.3847/1538-4365/aac832}, 237, 23

\bibitem[\protect\citeauthoryear{Pontzen, Ro{\v s}kar, Stinson  \& Woods}{Pontzen et~al.}{2013}]{Pontzen2013}
Pontzen A.,  Ro{\v s}kar R.,  Stinson G.,   Woods R.,  2013, Astrophysics Source Code Library, p. ascl:1305.002

\bibitem[\protect\citeauthoryear{Pontzen, Tremmel, Roth, Peiris, Saintonge, Volonteri, Quinn  \& Governato}{Pontzen et~al.}{2017}]{Pontzen2017}
Pontzen A.,  Tremmel M.,  Roth N.,  Peiris H.~V.,  Saintonge A.,  Volonteri M.,  Quinn T.,   Governato F.,  2017, \mn@doi [MNRAS] {10.1093/mnras/stw2627}, 465, 547

\bibitem[\protect\citeauthoryear{Pontzen, Rey, Cadiou, Agertz, Teyssier, Read  \& Orkney}{Pontzen et~al.}{2021}]{Pontzen2021}
Pontzen A.,  Rey M.~P.,  Cadiou C.,  Agertz O.,  Teyssier R.,  Read J.,   Orkney M. D.~A.,  2021, \mn@doi [MNRAS] {10.1093/mnras/staa3645}, 501, 1755

\bibitem[\protect\citeauthoryear{Power, Navarro, Jenkins, Frenk, White, Springel, Stadel  \& Quinn}{Power et~al.}{2003}]{Power2003}
Power C.,  Navarro J.~F.,  Jenkins A.,  Frenk C.~S.,  White S. D.~M.,  Springel V.,  Stadel J.,   Quinn T.,  2003, \mn@doi [MNRAS] {10.1046/j.1365-8711.2003.05925.x}, 338, 14

\bibitem[\protect\citeauthoryear{Prgomet, Rey, Andersson, Segovia~Otero, Agertz, Renaud, Pontzen  \& Read}{Prgomet et~al.}{2022}]{Prgomet2022}
Prgomet M.,  Rey M.~P.,  Andersson E.~P.,  Segovia~Otero A.,  Agertz O.,  Renaud F.,  Pontzen A.,   Read J.~I.,  2022, \mn@doi [MNRAS] {10.1093/mnras/stac1074}, 513, 2326

\bibitem[\protect\citeauthoryear{{Ragan-Kelley}, Perez, Granger, Kluyver, Ivanov, Frederic  \& Bussonnier}{{Ragan-Kelley} et~al.}{2014}]{Ragan-Kelley2014}
{Ragan-Kelley} M.,  Perez F.,  Granger B.,  Kluyver T.,  Ivanov P.,  Frederic J.,   Bussonnier M.,  2014, Am. Geophys. Un., 2014, H44D

\bibitem[\protect\citeauthoryear{Raiteri, Villata  \& Navarro}{Raiteri et~al.}{1996}]{Raiteri1996}
Raiteri C.~M.,  Villata M.,   Navarro J.~F.,  1996, A\&A, 315, 105

\bibitem[\protect\citeauthoryear{Rasera \& Teyssier}{Rasera \& Teyssier}{2006}]{Rasera2006}
Rasera Y.,  Teyssier R.,  2006, \mn@doi [A\&A] {10.1051/0004-6361:20053116}, 445, 1

\bibitem[\protect\citeauthoryear{Read, Iorio, Agertz  \& Fraternali}{Read et~al.}{2017}]{Read2017}
Read J.~I.,  Iorio G.,  Agertz O.,   Fraternali F.,  2017, \mn@doi [MNRAS] {10.1093/mnras/stx147}, 467, 2019

\bibitem[\protect\citeauthoryear{Reines, Greene  \& Geha}{Reines et~al.}{2013}]{Reines2013}
Reines A.~E.,  Greene J.~E.,   Geha M.,  2013, \mn@doi [ApJ] {10.1088/0004-637X/775/2/116}, 775, 116

\bibitem[\protect\citeauthoryear{Reines, Condon, Darling  \& Greene}{Reines et~al.}{2020}]{Reines2020}
Reines A.~E.,  Condon J.~J.,  Darling J.,   Greene J.~E.,  2020, \mn@doi [ApJ] {10.3847/1538-4357/ab4999}, 888, 36

\bibitem[\protect\citeauthoryear{Revaz \& Jablonka}{Revaz \& Jablonka}{2018}]{Revaz2018}
Revaz Y.,  Jablonka P.,  2018, \mn@doi [A\&A] {10.1051/0004-6361/201832669}, 616, A96

\bibitem[\protect\citeauthoryear{Rey \& Pontzen}{Rey \& Pontzen}{2018}]{Rey2018}
Rey M.~P.,  Pontzen A.,  2018, \mn@doi [MNRAS] {10.1093/mnras/stx2744}, 474, 45

\bibitem[\protect\citeauthoryear{Rey, Pontzen  \& Saintonge}{Rey et~al.}{2019a}]{Rey2019VarianceDMOs}
Rey M.~P.,  Pontzen A.,   Saintonge A.,  2019a, \mn@doi [MNRAS] {10.1093/mnras/stz552}, 485, 1906

\bibitem[\protect\citeauthoryear{Rey, Pontzen, Agertz, Orkney, Read, Saintonge  \& Pedersen}{Rey et~al.}{2019b}]{Rey2019UFDScatter}
Rey M.~P.,  Pontzen A.,  Agertz O.,  Orkney M. D.~A.,  Read J.~I.,  Saintonge A.,   Pedersen C.,  2019b, \mn@doi [ApJL] {10.3847/2041-8213/ab53dd}, 886, L3

\bibitem[\protect\citeauthoryear{Rey, Pontzen, Agertz, Orkney, Read  \& Rosdahl}{Rey et~al.}{2020}]{Rey2020}
Rey M.~P.,  Pontzen A.,  Agertz O.,  Orkney M. D.~A.,  Read J.~I.,   Rosdahl J.,  2020, \mn@doi [MNRAS] {10.1093/mnras/staa1640}, 497, 1508

\bibitem[\protect\citeauthoryear{Rey, Pontzen, Agertz, Orkney, Read, Saintonge, Kim  \& Das}{Rey et~al.}{2022}]{Rey2022EDGEHI}
Rey M.~P.,  Pontzen A.,  Agertz O.,  Orkney M. D.~A.,  Read J.~I.,  Saintonge A.,  Kim S.~Y.,   Das P.,  2022, \mn@doi [MNRAS] {10.1093/mnras/stac502}, 511, 5672

\bibitem[\protect\citeauthoryear{Rey, Katz, Cameron, Devriendt  \& Slyz}{Rey et~al.}{2024a}]{Rey2024Outflows}
Rey M.~P.,  Katz H.~B.,  Cameron A.~J.,  Devriendt J.,   Slyz A.,  2024a, \mn@doi [MNRAS] {10.1093/mnras/stae388}, 528, 5412

\bibitem[\protect\citeauthoryear{Rey et~al.,}{Rey et~al.}{2024b}]{Rey2024EDGERCs}
Rey M.~P.,  et~al., 2024b, \mn@doi [MNRAS] {10.1093/mnras/stae718}, 529, 2379

\bibitem[\protect\citeauthoryear{Richstein et~al.,}{Richstein et~al.}{2022}]{Richstein2022}
Richstein H.,  et~al., 2022, \mn@doi [ApJ] {10.3847/1538-4357/ac7226}, 933, 217

\bibitem[\protect\citeauthoryear{Ritter, Herwig, Jones, Pignatari, Fryer  \& Hirschi}{Ritter et~al.}{2018}]{Ritter2018}
Ritter C.,  Herwig F.,  Jones S.,  Pignatari M.,  Fryer C.,   Hirschi R.,  2018, \mn@doi [MNRAS] {10.1093/mnras/sty1729}, 480, 538

\bibitem[\protect\citeauthoryear{Romeo}{Romeo}{2020}]{Romeo2020}
Romeo A.~B.,  2020, \mn@doi [MNRAS] {10.1093/mnras/stz3367}, 491, 4843

\bibitem[\protect\citeauthoryear{Rosdahl \& Blaizot}{Rosdahl \& Blaizot}{2012}]{Rosdahl2012}
Rosdahl J.,  Blaizot J.,  2012, \mn@doi [MNRAS] {10.1111/j.1365-2966.2012.20883.x}, 423, 344

\bibitem[\protect\citeauthoryear{Rosdahl \& Teyssier}{Rosdahl \& Teyssier}{2015}]{Rosdahl2015RAMSESRT}
Rosdahl J.,  Teyssier R.,  2015, \mn@doi [MNRAS] {10.1093/mnras/stv567}, 449, 4380

\bibitem[\protect\citeauthoryear{Rosdahl, Blaizot, Aubert, Stranex  \& Teyssier}{Rosdahl et~al.}{2013}]{Rosdahl2013}
Rosdahl J.,  Blaizot J.,  Aubert D.,  Stranex T.,   Teyssier R.,  2013, \mn@doi [MNRAS] {10.1093/mnras/stt1722}, 436, 2188

\bibitem[\protect\citeauthoryear{Rosdahl et~al.,}{Rosdahl et~al.}{2018}]{Rosdahl2018}
Rosdahl J.,  et~al., 2018, \mn@doi [MNRAS] {10.1093/mnras/sty1655}, 479, 994

\bibitem[\protect\citeauthoryear{Rosen \& Bregman}{Rosen \& Bregman}{1995}]{Rosen1995}
Rosen A.,  Bregman J.~N.,  1995, \mn@doi [ApJ] {10.1086/175303}, 440, 634

\bibitem[\protect\citeauthoryear{Roth, Pontzen  \& Peiris}{Roth et~al.}{2016}]{Roth2016}
Roth N.,  Pontzen A.,   Peiris H.~V.,  2016, \mn@doi [MNRAS] {10.1093/mnras/stv2375}, 455, 974

\bibitem[\protect\citeauthoryear{Sales, Wetzel  \& Fattahi}{Sales et~al.}{2022}]{Sales2022}
Sales L.~V.,  Wetzel A.,   Fattahi A.,  2022, \mn@doi [Nat Ast] {10.1038/s41550-022-01689-w}, 6, 897

\bibitem[\protect\citeauthoryear{Sanati, Jeanquartier, Revaz  \& Jablonka}{Sanati et~al.}{2023}]{Sanati2023}
Sanati M.,  Jeanquartier F.,  Revaz Y.,   Jablonka P.,  2023, \mn@doi [A\&A] {10.1051/0004-6361/202244309}, 669, A94

\bibitem[\protect\citeauthoryear{Sand, Spekkens, Crnojevi{\'c}, Hargis, Willman, Strader  \& Grillmair}{Sand et~al.}{2015}]{Sand2015}
Sand D.~J.,  Spekkens K.,  Crnojevi{\'c} D.,  Hargis J.~R.,  Willman B.,  Strader J.,   Grillmair C.~J.,  2015, \mn@doi [ApJL] {10.1088/2041-8205/812/1/L13}, 812, L13

\bibitem[\protect\citeauthoryear{Sand et~al.,}{Sand et~al.}{2022}]{Sand2022}
Sand D.~J.,  et~al., 2022, \mn@doi [ApJL] {10.3847/2041-8213/ac85ee}, 935, L17

\bibitem[\protect\citeauthoryear{Schmidt}{Schmidt}{1959}]{Schmidt1959}
Schmidt M.,  1959, \mn@doi [ApJ] {10.1086/146614}, 129, 243

\bibitem[\protect\citeauthoryear{Scholte et~al.,}{Scholte et~al.}{2025}]{Scholte2025}
Scholte D.,  et~al., 2025, \mn@doi [MNRAS] {10.1093/mnras/stae2477}, 535, 2341

\bibitem[\protect\citeauthoryear{Seitenzahl et~al.,}{Seitenzahl et~al.}{2013}]{Seitenzahl2013}
Seitenzahl I.~R.,  et~al., 2013, \mn@doi [MNRAS] {10.1093/mnras/sts402}, 429, 1156

\bibitem[\protect\citeauthoryear{Sharma, Brooks, Tremmel, Bellovary  \& Quinn}{Sharma et~al.}{2023}]{Sharma2023}
Sharma R.~S.,  Brooks A.~M.,  Tremmel M.,  Bellovary J.,   Quinn T.~R.,  2023, \mn@doi [ApJ] {10.3847/1538-4357/ace046}, 957, 16

\bibitem[\protect\citeauthoryear{Simon}{Simon}{2018}]{Simon2018}
Simon J.~D.,  2018, \mn@doi [ApJ] {10.3847/1538-4357/aacdfb}, 863, 89

\bibitem[\protect\citeauthoryear{Simon}{Simon}{2019}]{Simon2019}
Simon J.~D.,  2019, \mn@doi [ARA\&A] {10.1146/annurev-astro-091918-104453}, 57, 375

\bibitem[\protect\citeauthoryear{Simon et~al.,}{Simon et~al.}{2023}]{Simon2023}
Simon J.~D.,  et~al., 2023, \mn@doi [ApJ] {10.3847/1538-4357/aca9d1}, 944, 43

\bibitem[\protect\citeauthoryear{Sk{\'u}lad{\'o}ttir et~al.,}{Sk{\'u}lad{\'o}ttir et~al.}{2023}]{Skuladottir2023a}
Sk{\'u}lad{\'o}ttir {\'A}.,  et~al., 2023, \mn@doi [The Messenger] {10.18727/0722-6691/5304}, 190, 19

\bibitem[\protect\citeauthoryear{Smith}{Smith}{2021}]{Smith2021IMF}
Smith M.~C.,  2021, \mn@doi [MNRAS] {10.1093/mnras/stab291}, 502, 5417

\bibitem[\protect\citeauthoryear{Smith et~al.,}{Smith et~al.}{2017}]{Smith2017}
Smith B.~D.,  et~al., 2017, \mn@doi [MNRAS] {10.1093/mnras/stw3291}, 466, 2217

\bibitem[\protect\citeauthoryear{Smith, Bryan, Somerville, Hu, Teyssier, Burkhart  \& Hernquist}{Smith et~al.}{2021}]{Smith2020PhotoRT}
Smith M.~C.,  Bryan G.~L.,  Somerville R.~S.,  Hu C.-Y.,  Teyssier R.,  Burkhart B.,   Hernquist L.,  2021, \mn@doi [MNRAS] {10.1093/mnras/stab1896}, 506, 3882

\bibitem[\protect\citeauthoryear{Smith et~al.,}{Smith et~al.}{2023}]{Smith2023BootesV}
Smith S. E.~T.,  et~al., 2023, \mn@doi [AJ] {10.3847/1538-3881/acdd77}, 166, 76

\bibitem[\protect\citeauthoryear{Smith et~al.,}{Smith et~al.}{2024}]{Smith2024Faintest}
Smith S. E.~T.,  et~al., 2024, \mn@doi [ApJ] {10.3847/1538-4357/ad0d9f}, 961, 92

\bibitem[\protect\citeauthoryear{Snaith, Park, Kim  \& Rosdahl}{Snaith et~al.}{2018}]{Snaith2018}
Snaith O.~N.,  Park C.,  Kim J.,   Rosdahl J.,  2018, \mn@doi [MNRAS] {10.1093/mnras/sty673}, 477, 983

\bibitem[\protect\citeauthoryear{Stanway, Eldridge  \& Becker}{Stanway et~al.}{2016}]{Stanway2016}
Stanway E.~R.,  Eldridge J.~J.,   Becker G.~D.,  2016, \mn@doi [MNRAS] {10.1093/mnras/stv2661}, 456, 485

\bibitem[\protect\citeauthoryear{Stopyra, Pontzen, Peiris, Roth  \& Rey}{Stopyra et~al.}{2021a}]{Stopyra2021}
Stopyra S.,  Pontzen A.,  Peiris H.,  Roth N.,   Rey M.~P.,  2021a, \mn@doi [ApJS] {10.3847/1538-4365/abcd94}, 252, 28

\bibitem[\protect\citeauthoryear{Stopyra, Peiris  \& Pontzen}{Stopyra et~al.}{2021b}]{Stopyra2021Voids}
Stopyra S.,  Peiris H.~V.,   Pontzen A.,  2021b, \mn@doi [MNRAS] {10.1093/mnras/staa3587}, 500, 4173

\bibitem[\protect\citeauthoryear{Str{\"o}mgren}{Str{\"o}mgren}{1939}]{Stromgren1939}
Str{\"o}mgren B.,  1939, \mn@doi [ApJ] {10.1086/144074}, 89, 526

\bibitem[\protect\citeauthoryear{Taibi, Battaglia, Rejkuba, Leaman, Kacharov, Iorio, Jablonka  \& Zoccali}{Taibi et~al.}{2020}]{Taibi2020}
Taibi S.,  Battaglia G.,  Rejkuba M.,  Leaman R.,  Kacharov N.,  Iorio G.,  Jablonka P.,   Zoccali M.,  2020, \mn@doi [A\&A] {10.1051/0004-6361/201937240}, 635, A152

\bibitem[\protect\citeauthoryear{Tan et~al.,}{Tan et~al.}{2025}]{Tan2025}
Tan C.~Y.,  et~al., 2025, \mn@doi [ApJ] {10.3847/1538-4357/ad9b0c}, 979, 176

\bibitem[\protect\citeauthoryear{Tarumi, Yoshida  \& Frebel}{Tarumi et~al.}{2021}]{Tarumi2021StellarHalo}
Tarumi Y.,  Yoshida N.,   Frebel A.,  2021, \mn@doi [ApJL] {10.3847/2041-8213/ac024e}, 914, L10

\bibitem[\protect\citeauthoryear{Teyssier}{Teyssier}{2002}]{Teyssier2002}
Teyssier R.,  2002, \mn@doi [A\&A] {10.1051/0004-6361:20011817}, 385, 337

\bibitem[\protect\citeauthoryear{Thielemann, Nomoto  \& Yokoi}{Thielemann et~al.}{1986}]{Thielemann1986}
Thielemann F.~K.,  Nomoto K.,   Yokoi K.,  1986, A\&A, 158, 17

\bibitem[\protect\citeauthoryear{Thornton, Gaudlitz, Janka  \& Steinmetz}{Thornton et~al.}{1998}]{Thornton1998}
Thornton K.,  Gaudlitz M.,  Janka H.~T.,   Steinmetz M.,  1998, \mn@doi [ApJ] {10.1086/305704}, 500, 95

\bibitem[\protect\citeauthoryear{Tollet et~al.,}{Tollet et~al.}{2016}]{Tollet2016}
Tollet E.,  et~al., 2016, \mn@doi [MNRAS] {10.1093/mnras/stv2856}, 456, 3542

\bibitem[\protect\citeauthoryear{Toro, Spruce  \& Speares}{Toro et~al.}{1994}]{Toro1994}
Toro E.~F.,  Spruce M.,   Speares W.,  1994, \mn@doi [Shock Waves] {10.1007/BF01414629}, 4, 25

\bibitem[\protect\citeauthoryear{Torrealba, Koposov, Belokurov  \& Irwin}{Torrealba et~al.}{2016}]{Torrealba2016}
Torrealba G.,  Koposov S.~E.,  Belokurov V.,   Irwin M.,  2016, \mn@doi [MNRAS] {10.1093/mnras/stw733}, 459, 2370

\bibitem[\protect\citeauthoryear{Torrealba et~al.,}{Torrealba et~al.}{2018}]{Torrealba2018}
Torrealba G.,  et~al., 2018, \mn@doi [MNRAS] {10.1093/mnras/sty170}, 475, 5085

\bibitem[\protect\citeauthoryear{Torrealba et~al.,}{Torrealba et~al.}{2019}]{Torrealba2019}
Torrealba G.,  et~al., 2019, \mn@doi [MNRAS] {10.1093/mnras/stz1624}, 488, 2743

\bibitem[\protect\citeauthoryear{Tweed, Devriendt, Blaizot, Colombi  \& Slyz}{Tweed et~al.}{2009}]{Tweed2009}
Tweed D.,  Devriendt J.,  Blaizot J.,  Colombi S.,   Slyz A.,  2009, \mn@doi [A\&A] {10.1051/0004-6361/200911787}, 506, 647

\bibitem[\protect\citeauthoryear{Vandenbroucke, Verbeke  \& De~Rijcke}{Vandenbroucke et~al.}{2016}]{VandenBroucke2016}
Vandenbroucke B.,  Verbeke R.,   De~Rijcke S.,  2016, \mn@doi [MNRAS] {10.1093/mnras/stw328}, 458, 912

\bibitem[\protect\citeauthoryear{Virtanen et~al.,}{Virtanen et~al.}{2020}]{Virtanen2020}
Virtanen P.,  et~al., 2020, \mn@doi [Nat Methods] {10.1038/s41592-019-0686-2}, 17, 261

\bibitem[\protect\citeauthoryear{Visbal, Bryan  \& Haiman}{Visbal et~al.}{2020}]{Visbal2020}
Visbal E.,  Bryan G.~L.,   Haiman Z.,  2020, \mn@doi [ApJ] {10.3847/1538-4357/ab994e}, 897, 95

\bibitem[\protect\citeauthoryear{Wang, Dutton, Stinson, Macci{\`o}, Penzo, Kang, Keller  \& Wadsley}{Wang et~al.}{2015}]{Wang2015}
Wang L.,  Dutton A.~A.,  Stinson G.~S.,  Macci{\`o} A.~V.,  Penzo C.,  Kang X.,  Keller B.~W.,   Wadsley J.,  2015, \mn@doi [MNRAS] {10.1093/mnras/stv1937}, 454, 83

\bibitem[\protect\citeauthoryear{Wang, Nadler, Mao, Adhikari, Wechsler  \& Behroozi}{Wang et~al.}{2021}]{Wang2021}
Wang Y.,  Nadler E.~O.,  Mao Y.-Y.,  Adhikari S.,  Wechsler R.~H.,   Behroozi P.,  2021, \mn@doi [ApJ] {10.3847/1538-4357/ac024a}, 915, 116

\bibitem[\protect\citeauthoryear{Wasleske \& Baldassare}{Wasleske \& Baldassare}{2024}]{Wasleske2024}
Wasleske E.~J.,  Baldassare V.~F.,  2024, \mn@doi [ApJ] {10.3847/1538-4357/ad5442}, 971, 68

\bibitem[\protect\citeauthoryear{Wheeler et~al.,}{Wheeler et~al.}{2019}]{Wheeler2019}
Wheeler C.,  et~al., 2019, \mn@doi [MNRAS] {10.1093/mnras/stz2887}, 490, 4447

\bibitem[\protect\citeauthoryear{Wojno, Gilbert, Kirby, Escala, Beaton, Tollerud, Majewski  \& Guhathakurta}{Wojno et~al.}{2020}]{Wojno2020}
Wojno J.,  Gilbert K.~M.,  Kirby E.~N.,  Escala I.,  Beaton R.~L.,  Tollerud E.~J.,  Majewski S.~R.,   Guhathakurta P.,  2020, \mn@doi [ApJ] {10.3847/1538-4357/ab8ccb}, 895, 78

\bibitem[\protect\citeauthoryear{Woosley \& Heger}{Woosley \& Heger}{2007}]{Woosley2007}
Woosley S.~E.,  Heger A.,  2007, \mn@doi [Phys. Rep.] {10.1016/j.physrep.2007.02.009}, 442, 269

\bibitem[\protect\citeauthoryear{Wright, Brooks, Weisz  \& Christensen}{Wright et~al.}{2019}]{Wright2019}
Wright A.~C.,  Brooks A.~M.,  Weisz D.~R.,   Christensen C.~R.,  2019, \mn@doi [MNRAS] {10.1093/mnras/sty2759}, 482, 1176

\bibitem[\protect\citeauthoryear{Xu et~al.,}{Xu et~al.}{2022}]{Xu2022EmpressOutflows}
Xu Y.,  et~al., 2022, \mn@doi [ApJ] {10.3847/1538-4357/ac5e32}, 929, 134

\bibitem[\protect\citeauthoryear{Xu, Zhu, Yu, Zhang, Liu, Ai  \& Jiang}{Xu et~al.}{2023}]{Xu2023FASTGasRichDwarf}
Xu J.-L.,  Zhu M.,  Yu N.,  Zhang C.-P.,  Liu X.-L.,  Ai M.,   Jiang P.,  2023, \mn@doi [ApJL] {10.3847/2041-8213/acb932}, 944, L40

\bibitem[\protect\citeauthoryear{Yates, Schady, Chen, Schweyer  \& Wiseman}{Yates et~al.}{2020}]{Yates2020}
Yates R.~M.,  Schady P.,  Chen T.-W.,  Schweyer T.,   Wiseman P.,  2020, \mn@doi [A\&A] {10.1051/0004-6361/201936506}, 634, A107

\bibitem[\protect\citeauthoryear{Zel'dovich}{Zel'dovich}{1970}]{Zeldovich1970}
Zel'dovich {\relax Ya}.~B.,  1970, A\&A, 5, 84

\bibitem[\protect\citeauthoryear{Zheng et~al.,}{Zheng et~al.}{2019}]{Zheng2019}
Zheng Y.,  et~al., 2019, \mn@doi [MNRAS] {10.1093/mnras/stz2563}, 490, 467

\bibitem[\protect\citeauthoryear{Zheng, Emerick, Putman, Werk, Kirby  \& Peek}{Zheng et~al.}{2020}]{Zheng2020}
Zheng Y.,  Emerick A.,  Putman M.~E.,  Werk J.~K.,  Kirby E.~N.,   Peek J. E.~G.,  2020, \mn@doi [ApJ] {10.3847/1538-4357/abc875}, 2020, 133

\bibitem[\protect\citeauthoryear{Zheng et~al.,}{Zheng et~al.}{2024}]{Zheng2024}
Zheng Y.,  et~al., 2024, \mn@doi [ApJ] {10.3847/1538-4357/acfe6b}, 960, 21

\bibitem[\protect\citeauthoryear{{van Cappellen} et~al.,}{{van Cappellen} et~al.}{2022}]{vanCappellen2022}
{van Cappellen} W.~A.,  et~al., 2022, \mn@doi [A\&A] {10.1051/0004-6361/202141739}, 658, A146

\bibitem[\protect\citeauthoryear{{van der Walt}, Colbert  \& Varoquaux}{{van der Walt} et~al.}{2011}]{vanderWalt2011}
{van der Walt} S.,  Colbert S.~C.,   Varoquaux G.,  2011, \mn@doi [Comput. Sci. Eng.] {10.1109/MCSE.2011.37}, 13, 22

\makeatother
\end{thebibliography}



\appendix

\section{Summary of EDGE1 and EDGE2 properties} \label{app:table}

Table~\ref{tab:simulations} summarizes the properties of the simulated dwarf galaxies with the \textsc{edge1} and \textsc{edge2} models at $z=0$.

\begin{table*}
  \centering
  \caption{Properties of simulated dwarf galaxies with the \textsc{edge1} and \textsc{edge2} models (E1 and E2 columns, respectively) at $z=0$ (see Section~\ref{sec:scalingrelations} for details on how each property is computed). Haloes with only an \textsc{edge2} column are new initial conditions that do not have an \textsc{edge1} counterpart (see Section~\ref{sec:sec:ics}). Two haloes were stopped at higher redshifts due to their computational cost (see Section~\ref{sec:sec:mstarmhalo}) -- their properties are indicated at the labelled redshift.}
  \resizebox{\textwidth}{!}{
   \begin{tabular}{|p{2.5cm}|c|c|c|c|c|c|c|c|c|c|c|c|c|c|c|}
    \hline
    Object & \multicolumn{1}{c|}{$\Mvir$ ($\Msol$)} & \multicolumn{2}{c|}{$\Mstar$ ($\Msol$)} & \multicolumn{2}{c|}{$\magv$} & \multicolumn{2}{c|}{$\rhalflight$ ($\kpc$)} & \multicolumn{2}{c|}{$\Mhi$ ($\Msol$)} & \multicolumn{2}{c|}{$\averagefeh$} & \multicolumn{2}{c|}{$\averageOmetallicity$} \\   \hline
    &  & E1 & E2 & E1 & E2 & E1 & E2 & E1 & E2 & E1 & E2 & E1 & E2 \\
    \hline
    Halo 1445 & $\xScientific{1.2}{9}$ & $\xScientific{7.9}{4}$ & $\xScientific{5.8}{4}$ & -6.35 & -5.36 & 0.20 & 0.29 & $\xScientific{1.2}{1}$ & $\xScientific{4.9}{1}$ & -2.58 & -2.24 & 6.31 & 7.25 \\
    Halo 1459: GM earlier & $\xScientific{1.4}{9}$ & $\xScientific{4.9}{5}$ & $\xScientific{4.4}{5}$ & -8.29 & -7.50 & 0.16 & 0.22 & $\xScientific{1.3}{1}$ & $\xScientific{4.3}{1}$ & -2.34 & -1.78 & 6.63 & 8.39 \\
    Halo 1459 & $\xScientific{1.4}{9}$ & $\xScientific{2.6}{5}$ & $\xScientific{3.6}{5}$ & -7.62 & -7.29 & 0.23 & 0.25 & $\xScientific{1.9}{1}$ & $\xScientific{4.9}{1}$ & -2.40 & -2.04 & 6.73 & 8.04 \\
    Halo 1459: GM later & $\xScientific{1.4}{9}$ & $\xScientific{1.2}{5}$ & $\xScientific{7.5}{4}$ & -6.84 & -5.60 & 0.31 & 0.28 & $\xScientific{1.7}{1}$ & $\xScientific{7.4}{1}$ & -2.58 & -2.03 & 6.55 & 7.25 \\
    Halo 1459: GM latest & $\xScientific{1.4}{9}$ & $\xScientific{4.5}{4}$ & $\xScientific{1.5}{4}$ & -5.72 & -3.94 & 1.00 & 2.63 & $\xScientific{2.0}{1}$ & $\xScientific{1.0}{2}$ & -2.93 & -2.53 & 6.44 & 6.35 \\
    Halo 600 & $\xScientific{3.4}{9}$ & $\xScientific{5.1}{5}$ & $\xScientific{6.2}{5}$ & -8.77 & -9.83 & 0.23 & 0.30 & $\xScientific{5.7}{5}$ & $\xScientific{1.8}{6}$ & -2.43 & -1.49 & 7.35 & 7.37 \\
    Halo 600: GM later & $\xScientific{3.2}{9}$ & $\xScientific{3.9}{5}$ & $\xScientific{6.4}{4}$ & -8.18 & -7.00 & 0.38 & 0.10 & $\xScientific{2.1}{5}$ & $\xScientific{3.9}{5}$ & -2.50 & -2.70 & 6.64 & 6.80 \\
    Halo 605 & $\xScientific{3.3}{9}$ & $\xScientific{1.7}{6}$ & $\xScientific{2.9}{6}$ & -9.73 & -9.91 & 0.26 & 0.32 & $\xScientific{3.2}{5}$ & $\xScientific{7.5}{5}$ & -2.21 & -1.35 & 7.30 & 7.73 \\
    Halo 624 & $\xScientific{2.5}{9}$ & $\xScientific{5.9}{5}$ & $\xScientific{4.5}{5}$ & -8.93 & -8.12 & 0.38 & 0.63 & $\xScientific{2.3}{5}$ & $\xScientific{3.1}{5}$ & -2.30 & -1.98 & 6.68 & 7.27 \\
    Halo 624: GM higher mass & $\xScientific{3.8}{9}$ & $\xScientific{1.5}{6}$ & $\xScientific{1.9}{6}$ & -9.63 & -9.83 & 0.31 & 0.25 & $\xScientific{2.3}{5}$ & $\xScientific{1.6}{6}$ & -2.17 & -1.60 & 7.42 & 7.50 \\
    Halo 383: GM earlier & $\xScientific{5.8}{9}$ & $\xScientific{4.4}{6}$ & $\xScientific{2.1}{7}$ & -10.88 & -12.19 & 0.33 & 0.49 & $\xScientific{3.8}{5}$ & $\xScientific{3.6}{6}$ & -1.99 & -0.96 & 7.70 & 8.03 \\
    Halo 383 & $\xScientific{5.7}{9}$ & $\xScientific{3.2}{6}$ & $\xScientific{9.4}{6}$ & -10.49 & -11.85 & 0.42 & 0.57 & $\xScientific{5.8}{2}$ & $\xScientific{7.7}{6}$ & -2.02 & -1.10 & 7.10 & 7.74 \\
    Halo 383: stochastic & $\xScientific{5.9}{9}$ & $\xScientific{9.6}{6}$ & $\xScientific{1.1}{7}$ & -11.76 & -12.23 & 0.17 & 0.35 & $\xScientific{5.5}{1}$ & $\xScientific{4.5}{6}$ & -1.87 & -1.05 & 7.38 & 7.84 \\
    Halo 383: GM later & $\xScientific{5.8}{9}$ & $\xScientific{3.3}{6}$ & $\xScientific{3.9}{6}$ & -10.85 & -11.53 & 0.06 & 0.78 & $\xScientific{5.9}{2}$ & $\xScientific{1.8}{5}$ & -2.07 & -1.26 & 7.78 & 7.38 \\
    Halo 383: GM higher mass ($z=0.9$) & $\xScientific{7.5}{9}$ & $\xScientific{1.1}{7}$ & $\xScientific{7.4}{7}$ & -12.45 & -14.00 & 0.81 & 0.55 & $\xScientific{7.8}{6}$ &$\xScientific{3.5}{7}$ & -1.68 & -1.60 & 7.57 & 7.12 \\
    Halo 339 & $\xScientific{5.2}{9}$ & -- & $\xScientific{5.8}{6}$ & -- & -11.10 & -- & 0.56 & -- & $\xScientific{3.5}{6}$ & -- & -1.05 & -- & 7.83 \\
    Halo 261: GM earlier & $\xScientific{6.9}{9}$ & -- & $\xScientific{1.9}{7}$ & -- & -12.40 & -- & 0.37 & -- & $\xScientific{5.1}{6}$ & -- & -1.02 & -- & 7.90 \\
    Halo 261  & $\xScientific{6.6}{9}$ & -- & $\xScientific{1.5}{7}$ & -- & -12.20 & -- & 0.51 & -- & $\xScientific{7.5}{6}$ & -- & -1.06 & -- & 7.82 \\
    Halo 261: GM later & $\xScientific{6.6}{9}$ & -- & $\xScientific{9.7}{6}$ & -- & -12.65 & -- & 0.46 & -- & $\xScientific{1.5}{7}$ & -- & -1.15 & -- & 7.59 \\
    Halo 153 (z=0.6) & $\xScientific{1.1}{10}$ & -- & $\xScientific{4.2}{7}$ & -- & -14.69 & -- & 0.56 & -- & $\xScientific{4.3}{7}$ & -- & -1.02 & -- & 7.89  \\
    \hline
  \end{tabular}
  }
  \label{tab:simulations}
\end{table*}

\section{Resolved feedback in \textsc{edge2} simulations} \label{app:resolvedfeedback}

The strength of the \textsc{edge} approach lies in its uncompromisingly high resolution to directly resolve key stellar feedback processes in the ISM of our dwarf galaxies. We quantify this statement in this appendix. 

\subsection{SN feedback}

\begin{figure}
  \centering
    \includegraphics[width=\columnwidth]{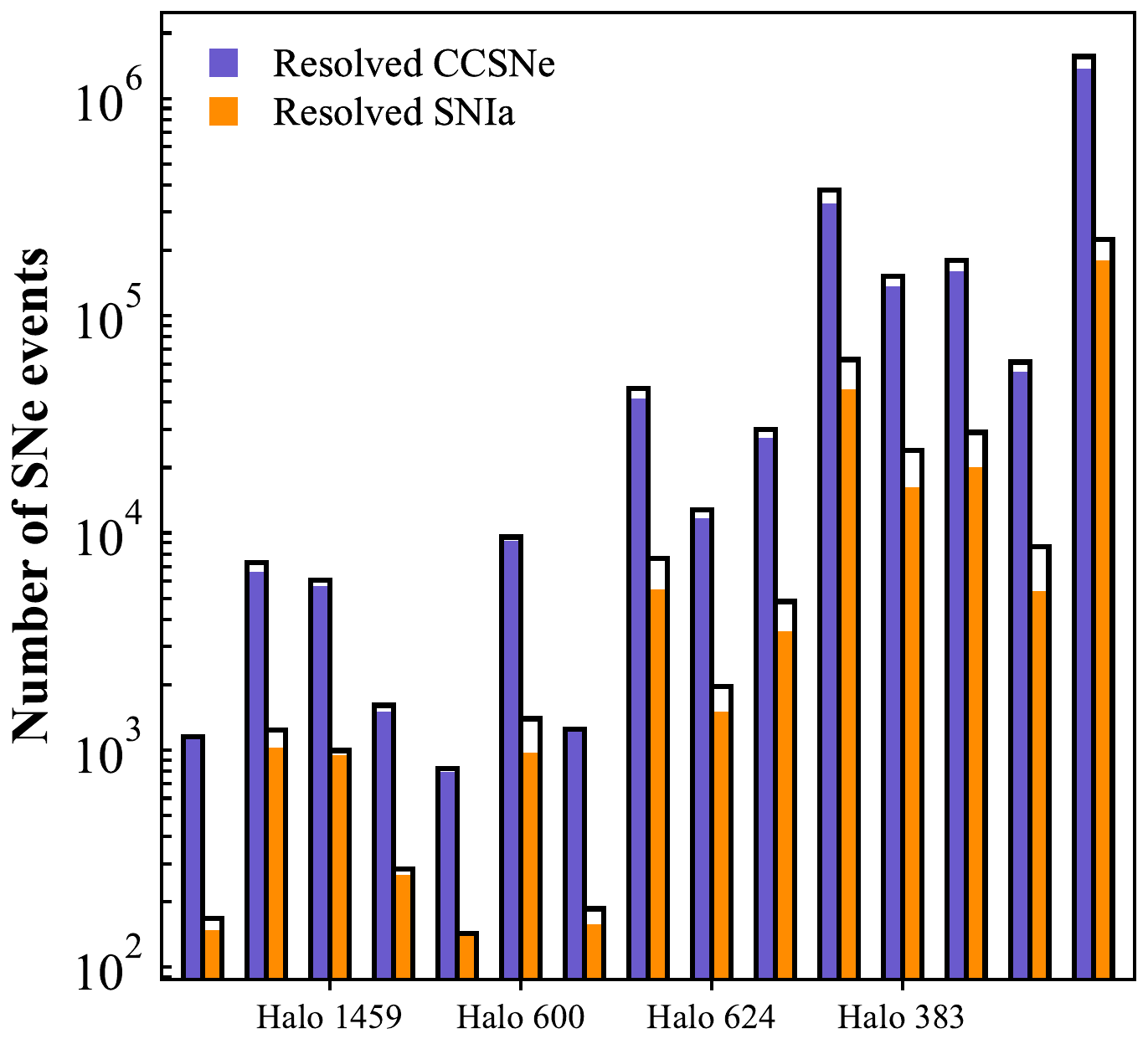}

    \caption{Fractions of resolved CCSNe (purple) and SNeIa (orange) explosions over the full cosmological history of each galaxy. Bars show the total number of SNe, with filling highlighting those for which we resolve the cooling radius (see the text). Each bar pair correspond to an individual dwarf galaxy, ordered with growing halo mass as in Table~\ref{tab:simulations}. Irrespective of the mass scale, $\geq 85\%$ of CCSNe, which dominate the SN feedback budget, are resolved.}
    \label{fig:SNstats}
\end{figure}

Resolving the cooling radius of SNe remnants is a key modelling milestone to accurately capture the momentum-build up during the Sedov-Taylor phase (e.g. \citealt{Kim2015}). Our injection scheme computes the cooling radius as 
\begin{equation}
  \label{eq:coolingradius}
  R_{\text{cool}}= 30.0 \, \pc \, (\frac{\nh}{\cmcube})^{-3/7} \, (\frac{E}{10^{51} \, \text{erg}})^{2/7} \, (\frac{Z}{\Zsol} + 0.01)^{1/7} \, ,
\end{equation}
where $\nh$ is the hydrogen density in the gas cell, $E$ is the total energy injected (this can be $\geq 10^{51} \, \text{erg}$ if more than one SN are exploding in the same timestep), and $Z$ is the gas metallicity (see \citealt{Hopkins2018}, appendix D for a derivation of this equation and e.g. \citealt{Cioffi1988, Thornton1998, Kim2015} for similar scalings).

Figure~\ref{fig:SNstats} shows the total number of SNe events in each simulated dwarf galaxy over their whole evolution (bar pair), divided between CCSNe (purple) and SNeIa (orange). The filling of the bar indicates `resolved' events where $R_{\text{cool}} \geq 6 \, \softening$ for which we directly inject thermal energy rather than momentum. 

All dwarf galaxies have more than $\geq 87\%$ of CCSNe events resolved, even as galaxy masses and ISM pressures increase (towards the right). A larger fraction of SNeIa are unresolved, most likely because these events can occur in lower-density environments, long after star formation when the adaptive resolution has been degraded. None the less, since CCSNe provide the overwhelming majority of the SN feedback budget, such statistics represent a significant achievement in the robustness of SN feedback modelling.  

\subsection{Radiative feedback}

Even if SN feedback is accurately captured as we have shown above, the coupling of this feedback with the surrounding ISM and the subsequent ability to drive galactic outflows depend on the gas conditions in which the explosions occur. These conditions in turn depend on other, pre-SN feedback channels, especially radiative feedback. 

For radiative feedback, the key length scale to resolve is the Str\"{o}mgren radius associated to the $\hii$ region around the star formation event. Requiring equilibrium between ionization and recombination of a hydrogen sphere, we obtain the Str\"{o}mgren radius (\citealt{Stromgren1939}) as 
\begin{equation}
  \label{eq:stromgrensphere}
  R_{\text{S}} = \left(\frac{3\, Q}{4\, \pi \nh^{2} \alpha_{B}}\right)^{1/3} \, .
\end{equation}
Here $Q$ is the production rate of hydrogen-ionizing (Lyman-Compton) photons and $\alpha_{B}$ is the case-B recombination rate of hydrogen. Note that this equation assumes spherical symmetry, a homogeneous gas distribution and neglects the respective contributions of helium and metallic ions to the ionization and recombination balance. None the less, this is enough to gain order-of-magnitude estimates of whether ionization fronts from photo-ionization feedback are resolved, since gas density is the strongest driver in equation~\eqref{eq:stromgrensphere}.

We can start by estimating an order of magnitude for $R_{\text{S}}$ when our star-formation algorithm spawns a stellar particle (see Section~\ref{sec:sec:sec:sf} for details). The density at which stellar particles are spawned peaks around $\approx 500 \, \mppercmcube$. Accounting for the depletion of $300\, \Msol$ into a stellar particle, and assuming a primordial mixture, we input $\nh = 200 \, \cmcube$ into equation~\eqref{eq:stromgrensphere}. Given our choice of SED (\citealt{Bruzual2003}), a zero-age $300\, \Msol$ stellar population has an ionizing rate of $Q \approx 10^{49} s^{-1}$ (see e.g. \citealt{Rosdahl2018}, appendix D for a visualization). Assuming the temperature-dependent $\alpha_{B}$ from \citet{Hui1997} evaluated at $T = \xScientific{2}{4}\, \K$ (see their appendix A), we obtain $R_{\text{S}} \approx 2 \, \pc$. This is already comparable to the maximum resolution of the simulation, $\softening \approx 3 \, \pc$. Pushing to $\nh = 10^3 \, \cmcube$ which is common for star formation events in our simulations, we find $R_{\text{S}} \approx 0.7 \, \pc$.

The choice of the SED (through $Q$), of the recombination rate ($\alpha_{B}$) or the assumptions of a primordial plasma will slightly modify these findings, but all these parameters enter Equation~\ref{eq:stromgrensphere} with weaker scalings than density. As a result, the Str\"{o}mgren sphere associated to the birth of a stellar particle is likely to be poorly resolved. This shortcoming is intrinsic to our star formation algorithm and cannot be avoided without modifying the very nature of the recipe (see e.g. \citealt{Katz2024MegPilot} for such a change). 

None the less, even if the Str\"{o}mgren sphere is unresolved at birth, the dynamics of the surrounding ISM could quickly alleviate this issue. Lowering the density to $\nh = 10 \, \cmcube$, for example, gives $R_{\text{S}} \approx 14 \, pc$ and a much better-resolved $\hii$ region. To test this, we identify young stellar particles with ages $\leq 5\, \Myr$ for each saved snapshot in the history of our individual galaxies. For each young stellar particle, we identify the density of their host gas cell and their ionizing output using our assumed SED at the particle's stellar age and metallicity. We then use Equation~\ref{eq:stromgrensphere} to compute $R_{\text{S}}$ (using the same $\alpha_{B}$ as previously) and consider a Str\"{o}mgren sphere to be resolved if $R_{\text{S}} \geq 6 \, \softening$. In this case, a spherical $\hii$ region would be captured by > 200 resolution elements, which is enough to accurately capture its ionization and temperature structure 
(e.g. \citealt{Deng2024RadFeedback}). We also compute the fraction of resolved Str\"{o}mgren spheres with $R_{\text{S}} \geq 4 \, \softening$, which is a more marginal case with $\approx$ 60 resolution elements per sphere.

\begin{figure}
  \centering
    \includegraphics[width=\columnwidth]{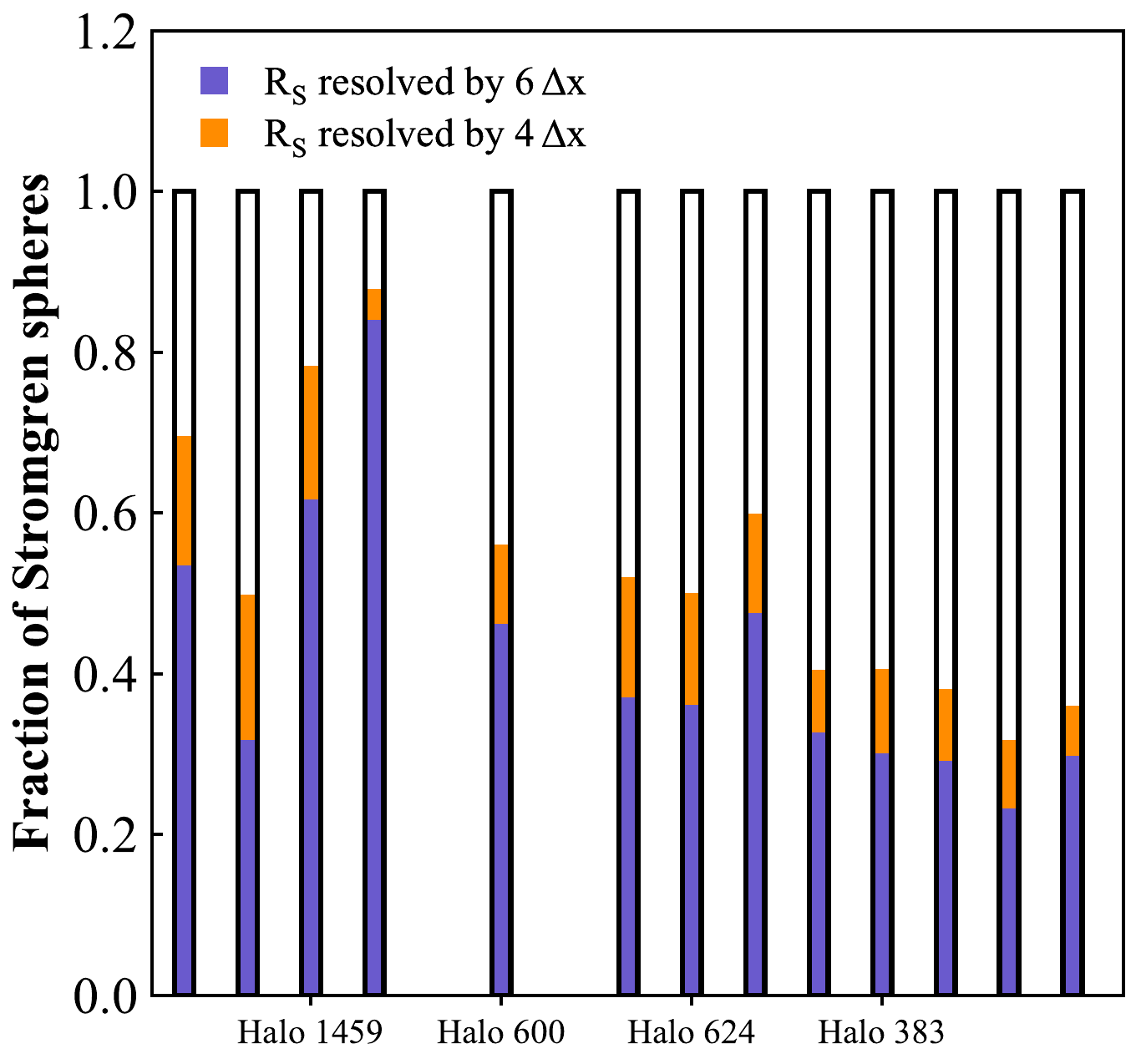}

    \caption{Fractions of resolved Str\"{o}mgren spheres for each dwarf galaxy along the cosmological history of each dwarf, with filling highlighting those for which we resolve the Str\"{o}mgren radius with six (blue) and four (orange) resolution elements (>200 and 60 elements per sphere, respectively). On average, $\approx 40\%$ of Str\"{o}mgren spheres are resolved by six $\softening$ across the suite, with fractions as low as $\approx 20\%$ for the most massive dwarf galaxies (towards the right). Two dwarf galaxies are absent as their saved snapshots do not sample young stars and $\hii$ regions.}
    \label{fig:sspheres}
\end{figure}

Figure~\ref{fig:sspheres} shows the obtained statistics, with bars showing the resolved fraction of Str\"{o}mgren spheres over the history of our dwarfs. Some galaxies are missing from the plot as their (short) formation histories and snapshot spacing does not allow sampling of $<5\, \Myr$ stars. Across the whole suite, we resolve very well $42\%$ of Str\"{o}mgren spheres on average, with the smallest fraction being $23\%$ for `Halo 383: GM early'. Assuming a less stringent requirement of $R_{\text{S}} \geq 4 \, \softening$; this average fraction climb to $52\%$ with a minimum of $31\%$. Similarly, shifting the age cut for `young' stars from 5 to 10 Myr leads to a $57\%$ average resolved fraction with a minimum $40\%$ fraction. 

Overall, unlike the situation for SNe for which the feedback modelling is nearly all resolved (Figure~\ref{fig:SNstats}), only a rough half of radiative feedback events are numerically well captured at our resolution. Furthermore, the fraction of resolved events falls rapidly as galaxies become more massive and their ISM denser. This is a clear area of improvement for radiation-hydrodynamics simulations aiming to explicitly model radiative feedback (see Section~\ref{sec:numerics:sspheres} for a discussion). 

\section{Burstiness of star formation} \label{app:sfhs}

\begin{figure}
  \centering
    \includegraphics[width=\columnwidth]{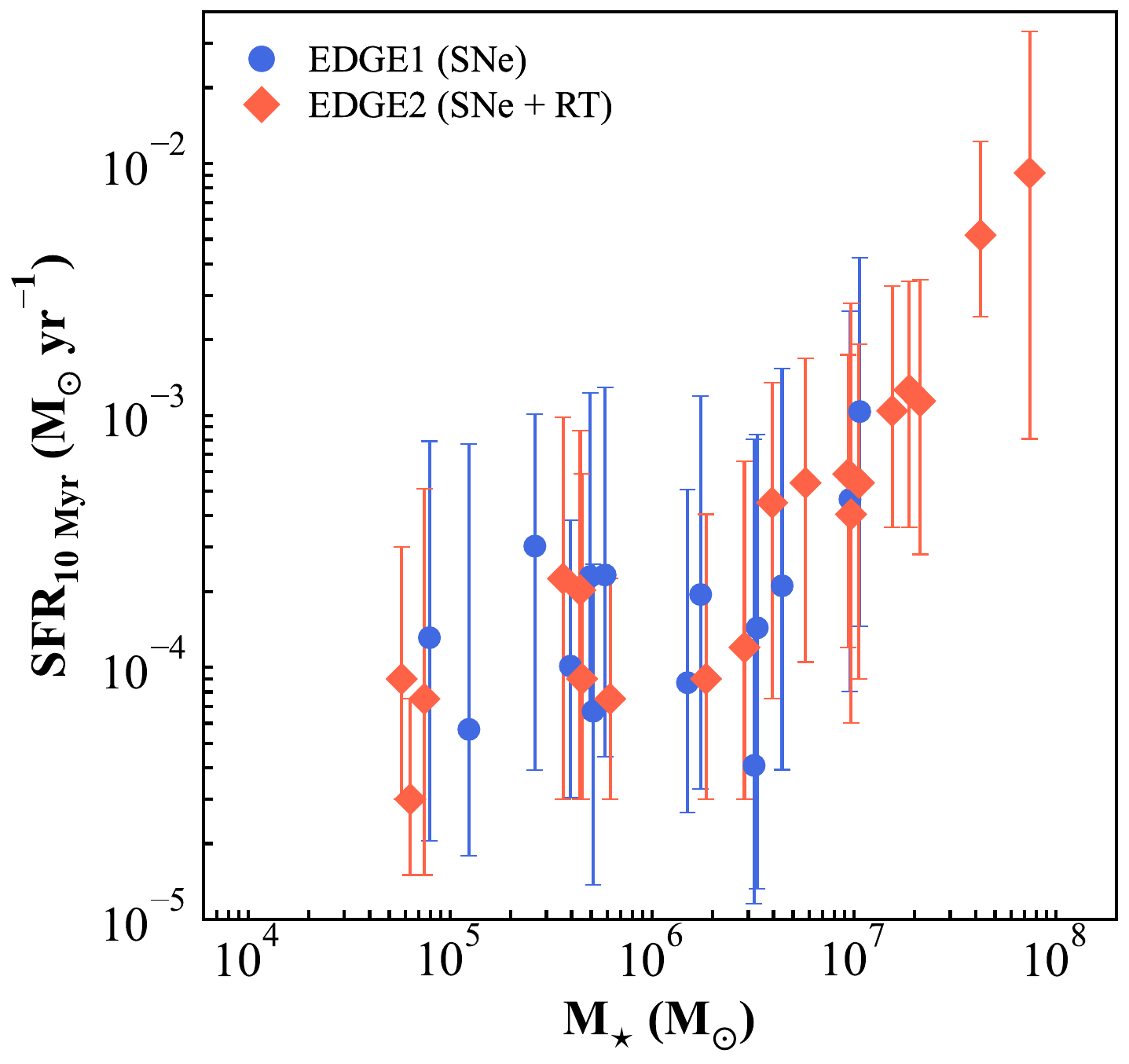}

    \caption{Median and 16-84 percentiles of the SFR over the full assembly history of each dwarf between $z=99$ and $z=0$. The Hubble-time-average SFRs are well converged between the \textsc{edge1} and \textsc{edge2} models (red and blue). With radiative feedback, \textsc{edge2} shows a less bursty star formation mode with a reduced scatter around the median.}
    \label{fig:sfrs}
\end{figure}

Figure~\ref{fig:sfrs} shows the distribution of SFRs for each dwarf galaxy over its full cosmological assembly history. Symbols show the median SFR, while the lines show the 16-84 percentiles of the distribution. We compute these statistics by computing the SFR averaged over 10 Myr for the Hubble time and eliminate vanishing SFRs. Only dwarf galaxies which have at least 10 snapshots with non-vanishing SFRs are shown on Figure~\ref{fig:sfrs}. (Some of our low-mass dwarf galaxies can have very short SFHs and billions of years of vanishing SFRs, leading to heavily biased median and 16-84 estimates.) 

The median SFRs are well converged between the \textsc{edge1} and \textsc{edge2} models (blue points and red diamonds, respectively) in line with the aligned $\Mstar$ reported in Section~\ref{sec:sec:mstarmhalo}. Computing the average extent of the 16-84 interval normalized by its median across each suite, \textsc{edge2} shows a reduction in scatter of $\approx 2$ compared to \textsc{edge1}. Such reduced burstiness for star formation is expected as a by-product of radiative feedback and aligns with the much calmer regulation mode in \textsc{edge2} (Section~\ref{sec:rtimpact}).

\section{Star-formation and supernova statistics} \label{app:sfstats}

\begin{figure*}
  \centering
    \includegraphics[width=\textwidth]{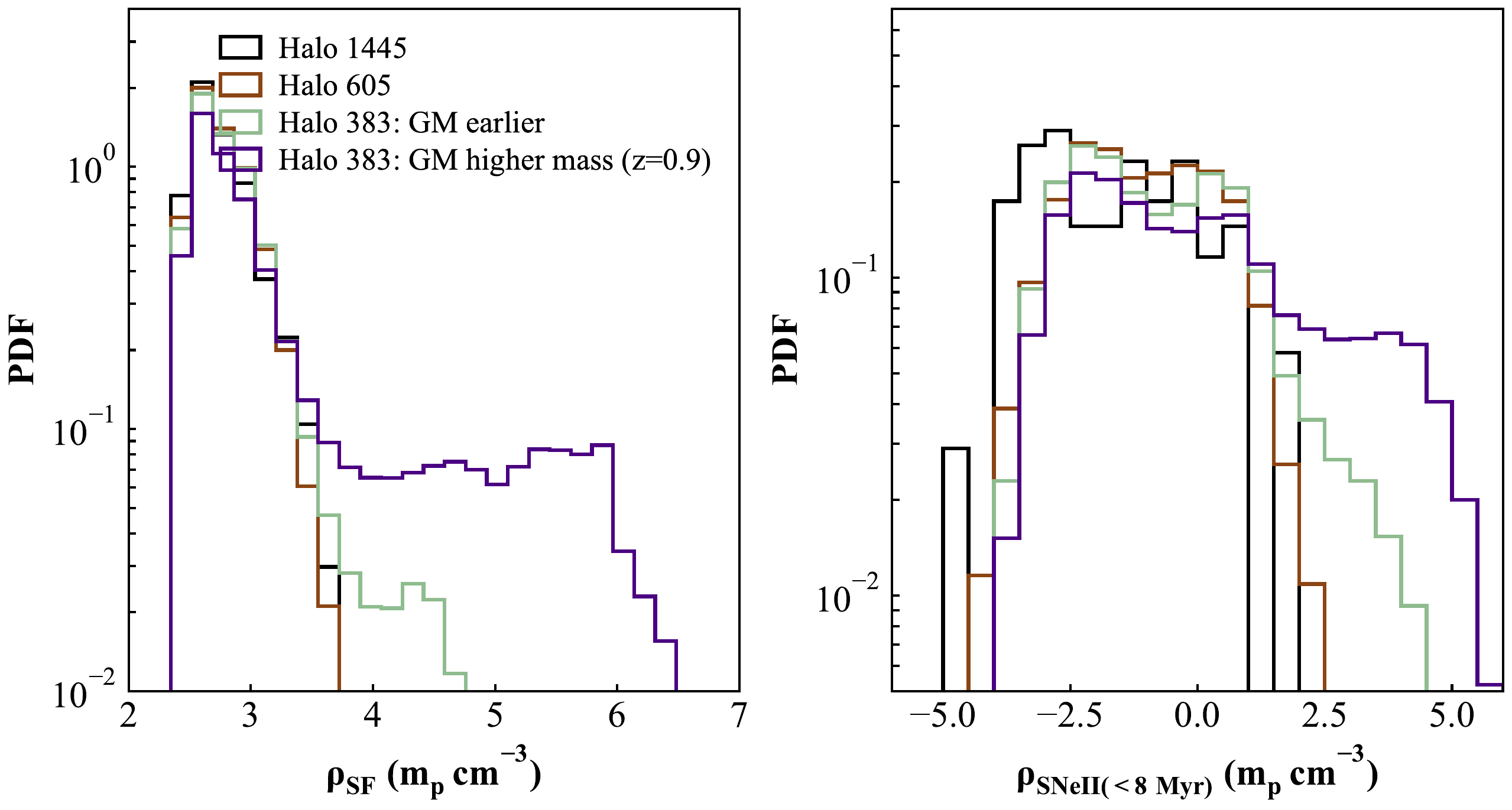}

    \caption{Density distribution of star-formation (left) and SNe (right) events across the history of four \textsc{edge2} dwarf galaxies. As objects become more massive, star formation and SN events at increasingly high densities start to appear.}
    \label{fig:sfsnstats}
\end{figure*}

\begin{figure*}
  \centering
    \includegraphics[width=\textwidth]{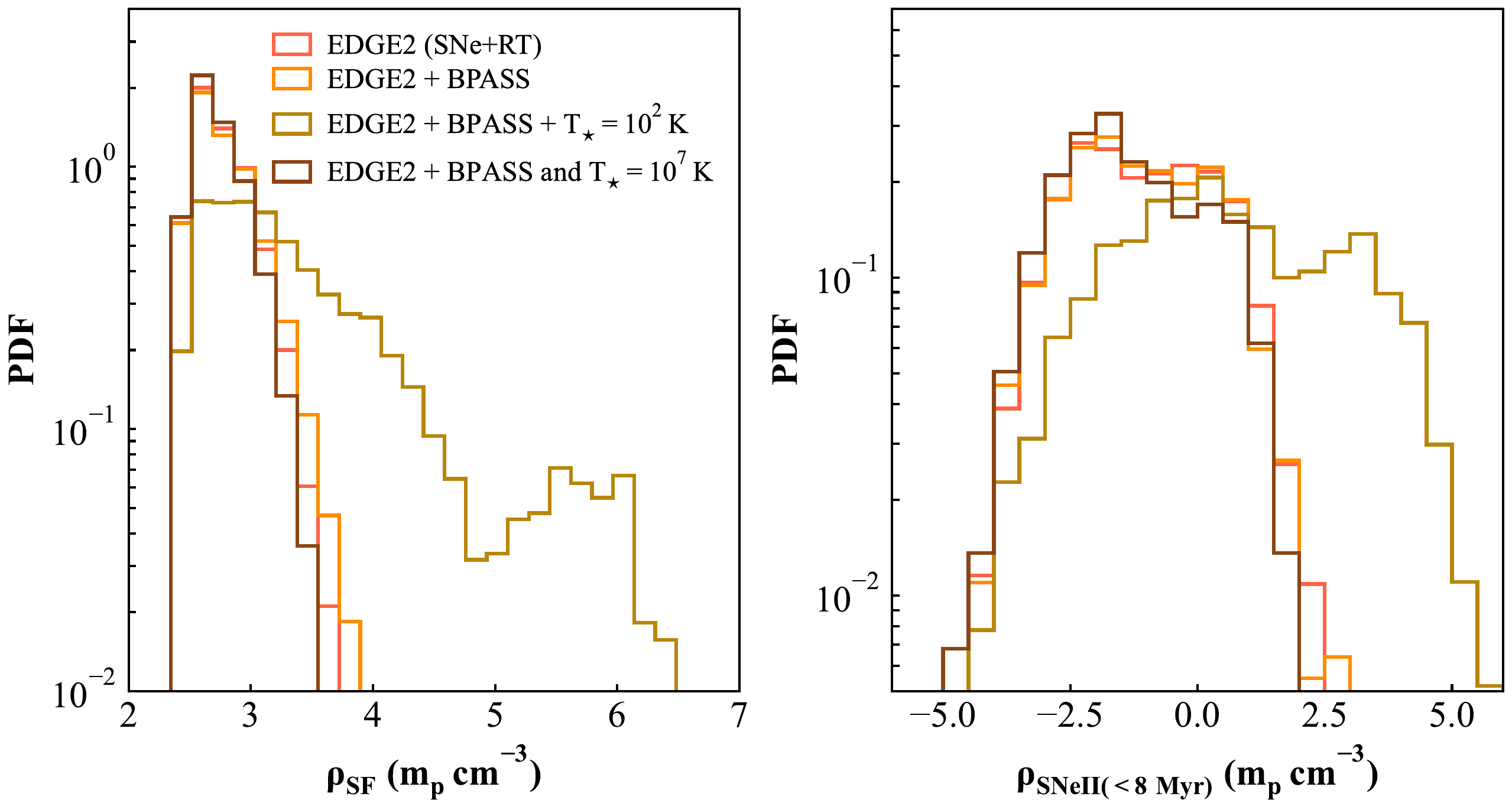}

    \caption{Same as Figure~\ref{fig:sfsnstats}, but for the four re-simulations of `Halo 605' that vary input parameters as described in Section~\ref{sec:numerics}. For the same galaxy, forbidding star formation in warm gas ($T_{\star} = 10^2\, \K$, gold) recreates a tail of high-density star formation events by reducing the clustering of star formation and subsequent SN feedback.}
    \label{fig:sfsnstats605}
\end{figure*}

To visualize the consequences of the much denser ISM of the \textsc{edge2} model highlighted in Section~\ref{sec:rtimpact}, Figure~\ref{fig:sfsnstats} shows the densities at which stellar particles are formed (left), and at which the CCSNe from a stellar particle explode (i.e. SNe younger than $ 8\, \, \Myr$ which corresponds to the main sequence lifetime of a $\approx 20\, \Msol$ massive star). We show the histograms of all SF and SNe events across four \textsc{edge2} galaxies that span the full range of $\Mstar$. (We verified that conclusions are unchanged if using other galaxies.) Unfortunately, star formation and SNe statistics were not recorded in the \textsc{edge1} model, so a direct comparison is not possible. 

As expected, no star formation event (left-hand panel) occurs below our density threshold for SF ($\geq 300 \, \mppercmcube$). For low-$\Mstar$ objects (black and brown), there is a clear peak around $\approx 10^3 \, \mppercmcube$ that quickly tapers off. As dwarfs galaxies increase in $\Mstar$ (green, and purple), however, the tail of star-forming events extends to higher densities reaching ($\geq 10^6 \, \mppercmcube$). 

The direct consequence is that SNe explosions occur in increasingly dense gas (right-hand panel). The amount of momentum emerging from the Sedov-Taylor phase of a SN remnant only weakly depends on the background density (e.g. \citealt{Kim2015}). But since our spatial scales are ultimately limited by numerical resolution, explosions in an increasing dense medium couple this momentum to an ever-increasing mass, thus yielding smaller velocities and weaker outflows\footnote{For a typical Sedov-Taylor momentum ($p_{SN} = 5 \times 10^5 \, \Msol \, \kmpers$) exploding in a $3\, \pc$-cell with $\rho = 10^5\, \, \mppercmcube$, the generated velocity is $7\, \kmpers$. This is insufficient to escape the gravitational potential well of even our small dwarfs ($\vcirc \geq 10 \, \kmpers$).}. 

Part of these effects are driven by our algorithms, which we illustrate now. In particular, our star formation algorithm requires gas to be both sufficiently dense and sufficiently cold to trigger star formation (Section~\ref{sec:sec:sec:sf}). Figure~\ref{fig:sfsnstats605} shows the same statistics as Figure~\ref{fig:sfsnstats} for the runs described in Section~\ref{sec:numerics}, in which the temperature criteria are raised and lowered. In particular, for a fixed object, a lower $T_{\star}$ drives an extended tail of star formation events and SNe at high densities (gold), while a higher $T_{\star}$ truncates both to lower densities. 

These results should be attributed to interactions between radiative feedback and our star formation prescription. Following a star formation event, the dense gas surrounding the newborn stellar particle is immediately heated to $T \geq T_{\star} = 10^3\, \K$ by radiative feedback. Any gas in the immediate vicinity of a star-forming event is thus prevented from forming new stars, staying warm and dense until the first SNe can clear it ($\approx 5\, \Myr$ later). This suppresses the clustering of star formation and SNe, making it harder to generate powerful outflows. 

Even though we isolated the $T_{\star}$ parameter here, we stress that this parameter is not the only driver of this behaviour. Rather, the combination of $T_{\star}$, the heating power of stellar radiation, and the surrounding ISM densities driving gas cooling and collapse all contribute towards setting the distribution of star-forming densities. For example, a harder SED will raise more gas over a fixed $T_{\star}$ and drive a slightly longer tail of star formation at higher densities (Figure~\ref{fig:sfsnstats605}, orange versus red). And, at fixed SED and $T_{\star}$, a more massive galaxy with a deeper gravitational potential well and higher ISM densities will retain high densities more effectively and compound this effect (Figure~\ref{fig:sfsnstats}, purple versus brown).  

Combined with under-resolved radiative feedback events (Appendix~\ref{app:resolvedfeedback}), these findings highlight some key uncertainties remaining in our radiation-hydrodynamics simulations. None the less, while these effects drive large changes on the statistics of star formation and SNe events, they have a much more moderate impact on the integrated $\Mstar$ and other observables (Sections~\ref{sec:scalingrelations} and~\ref{sec:numerics}). Observables such as rest-frame optical emission lines are likely more sensitive to changes in the structure of the ISM and its star-forming regions. In future iterations of the \textsc{edge} model, we will continue to improve our implementations and compare with these vital observational clues. 


\bsp	
\label{lastpage}
\end{document}